\documentclass[a4paper,11pt]{article}%
\pdfoutput=1

\usepackage{amsmath,amssymb,amsthm,latexsym}
\usepackage{hyperref}
\usepackage{epsfig,subfigure}
\usepackage{fullpage}
\usepackage{amsmath}
\usepackage{amsfonts}
\usepackage{amssymb,bbold}
\usepackage{graphicx}%
\providecommand{\U}[1]{\protect\rule{.1in}{.1in}}
\theoremstyle{plain}
\newtheorem{thm}{Theorem}[section]

\newtheorem{remark}[thm]{Remark}

\newcommand{\R}{{\mathbb R}}

\newcommand{\Rn}{\R^n}
\newcommand{\bfx}{\mathbf{x}}
\newcommand{\bfy}{\mathbf{y}}

\newcommand{\dt}{ \, {\rm d} t}
\newcommand{\dx}{ \, {\rm d} x}
\newcommand{\dy}{ \, {\rm d} y}
\newcommand{\dtau}{ \, {\rm d} \tau}

\newcommand{\dbfy}{ \, {\rm d} \bfy}
\newcommand{\One}{\mathbb{1}}

\newcommand{\Om}{\Omega}
\newcommand{\barrho}{\bar{\rho}}
\newcommand{\trho}{\tilde{\rho}}
\newcommand{\Omrho}{\Om_{\barrho}}
\newcommand{\D}{D}

\newcommand{\g}{g}
\newcommand{\f}{f}

\newcommand{\vs}{L}
\newcommand{\dl}{d_1}
\newcommand{\dr}{d_2}
\newcommand{\mr}{r_M}
\newcommand{\gc}{g_c}
\newcommand{\tgc}{\tilde{g}_c}
\newcommand{\ra}{\alpha}
\newcommand{\rb}{\beta}
\newcommand{\rc}{\gamma}
\newcommand{\CM}{C_2}
\newcommand{\md}{m_d}
\newcommand{\dli}{\dl^{(i)}}
\newcommand{\drj}{\dr^{(j)}}

\newcommand{\CalA}{{\mathcal{A}}}

\newcommand{\CalP}{{\mathcal{P}}}

\newcommand*\samethanks[1][\value{footnote}]{\footnotemark[#1]}

\begin{document}

\title{Swarm equilibria in domains with boundaries}

\author{R.C. Fetecau \thanks{Department of Mathematics, Simon Fraser University, 8888 University Dr., Burnaby, BC V5A 1S6, Canada}
\and M. Kovacic \samethanks
}

 
\maketitle

\begin{center}
\textbf{Abstract}
\end{center}

\begin{quote}
We study equilibria in domains with boundaries for a first-order aggregation model that includes social interactions and exogenous forces. Such equilibrium solutions can be connected or disconnected, the latter consisting in a delta concentration on the boundary and a free swarm component in the interior of the domain. Equilibria are stationary points of an energy functional, and stable configurations are local minimizers of this functional. We find a one-parameter family of disconnected equilibrium configurations which are not energy minimizers; the only stable equilibria are the connected states. Nevertheless,  we demonstrate that in certain cases the dynamical evolution, along the gradient flow of the energy functional, tends to overwhelmingly favour the formation of (unstable) disconnected equilibria.
\end{quote}



\section{Introduction}

Research in mathematical modelling for self-organizing behaviour or swarming has surged in recent years. An aggregation model that has attracted a great amount of interest is given by the following integro-differential equation in $\Rn$:
\label{sect:intro}
\begin{subequations}
\label{eqn:model}
\begin{gather}
\rho_{t}+\nabla\cdot(\rho v)=0\label{eqn:cont}\\
v=-\nabla K\ast\rho - \nabla V. \label{eqn:v}
\end{gather}
\end{subequations}
Here  $\rho$ represents the density of the aggregation, $K$ is an interaction potential, and $V$ is an external potential. The asterisk $\ast$ denotes convolution. Typically, the interaction potential $K$ models symmetric inter-individual social  interactions such as long-range attraction and short-range repulsion.

Model \eqref{eqn:model} appears in various contexts related to swarming and social aggregations, and the associated literature is vast and covers a wide range of topics:  modelling and pattern formation \cite{M&K,TBL,KoSuUmBe2011,LeToBe2009,FeHuKo11}, well-posedness of solutions \cite{BodnarVelasquez2, BertozziLaurent, BertozziCarilloLaurent}, long-time 
behaviour of solutions \cite{FeHu13,LeToBe2009}, blow-up (in finite or infinite time) by mass concentration \cite{FeRa10,BertozziCarilloLaurent, HuBe2010}. The equation also arises in a number of other applications such as granular media~\cite{Toscani2000,CaMcVi2006}, self-assembly of nanoparticles~\cite{HoPu2005,HoPu2006}, Ginzburg-Landau vortices~\cite{E1994,DuZhang03}, molecular dynamics simulations of matter~\cite{Haile1992} and opinion dynamics \cite{MotschTadmor2014}. 

In this paper we study the aggregation model \eqref{eqn:model} in domains with boundaries. Despite the extensive literature on model \eqref{eqn:model} in free space, there has been only a handful of works that consider the presence of boundaries  \cite{BeTo2011,WuSlepcev2015, CarrilloSlepcevWu2016}. These papers are motivated by physical/biological scenarios where the environment involves an obstacle or an impenetrable wall; in the locust model from \cite{ToDoKeBe2012} for example, such an obstacle is the ground. We assume in this work that the presence of  boundaries limits the movement in the following way \cite{WuSlepcev2015, CarrilloSlepcevWu2016}: once particles/individuals meet the boundary, they do not exit the domain, but move instead freely along it. The precise mathematical formalism of this ``slip, no-flux" boundary condition is elaborated below.

Consider the aggregation model \eqref{eqn:model} confined to a closed domain $\Om \subset \Rn$. Suppose that $\Om$ has a smooth $C^1$ boundary with outward normal vector $\nu_x$ at $x \in \partial \Om$. The geometric confinement constrains the velocity field as follows. At points in the interior of $\Om$, or at points on the boundary where the velocity vector, computed with \eqref{eqn:v}, points inward  ($v \cdot \nu_x  \leq 0$), no modification is needed and the velocity is given by \eqref{eqn:v}. On the other hand, for points on the boundary where the velocity computed with \eqref{eqn:v} points {\em outward} ($v \cdot \nu_x >0$), its projection on the tangent plane to the boundary is considered instead. 

The model in domains with boundaries is then given by:
\begin{subequations}
\label{eqn:modelb}
\begin{gather}
\rho_{t}+\nabla\cdot(\rho v)=0\label{eqn:contp}\\
v=P_x(-\nabla K\ast\rho - \nabla V), \label{eqn:vp}
\end{gather}
\end{subequations}
where
\begin{equation}
P_x \xi=
\begin{cases}
\xi & \text{ if } x \not\in \partial \Omega \;\text{ or } \; x \in \partial \Omega \text{ and } \xi \cdot \nu_x \leq 0 \\
\Pi_{\partial \Om} \, \xi, & \text{ otherwise}.
\end{cases}
\label{eqn:proj}
\end{equation}
Here $\Pi_{\partial \Om}$ denotes the projection on the tangent plane to the boundary. Note that  solutions to \eqref{eqn:modelb} conserve the total mass, however the linear momentum is no longer preserved (as opposed to the model in free space). The latter observation has important implications for the long time behaviour of the solutions, as discussed later in the paper.
  
The well-posedness of weak measure solutions of \eqref{eqn:modelb} has been investigated recently in \cite{WuSlepcev2015, CarrilloSlepcevWu2016} in the framework of gradient flows in spaces of probability measures \cite{AGS2005, Figalli_etal2011}. The setting of {\em measure}-valued solutions in these works is absolutely essential in this context, for various reasons. First,  mass accumulates on the boundary of the domain and solutions develop Dirac delta singularities there. Second, the measure framework is the appropriate setup for connecting the PDE model with its discrete/particle approximation. In regard to the latter, by approximating the initial density $\rho_0$ with a finite number of delta masses, \eqref{eqn:modelb} reduces to an ODE system, which then can be studied on its own. In \cite{CarrilloSlepcevWu2016}, the authors establish several important properties of such particle approximations. One is the well-posedness of the approximating particle system where, due to the discontinuities of the velocity field at the boundary, the theory of differential inclusions \cite{Filippov1988,Cortes2008} is being employed. Another is the rigorous limit of the discrete approximation as the number of particles approach infinity; this limit is shown to be a weak measure solution of the PDE model \eqref{eqn:modelb}. 

The focus of the present paper is equilibrium configurations of model \eqref{eqn:modelb}. A density $\barrho$ is an equilibrium if the velocity \eqref{eqn:vp} vanishes everywhere on its support:
\begin{equation}
\label{eqn:ss}
P_x(-\nabla K\ast\barrho - \nabla V) = 0 \qquad \text{ in } \text{supp}(\barrho).
\end{equation}
We note however that at points on the boundary, the unprojected velocity (i.e., $-\nabla K*\barrho - \nabla V$) may not be zero; by \eqref{eqn:proj} it can have a nonzero normal component that is pointing outward. This scenario is akin to a falling object hitting a surface, when there is still a force acting on it but there is nowhere to go. 

Model \eqref{eqn:modelb} is a gradient flow and its equilibria are stationary points of the energy functional. We use the framework developed by Bernoff and Topaz  \cite{BeTo2011} to look for these stationary densities.
We also investigate their stability; given the variational formulation,  stable equilibria can be characterized as local minima of the energy. 
Most of the paper concerns a specific interaction potential, consisting of Newtonian repulsion and quadratic attraction \cite{FeHuKo11,FeHu13}. The main advantage of using this potential is that the equilibria must have constant densities away from the boundary, which restricts the possible equilibrium configurations and simplifies the calculations. 

The present paper contains the first systematic study of equilibria for model \eqref{eqn:modelb} in two dimensions; we note here that the results in  \cite{BeTo2011} consider only one and quasi-two dimension cases. Of particular relevance is a family of two-component equilibria that we found in our study (in both one and two dimensions), consisting of one swarm component on the boundary and another in the interior of the domain. These two-component equilibria can be further differentiated as connected and disconnected, depending whether the two components are adjacent or not. We find that none of the disconnected equilibria are local minima of the energy. In contrast, some connected configurations can be shown to be local (and in some cases global) energy minimizers.


Nevertheless, we show that starting from a large class of initial densities, solutions to \eqref{eqn:modelb} do evolve into such (unstable) disconnected equilibria that are not local energy minimizers. While unusual, this behaviour has been observed in continuum mechanics systems wherein singularities form which act as barriers preventing further energy decrease  \cite{Ball1992,BaHoJaPeSw1991,SwartHolmes1992}. Describing and understanding this behaviour for model \eqref{eqn:modelb} is one of the main goals of this paper.

 

The summary of the paper is as follows. Section \ref{sect:prelim} presents some background on model \eqref{eqn:modelb}. In Section \ref{sect:oned} we study the one-dimensional problem on half-line. We find explicit expressions for the equilibria and make various investigations of the dynamical model to quantify on how these equilibria are being reached. Section \ref{sect:twod} considers the two dimensional problem on half-plane. We compute the connected and disconnected equilibria and investigate their stability. Finally, we present details on the numerical implementations.


\section{Preliminaries}
\label{sect:prelim}

\paragraph{Well-posedness and gradient flow formulation.} The well-posedness of weak measure solutions to model \eqref{eqn:modelb} has been established recently in \cite{WuSlepcev2015} and \cite{CarrilloSlepcevWu2016}. The functional setup in these works consists in the space $\CalP_2(\Om)$ of probability measures on $\Om$ with finite second moment, endowed with the $2$-Wasserstein metric.  Under appropriate assumptions on the domain $\Om$ and on the potentials $K$ and $V$, it is shown that the initial value problem for \eqref{eqn:modelb} admits a weak measure solution $\rho(t)$ in $\CalP_2(\Om)$. We refer to \cite{WuSlepcev2015, CarrilloSlepcevWu2016} for specific details on the well-posedness theorems and proofs, we only highlight here the facts that are relevant for our work.

It is a well-established result that  the aggregation model in free space (model \eqref{eqn:model}) can be formulated as a gradient flow on the space of probability measures $\CalP_2(\Om)$ equipped with the $2$-Wasserstein metric \cite{AGS2005}.  A key result in \cite{WuSlepcev2015, CarrilloSlepcevWu2016} is that such an interpretation exists for model \eqref{eqn:modelb} as well. Specifically, consider the energy functional
\begin{equation}
\label{eqn:energy}
E[\rho] = \frac{1}{2} \int_\Om \int_\Om K(x-y) \rho(x) \rho(y) \dx  \dy + \int_\Om V(x) \rho(x) \dx,
\end{equation}
where the first term represents the interaction energy and the second is the potential energy\footnote{Note that throughout the present paper  $\int \varphi(x) \rho(x) \dx$ denotes the integral of $\varphi$  with respect to the measure $\rho$, regardless of whether $\rho$ is absolutely continuous with respect to the Lebesgue measure.}. 

The weak measure solution $\rho(x,t)$ to 
model \eqref{eqn:modelb} is shown to satisfy the following energy dissipation equality \cite{CarrilloSlepcevWu2016}:
\begin{equation}
\label{eqn:gflow}
E[\rho(t)] - E[\rho(s)] = - \int_s^t \int_{\Om} |P_x(-\nabla K \ast \rho(x,\tau) - \nabla V(x))|^2 \rho(x,\tau) \dx,
\end{equation}
for all $0 \leq s \leq t < \infty$. Equation \eqref{eqn:gflow} is a generalization of the energy dissipation for the model in free space \cite{Figalli_etal2011}. Characterization of equilibria of \eqref{eqn:model} as ground states of the interaction energy \eqref{eqn:energy} has been a very active area of research lately \cite{Balague_etalARMA,ChFeTo2015,CaCaPa2015,SiSlTo2015}.

The authors in \cite{CarrilloSlepcevWu2016} use particle approximations of the continuum model \eqref{eqn:modelb} as an essential tool to show the existence of gradient flow solutions. The method consists in approximating an initial density $\rho_0$ by a sequence $\rho_0^N$ of delta masses supported at a discrete set of points. For $N$ fixed, the evolution of model \eqref{eqn:modelb} with discrete initial data $\rho_0^N$ reduces to a system of ordinary differential equations, for which ODE theory can be applied. The ODE system governs the evolution of the characteristic paths (or particle trajectories) which originate from the points in the discrete support of $\rho_0^N$.  Hence, the solution $\rho^N(t)$ consists of delta masses supported at a discrete set of characteristic paths. The key ingredient in the analysis is to find a stability property of solutions $\rho^N$ with respect to initial data $\rho_0^N$ and show that in the limit $N\to \infty$, $\rho^N$ converges (in the Wasserstein distance) to a weak measure solution of \eqref{eqn:modelb} with initial data $\rho_0$. This is one of the major results established in \cite{CarrilloSlepcevWu2016}.

\paragraph{Equilibria and energy minimizers.} The authors in \cite{BeTo2011} study the energy functional \eqref{eqn:energy} and find conditions for critical points to be energy minimizers. We review briefly the setup there.

First note that the dynamics of model \eqref{eqn:modelb} conserves mass:
 \begin{equation}
\label{eqn:massM}
\int_{\Om} \rho(x,t) \dx = M \qquad \text{ for all } t\geq 0.
\end{equation}
Hence, in what follows it is sufficient to consider zero-mass perturbations of a fixed equilibrium.

Consider an equilibrium solution $\barrho$ with mass $M$ and support $\Omrho \subset \Om$, and take a small perturbation $\epsilon \trho$ of zero mass:
\[
\rho(x) = \barrho(x) + \epsilon \trho(x),
\]
where
\begin{subequations}
\label{eqn:massc}
\begin{gather}
\int_{\Om} \barrho(x) \dx =M\label{eqn:massrho},\\
\int_{\Om} \trho(x) \dx =0. \label{eqn:masstrho}
\end{gather}
\end{subequations}

Since the energy functional is quadratic in $\rho$, one can write:
\[
E[\rho] = E[\barrho] + \epsilon E_1[\barrho,\trho] + \epsilon^2 E_2[\trho,\trho],
\]
where $E_1$ denotes the first variation:
\begin{equation}
\label{eqn:1stvar}
E_1[\barrho,\trho] =  \int_\Om \left[ \int_\Om K(x-y) \barrho(y) \dy + V(x) \right]  \trho(x) \dx,
\end{equation}
and $E_2$ the second variation:
\begin{equation}
\label{eqn:2ndvar}
E_2[\trho,\trho] =  \frac{1}{2} \int_\Om \int_\Om K(x-y) \trho(x) \trho(y) \dx \dy.
\end{equation}
Using the notation 
\begin{equation}
\label{eqn:Lambda}
\Lambda(x) = \int_{\Omrho} K(x-y) \barrho(y) \dy + V(x), \qquad \text{ for } x \in \Om,
\end{equation}
one can also write the first variation as 
\begin{equation}
\label{eqn:1stvarL}
E_1[\barrho,\trho] =  \int_\Om  \Lambda(x)  \trho(x) \dx. 
\end{equation}

Two classes of perturbations are considered in \cite{BeTo2011}: perturbations $\trho$ supported in $\Omrho$ (first class), and general perturbations $\barrho$ in the domain $\Om$ (second class). Perturbations of the first class are a subset of the perturbations of second class.
\smallskip

Start by taking perturbations of first class.  Since $\trho$ changes sign in $ \Omrho$, for $\barrho$ to be a critical point of the energy, the first variation must vanish. From \eqref{eqn:1stvarL}, given that perturbations $\trho$ are arbitrary and satisfy  \eqref{eqn:masstrho}, one finds that $E_1$ vanishes provided $\Lambda$ is constant in $\Omrho$, i.e., 
\begin{equation}
\label{eqn:equilsup}
\Lambda(x) = \lambda, \qquad \text{ for } x \in \Omrho.
\end{equation}
The (Lagrange) multiplier $\lambda$ is given a physical interpretation in \cite{BeTo2011}: it represents the energy per unit mass felt by a test mass at position $x$ due to interaction with the swarm in $\barrho$ and the exogenous potential. Indeed this interpretation is valid for all points $x$ by considering $\Lambda(x)$ as the energy per unit mass felt by a test mass at position $x$. This interpretation is critical for the study in \cite{BeTo2011}, as well as for the present paper.

Equation \eqref{eqn:equilsup} represents a necessary condition for $\barrho$ to be an equilibrium. For $\barrho$ that satisfies \eqref{eqn:equilsup} to be a local minimizer with respect to the first class of perturbations, the second variation \eqref{eqn:2ndvar} must be positive. In general, the sign of $E_2$ cannot be assessed easily.

Consider now perturbations of the second class. Since perturbations $\trho$ must be non-negative in the complement $\Omrho^c = \Om \setminus \Omrho$, it is shown in \cite{BeTo2011} that a necessary and sufficient condition for $E_1 \geq 0$ is 
\begin{equation}
\label{eqn:equilcomp}
\Lambda(x) \geq \lambda, \qquad \text{ for } x \in \Omrho^c.
\end{equation}
The interpretation of \eqref{eqn:equilcomp} is that transporting mass from $\Omrho$ into its complement $\Omrho^c$ increases the total energy \cite{BeTo2011}.

In summary, a critical point $\barrho$ for the energy satisfies the Fredholm integral equation \eqref{eqn:equilsup} on its support. Also, $\barrho$ is a local minimizer (with respect to the general, second class perturbations) if it satisfies \eqref{eqn:equilcomp}. 

As discussed in \cite{BeTo2011}, the support $\Omrho$ of an equilibrium density has in general multiple disconnected components. Assuming $m$ disjoint, closed and connected components $\Om_i$, $i=1,\dots,m$, one can write 
\begin{equation}
\label{eqn:multiComp}
\Omrho = \Om_1 \cup \Om_2 \cup \dots \cup \Om_m, \qquad \Om_i \cap \Om_j = \emptyset, \quad i \neq j.
\end{equation}
In \cite{BeTo2011}, a swarm equilibrium is defined as a configuration in which $\Lambda$ is constant in every component of the swarm, i.e.,
\begin{equation}
\label{eqn:equilsup-gen}
\Lambda(x) = \lambda_i, \qquad \text{ for } x \in \Om_i, \quad i=1,\dots,m.
\end{equation}
Moreover, a swarm minimizer is defined there as a swarm equilibrium which satisfies
\begin{equation}
\label{eqn:equilcomp-gen}
\Lambda(x) \geq \lambda_i, \qquad \text{ in some neighbourhood of each } \Om_i.
\end{equation}
Following the interpretation of $\Lambda$ given above, \eqref{eqn:equilcomp-gen} means that an infinitesimal redistribution of mass in a neighbourhood of $\Om_i$ increases the energy.

Multiple connected equilibria of model \eqref{eqn:modelb} is a major focus of the present study. To find such equilibria we look for solutions of \eqref{eqn:equilsup-gen}. To set the ideas right however, the following remark is in order.
\begin{remark}
\label{rmk:Lambda-const}
We point out that condition \eqref{eqn:equilsup-gen} is only a {\em necessary} condition for $\barrho$ to be an equilibrium of \eqref{eqn:modelb}. Indeed, consider a density $\barrho$ that satisfies \eqref{eqn:equilsup-gen} and check whether it satisfies the equilibrium condition \eqref{eqn:ss}. By \eqref{eqn:equilsup-gen}, equation \eqref{eqn:ss} is indeed satisfied in every component $\Om_i$ that lies in the interior of $\Om$ (the projection plays no role there). However, consider a component $\Om_i$ of the swarm that lies {\em on} the boundary of the physical domain $\Om$. The component $\Om_i$ can be for instance a codimension one manifold, such as a line in $\R^2$; our numerical investigations in Section \ref{sect:twod} focus on this example in fact. Since $\Lambda(x)$ is constant on $\Om_i \subset \partial \Om$, we infer that the tangential component to $\partial \Om$ of  $\nabla \Lambda$ is zero at any point $x \in \Om_i$. Consequently, by  \eqref{eqn:Lambda}, we conclude that the unprojected velocity at $x$ (c.f., \eqref{eqn:v}) is {\em normal} to $\partial \Om$. For an equilibrium solution, this normal component must point {\em into} $\partial \Om$ ($v \cdot \nu_x >0 $) -- see \eqref{eqn:vp} and \eqref{eqn:proj}, however one cannot infer this condition from \eqref{eqn:equilsup-gen}. Section \ref{sect:twod} provides examples where solutions to \eqref{eqn:equilsup-gen} do not yield equilibria, precisely because the velocity at some points on the boundary is directed toward the interior of $\Om$, and thus the steady state condition \eqref{eqn:ss} fails.
\end{remark}


\paragraph{Newtonian repulsion and quadratic attraction.}
The present study focuses on a specific interaction potential $K$ given by 
\begin{equation}
\label{eqn:intpot}
K(x) = \phi(x) + \frac{1}{2} |x|^{2},
\end{equation}
where $\phi(x)$ is the free-space Green's function for the negative Laplace
operator $-\Delta$:
\begin{equation}
\phi(x)=
\begin{cases}
-\frac{1}{2}|x|, & n=1\\
-\frac{1}{2\pi}\ln|x|, & n=2.
\end{cases}
\label{eqn:phi}
\end{equation}
Potentials in the form \eqref{eqn:intpot}, consisting of Newtonian repulsion and quadratic attraction, have been considered in various recent works \cite{FeRa10, FeHuKo11, FeHu13, HuFe2013}. The remarkable property of such potentials is that they lead to compactly supported equilibrium states of constant densities \cite{FeRa10, FeHuKo11}. This property will be further elaborated below. 

We note that the analysis in \cite{WuSlepcev2015,CarrilloSlepcevWu2016} requires assumptions on $K$ which the potential \eqref{eqn:intpot} does not satisfy. In particular, the interaction potential is required there to be $C^1$ and $\lambda$-geodesically convex. Consequently, the results in \cite{WuSlepcev2015,CarrilloSlepcevWu2016} do not immediately apply to our study. Nevertheless we consider the framework developed in these papers, in particular the gradient flow and the energy dissipation (see \eqref{eqn:gflow}), and the particle approximation method which can be turned into a very valuable computational tool. Indeed, to validate our equilibrium calculations we use a particle method to simulate solutions to \eqref{eqn:modelb}.


\paragraph{Transport along characteristics.} In the absence of an exogenous potential ($V=0$), the aggregation model \eqref{eqn:model} with interaction potential \eqref{eqn:intpot} evolves into constant, compactly supported steady states. This can be inferred from a direct calculation using the specific form of the potential \eqref{eqn:intpot}. Indeed, expand 
$$\nabla\cdot(\rho v)=v\cdot\nabla\rho+\rho\nabla\cdot v,$$
and write the aggregation equation \eqref{eqn:model} as
\begin{equation}
\label{eqn:exp}
\rho_{t}+v\cdot\nabla\rho=-\rho \nabla \cdot v.
\end{equation}
From \eqref{eqn:v} and \eqref{eqn:intpot}, using $-\Delta \phi = \delta$ and the mass constraint \eqref{eqn:massM}, one gets
\begin{align}
\nabla\cdot v &= - \Delta K \ast \rho \nonumber \\
&= \rho - nM.  
\label{eqn:divv}
\end{align}

This calculation shows that $\nabla\cdot v$ is a {\em local} quantity. By using \eqref{eqn:divv} in \eqref{eqn:exp}, one finds that along characteristic paths $X(\alpha,t)$, defined by
\begin{equation}
\frac{d}{dt}X(\alpha,t)=v(X(\alpha,t),t),\qquad X(\alpha,0)=\alpha,
\label{eqn:char}%
\end{equation}
$\rho(X(\alpha,t),t)$ satisfies:
\begin{equation}
\frac{D}{Dt}\rho=-\rho(\rho-nM). \label{eqn:Drho-c}%
\end{equation}

The remarkable property of the interaction potential \eqref{eqn:intpot}, as seen from equation \eqref{eqn:Drho-c}, is that the evolution of the density along a certain characteristic path $X(\alpha,t)$ satisfies a decoupled, stand-alone, ordinary differential equation. Hence, as inferred from \eqref{eqn:Drho-c}, $\rho(X(\alpha,t),t)$ approaches the value $nM$ as $t\rightarrow\infty$, along \emph{all} characteristic paths $X(\alpha,t)$ that transport non-zero densities. 

It has been demonstrated in \cite{BertozziLaurentLeger, FeHuKo11} that solutions to equation \eqref{eqn:model}, with $K$ given by \eqref{eqn:intpot}, approach asymptotically a radially symmetric equilibrium that consists in a ball of constant density $nM$. In domains with boundaries, which is the focus of the present study, accumulation on boundaries can occur and also, the equilibrium swarms in the interior are not expected to be radially symmetric. Nevertheless, by transport along characteristics, given by equation \eqref{eqn:Drho-c}, aggregation patches that form asymptotically {\em away} from the boundaries have constant density $nM$ in their support, and this is the key observation used in Sections \ref{sect:oned} and \ref{sect:twod} to investigate equilibria for model \eqref{eqn:modelb}.

In the presence of an external potential, calculation of $\nabla \cdot v$ from \eqref{eqn:v} and \eqref{eqn:intpot} (see also \eqref{eqn:divv}) yields:
\[
\nabla\cdot v = \rho - nM - \Delta V.
\] 
Along characteristics (c.f., \eqref{eqn:exp} and \eqref{eqn:char}), densities $\rho(X(\alpha,t),t)$ evolve according to 
\begin{equation}
\label{eqn:evol}
\frac{D}{Dt}\rho=-\rho(\rho-nM - \Delta V). 
\end{equation}

In general, equation \eqref{eqn:evol} is not a stand-alone equation for the evolution of $\rho(X(\alpha,t),t)$, as $\Delta V$ is evaluated along the characteristic path $X(\alpha,t)$. Consequently, one also needs to solve \eqref{eqn:char} for  the characteristic paths, but since the velocity field in \eqref{eqn:char} is nonlocal, \eqref{eqn:char} represents a fully coupled family of ordinary differential equations. In Sections \ref{sect:oned} and \ref{sect:twod} we work with a linear gravitational potential $V$ for which $\Delta V =0$ and hence, \eqref{eqn:evol} reduces to \eqref{eqn:Drho-c}. However these cases differ slightly from the arguments above in that we have not dealt with the accumulation on the wall and its affect. Section  \ref{subsubsect:1d-CoMDyn} provides further evidence that we retain the local character for the free swarm, and also, that we only have to consider equilibria of constant density $nM$ away from boundaries.


\section{One dimension: equilibria on half-line}
\label{sect:oned}

In Sections \ref{subsect:1dnoV}-\ref{subsect:1d-dyn} we consider the one dimensional problem on $\Om = [0,\infty)$, with interaction kernel given by \eqref{eqn:intpot}-\eqref{eqn:phi}. In Section \ref{subsect:1d-Morse} we consider a different interaction kernel, namely a Morse-type kernel as investigated in \cite{BeTo2011}. We study the existence and stability of both connected and disconnected equilibria throughout. 

\subsection{No exogenous potential}
\label{subsect:1dnoV}

We consider first the case $V(x) =0$ (no exogenous forces). In the absence of boundaries, the time evolution \eqref{eqn:evol} yields an equilibrium solution that has constant density $M$ in its support. Therefore, away from boundaries, the equilibria for model \eqref{eqn:model} also consist of constant densities on the support. Moreover, at the boundary (the origin in this case) one can expect a delta-aggregation build up \cite{BeTo2011}. 

Based on these considerations (see also Remark \ref{rmk:1d-gloMin} below), we look for equlibria in the form of a delta accumulation of strength $S$ at the origin and a constant density in an interval $(\dl,\dl+\dr)$, with $\dl \geq0$, $\dr>0$:
\begin{equation}
\label{eqn:ss-oned}
\barrho(x) = S \delta(x) + M \One_{(\dl, \dl+\dr)}.
\end{equation}
The support $\Omrho$ of $\barrho$ consists of two (possibly disconnected) components: 
\[
\Om_1 = \{0\} \quad \text{ and } \quad \Om_2 = [\dl,\dl+\dr].
\] 

First observe that by the constant mass condition \eqref{eqn:massrho}, we have
\begin{equation}
\label{eqn:condx2}
S + M \dr = M.
\end{equation}

A necessary condition for $\barrho$ to be an equilibrium is to satisfy \eqref{eqn:equilsup-gen}. Equation \eqref{eqn:equilsup-gen} is satisfied provided $\Lambda(x)$ is constant on each component of $\Omrho$:
\begin{equation}
\label{eqn:Lambda-const}
\Lambda(0) = \lambda_1, \quad \text{ and } \quad \Lambda(x) = \lambda_2 \quad \textrm{ in } [\dl,\dl+\dr].
\end{equation}
The calculation of $\Lambda(x)$ from \eqref{eqn:Lambda} yields:
\begin{equation}
\label{eqn:Lambda1d}
\Lambda(x) = S \left( \frac{1}{2} x^2 - \frac{1}{2} x\right) + \int_{\dl}^{\dl+\dr} \left( \frac{1}{2} (x-y)^2 - \frac{1}{2}|x-y|\right) M dy.
\end{equation}
For $x \in (\dl,\dl+\dr)$,  an elementary calculation of $\Lambda(x)$ gives
\[
\Lambda(x) = \frac{1}{2}(S+M\dr -M) x^2 + \frac{1}{2} \left(-S + M(2\dl + \dr)(1-\dr)\right) x + \frac{M}{6} (3 \dl^2 \dr + 3 \dl \dr^2 + \dr^3)- \frac{M}{4} (2 \dl^2 + 2 \dl \dr + \dr^2).
\]
The second condition in \eqref{eqn:Lambda-const} is satisfied only if the coefficients of $x^2$ and $x$ of the polynomial above are zero . Setting the coefficient of $x^2$ to zero yields the mass constraint condition \eqref{eqn:condx2}, while the coefficient of $x$ vanishes provided
\begin{equation}
\label{eqn:condx}
S = M(2\dl + \dr)(1-\dr). 
\end{equation}
Combining the two conditions \eqref{eqn:condx2} and \eqref{eqn:condx} we arrive at:
\begin{equation}
\label{eqn:cond-d2}
S = M(1-\dr), \qquad \dl = \frac{1-\dr}{2}.
\end{equation}

Hence, there is a family of solutions to \eqref{eqn:Lambda-const} in the form \eqref{eqn:ss-oned} with parameter $d_2 \in (0,1]$. Note that $\dl+ \frac{\dr}{2} = \frac{1}{2}$, implying that for all the equilibria in this family, the centre of mass of the free swarm is at $\frac{1}{2}$.

\begin{figure}[thb]
  \begin{center}
 \includegraphics[width=0.32\textwidth]{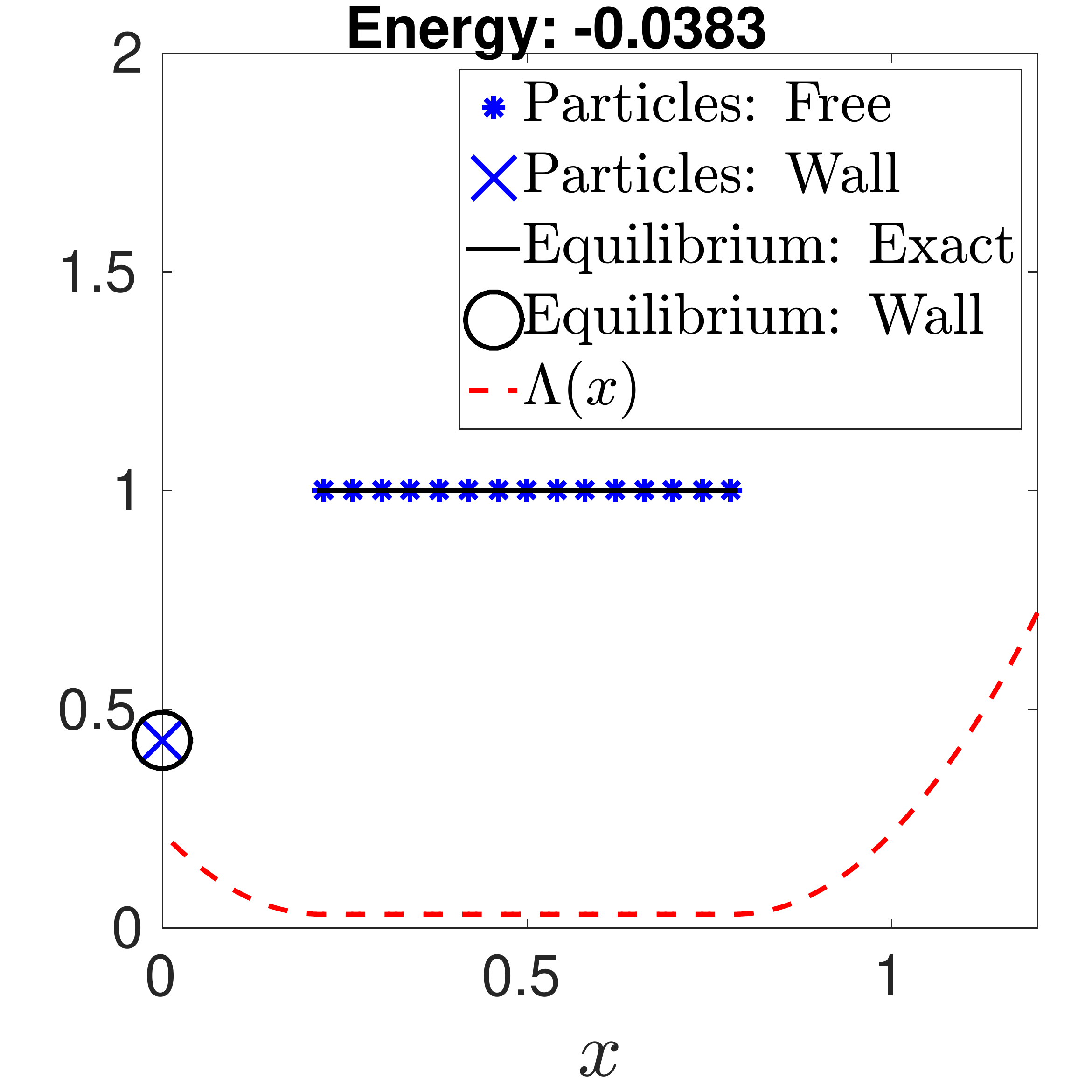} 
 \includegraphics[width=0.32\textwidth]{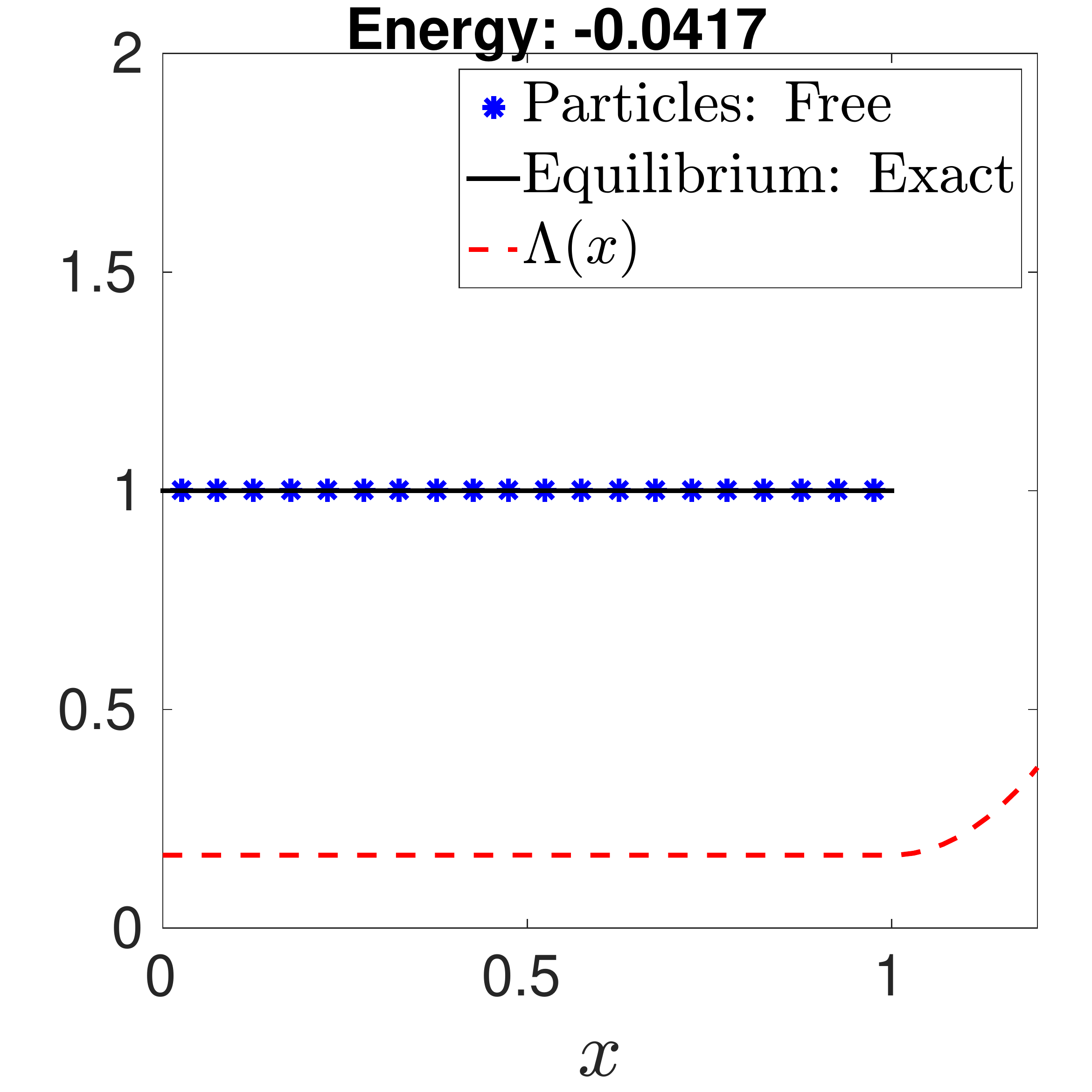} 
 \includegraphics[width=0.32\textwidth]{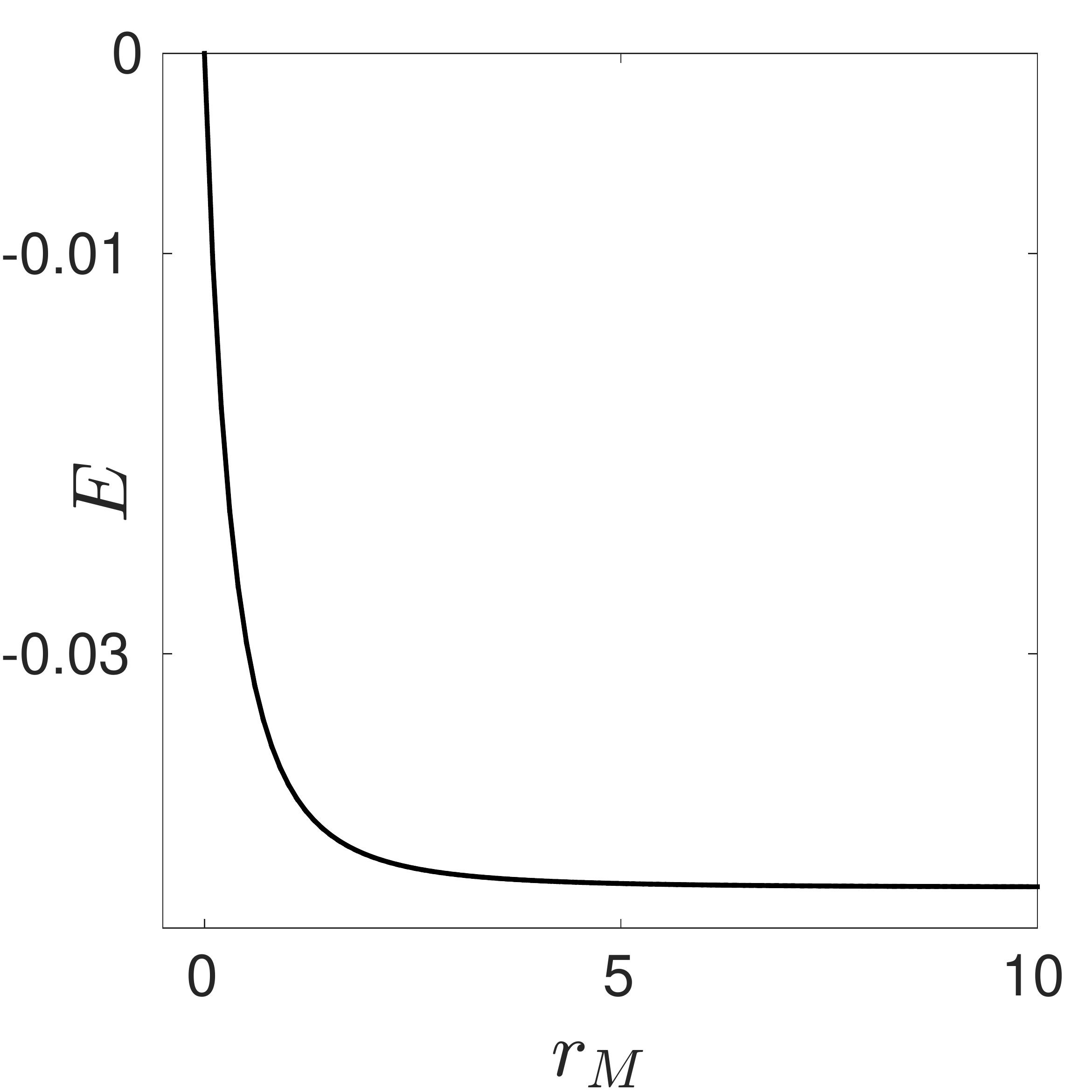}
\end{center}
\hspace{0.165\textwidth} (a) \hspace{0.28\textwidth} (b) \hspace{0.29\textwidth} (c)
\caption{Equilibria \eqref{eqn:ss-oned} on half-line for $V=0$ (no exogenous potential). (a) Disconnected equilibrium consisting in a free swarm of constant density and a delta aggregation at the origin. (b) Connected equilibrium of constant density in $(0,1)$. (c) Energy of equilibria \eqref{eqn:ss-oned} as a function of the mass ratio; the lowest energy state corresponds to the connected equilibrium ($\mr=\infty$). Note that for a better visualization $\Lambda(x)$ has been shifted and stretched vertically.}
\label{fig:oned-zerog}
\end{figure}

By expressing everything in terms of $\dr$ only, $\Lambda$ takes the following values on the two components $\Om_1$ and $\Om_2$ of $\Omrho$, respectively:
\begin{subequations}
\label{eqn:lambda12}
\begin{align}
\lambda_1 &= -\frac{M}{24}(1-\dr)^3 + \frac{M}{8}(1-\dr)^2 - \frac{M}{12}, \\
\lambda_2 &= -\frac{M}{24}(1-\dr)^3  - \frac{M}{12}.
\end{align}
\end{subequations}
Note that $\lambda_1>\lambda_2$, unless $\dr=1$, in which case $\lambda_1=\lambda_2$. Based on this observation, we distinguish between two qualitatively different equilibria:

\medskip
{\em i) Disconnected equilibria  ($\dl>0$).} A generic disconnected solution to \eqref{eqn:Lambda-const} of form \eqref{eqn:ss-oned} is shown in Figure \ref{fig:oned-zerog}(a); the solid line indicates the constant density in the free swarm and the circle on the vertical axis indicates the strength $S$ of the delta-aggregation at the origin. Note that in {\em all} numerical simulations presented in this paper we take $M=1$. 

To check that these solutions to \eqref{eqn:Lambda-const}  are in fact equilibria reduces to show that the velocity (see \eqref{eqn:vp} and \eqref{eqn:proj}) vanishes at points in the support $\Omrho = \Om_1 \cup \Om_2$. Since $\Lambda(x)$ is constant in $[\dl,\dl+\dr]$, it follows that the velocity vanishes everywhere in $\Om_2$. The more delicate part is evaluating the velocity at the origin. By \eqref{eqn:vp}, the velocity at the origin is computed by accounting (via a spatial convolution) for all the attractive and repulsive effects of points that lie in $\Omrho$. The key observation is that the point at the origin (the only point in $\Om_1$) does not have any interaction effects on the origin itself; in a discrete setting this amounts to the fact that particles sitting on top of each other do not exert interactions (attractive or repulsive) among themselves. Therefore, the velocity $v(0)$ calculated from \eqref{eqn:vp} reduces to an integral over $\Om_2$ only:
\begin{equation}
\label{eqn:v0}
v(0) = P_0 \left( - \int_{\Om_2}  K'(-y)  \barrho(y) dy \right).
\end{equation}
An elementary calculation, using $K'(y) = y - \operatorname{sgn}(y)$ and $\barrho(y) = M$ in $\Om_2=(\dl,\dl+\dr)$, yields:
\begin{equation}
\label{eqn:v0-Om2}
- \int_{\Om_2}  K'(-y)  \barrho(y) dy =\frac{M}{2} \dr (2\dl + \dr -1).
\end{equation}
Finally, by \eqref{eqn:cond-d2}, 
\[
v(0)= P_0(0)=0,
\]
so the disconnected state is indeed an equilibrium.

We now check whether the disconnected equilibria are energy minimizers. By an elementary calculation, for all  $0<\dr <1$ (or equivalently $0<\dl < \frac{1}{2}$),  $\Lambda(x)$ (given by \eqref{eqn:Lambda1d}) can be shown to be strictly decreasing in $(0,\dl)$ and strictly increasing in $(\dl+\dr,\infty)$ --- see the dashed line in Figure \ref{fig:oned-zerog}(a). This calculation shows that  disconnected equilibria $\barrho$ in the form \eqref{eqn:ss-oned} are {\em not} local minima (swarm minimizers), as \eqref{eqn:equilcomp} is not satisfied near the origin; since $\Lambda$ is strictly decreasing in $(0,\dl)$, an infinitesimal perturbation of mass from the origin would bring that mass into the free swarm, which is a more energetically favourable state. 

Nevertheless, $\barrho$ are steady states and, as demonstrated in Section \ref{subsubsect:1d-CoMDyn}, are asymptotically stable with respect to {\em certain} perturbations; given the considerations above, it is clear that such perturbations must only be with respect to the aggregation in the free swarm.  Also shown in Section \ref{subsubsect:1d-NumIniMat}, the dynamic evolution of model \eqref{eqn:modelb} consistently achieves  (asymptotically) disconnected steady states starting from a diverse set of initial densities, which make such equilibria very relevant for the dynamics. Figure \ref{fig:oned-zerog}(a) shows in fact the disconnected equilibrium \eqref{eqn:ss-oned} achieved via particle simulations: stars represent particles and the cross indicates a superposition of particles at the origin.

\medskip
{\em ii) Connected equilibria.} There are two possible connected equilibria. The first is a degenerate case of  \eqref{eqn:ss-oned}, where $\dl=\dr=0$, and all mass lies at the origin (or by translation, at any point in $(0,\infty)$):
\begin{equation}
\label{eqn:linpot2}
\barrho(x) = M \delta(x).
\end{equation} 
While \eqref{eqn:linpot2} is an equilibrium solution, it is not an energy minimizer, as can be inferred from the expression of $\Lambda$:
\[
\Lambda(x) = -\frac{1}{2}M|x| + \frac{1}{2} Mx^2,
\]
by noting that \eqref{eqn:equilcomp} is not satisfied for $x\in (0,1)$. Any perturbation, which does not create a delta accumulation in the interior of $\Om$, from this trivial equilibrium would result either in a disconnected state or in the connected equilibrium discussed below.

The other connected equilibrium can be obtained as a limiting case $d_1\to 0$ of the disconnected equilibria \eqref{eqn:ss-oned} (see also \eqref{eqn:cond-d2}). In this limit, there is no delta aggregation on the wall ($S=0$), $\dr=1$, and the solution consists in a constant density in the interval $(0,1)$ --- see solid line in Figure \ref{fig:oned-zerog}(b). Alternatively, one can consider an entire family of such solutions, by taking arbitrary translations of the constant swarm to the right; this corresponds in fact to the equilibrium solution in the absence of boundaries, as discussed in Section \ref{sect:prelim}. The connected state is a swarm minimizer, as can be inferred by a direct calculation of $\Lambda(x)$; for an illustration, see the dashed line in Figure \ref{fig:oned-zerog}(b).

The energy corresponding to the equilibria \eqref{eqn:ss-oned} can be easily computed from \eqref{eqn:energy}, \eqref{eqn:Lambda}, and \eqref{eqn:Lambda-const}, by noting that in the absence of an external potential,
\[
E[\barrho] = \frac{1}{2}\int_{\Omrho} \Lambda(x) \barrho(x) dx = \frac{\lambda_1}{2} \int_{\Om_1} \barrho(x) dx + \frac{\lambda_2}{2} \int_{\Om_2} \barrho(x) dx.
\]
After a simple calculation, using the explicit expressions of $\lambda_1$ and $\lambda_2$ from \eqref{eqn:lambda12}, one finds:
\begin{equation}
\label{eqn:energy-ss}
E[\barrho] = \frac{M^2}{3} \left( \dl^3- \frac{1}{8}\right) = \frac{M^2}{24} \dr (-3 + 3 \dr -\dr^2).
\end{equation}
Note that $E[\barrho]$ has the lowest energy for $\dl=0$ (or equivalently, $\dr=1$), which corresponds to the (limiting) connected equilibrium.

\begin{remark}
\label{rmk:mr-oned-zerog}
The equilibria discussed above can be alternatively parametrized by $\mr$, defined as the mass ratio between the mass in the free swarm and the mass accumulated at the boundary of the domain (the origin in this case). This is in fact the parametrization used for the two-dimensional study in Section \ref{sect:twod} (see \eqref{eqn:mr}). In one dimension, the mass ratio of the two components (c.f., \eqref{eqn:ss-oned} and \eqref{eqn:cond-d2}) is given by
\begin{equation*}
\mr = \frac{M \dr}{S} = \frac{\dr}{1-\dr}.
\end{equation*}
The parameter $\dr$ ranges in $(0, 1)$ for the disconnected equilibria in part i), while the connected equilibria in part ii) correspond to $\dr=0$ and $\dr=1$, respectively. Consequently, in the absence of an exogenous potential, an equilibrium exists for any $\mr \in \big[0, \infty)$, as well as $\mr =\infty$. However, the only equilibrium that is an energy minimizer, and hence stable, is the one with infinite mass ratio, corresponding to the connected steady state which has all mass in the free swarm -- Figure \ref{fig:oned-zerog}(b). 
\end{remark}

Figure \ref{fig:oned-zerog}(c) shows a plot of the energy $E[\barrho]$ calculated in \eqref{eqn:energy-ss}, as a function of mass ratio $\mr$. We find a monotonically decreasing profile with the lowest energy state corresponding to the connected equilibrium with all mass in the free swarm ($\mr = \infty$). The connected equilibrium ($\dr = 1$ and $\dl = 0$) is in fact the global minimizer in this case, as it can be inferred from the remark below.

\begin{remark}
\label{rmk:1d-gloMin}
To conclude that the connected equilibrium is the global minimizer one needs to consider other possible minimizers and show that their energies are larger. We have already shown that disconnected equilibria of form \eqref{eqn:ss-oned} are not minimizers. One can also show that a multi-component free swarm is not an energy minimizer either. The argument essentially comes from \cite{BeTo2011} where we wish to show that $\Lambda(x)$ is convex between free swarm components, which is a sufficient condition to show it is not an energy minimizer as \eqref{eqn:equilcomp-gen} does not hold. 

Assume a disconnected equilibrium of the form
\begin{equation}
\label{eqn:ss-multiple}
\rho(x) = S\delta(x) + \sum_{i=1}^m \rho_i(x),
\end{equation}
where $\rho_i$ are supported on $\Om_i$ ($\Om_i$ are disjoint from each other and do not include the origin). Then \eqref{eqn:Lambda} becomes
\[
\Lambda(x) = S\left(\frac{1}{2}x^2 - \frac{1}{2}x\right) + \sum_{i=1}^m \int_{\Om_i} K(x-y)\rho_i(y)\dy,
\]
and for $x \notin \cup \Om_i$ one gets
\[
\Lambda''(x) = S + \sum_{i=1}^m \int_{\Om_i} \rho_i(y)\dy > 0.
\]
Therefore, $\Lambda(x)$ is indeed convex between free swarm components and \eqref{eqn:ss-multiple} cannot be a minimizer.
\end{remark}

\subsection{Linear exogenous potential}
\label{subsect:1d-linV}

Consider the exogenous gravitational potential $V(x) = g x$, with $g>0$. The domain of the problem is again, the half-line $\Om = [0,\infty)$. From equation \eqref{eqn:evol}, as $V''(x)=0$, we infer that in the absence of boundaries the equilibrium density is constant on its support. This implies that away from the boundary, which in this problem consists of just the origin, an equilibrium solution has constant density $M$ in the support. 

We focus, as in Section \ref{subsect:1dnoV}, on equilibria that have possibly disconnected components, and look for steady states in the form \eqref{eqn:ss-oned}, consisting of a delta aggregation at the origin, and a constant density $M$ in the interval $(\dl,\dl+\dr)$, where $\dl \geq 0$, $\dr>0$. Similar to above, the support $\Omrho$ consists of two components, $\Om_1 = \{0\}$ and $\Om_2 = [\dl,\dl+\dr]$, and the constant mass condition yields \eqref{eqn:condx2}.

Equilibria \eqref{eqn:ss-oned} must satisfy the necessary condition \eqref{eqn:equilsup-gen}, which in this case reduces to \eqref{eqn:Lambda-const}. By direct calculation,
\begin{equation}
\label{eqn:Lambda1d-g}
\Lambda(x) = S \left( \frac{1}{2} x^2 - \frac{1}{2} x\right) + \int_{\dl}^{\dl+\dr} \left( \frac{1}{2} (x-y)^2 - \frac{1}{2}|x-y|\right) M dy + gx.
\end{equation}

By evaluating at $x\in(\dl,\dr)$ and requiring that $\Lambda(x)$ is constant in this interval, we arrive at the following constraints on the parameters. First, by setting to zero the coefficient of $x^2$ we find equation \eqref{eqn:condx2} that represents the mass constraint condition. Then we note that the coefficient of $x$ vanishes provided
\[
S = M(2\dl + \dr)(1-\dr) + 2g. 
\]
Combine this equation with the mass constraint \eqref{eqn:condx2} to find
\begin{equation}
\label{eqn:cond-d2g}
S = M(1-\dr), \qquad \dl = -\frac{g}{M(1-\dr)} + \frac{1-\dr}{2}.
\end{equation}
Note that since $\dl\geq0$ and $0<\dr<1$, then necessarily $g<\frac{M}{2}$ and $0<\dr \leq 1 - \sqrt{\frac{2g}{M}}$. 

Denote by $\gc = \frac{M}{2}$ this critical value of $g$. From the above we conclude that for any $g<\gc$, we have a family of solutions to \eqref{eqn:equilsup-gen} of the form \eqref{eqn:ss-oned} with parameter $\dr \in \bigl( 0, 1 - \sqrt{\frac{2g}{M}} \bigr]$. For any $\dr$ in the open interval $\bigl( 0, 1 - \sqrt{\frac{2g}{M}} \bigr)$, $\dl>0$  and hence these states are disconnected. For $\dr = 1 - \sqrt{\frac{2g}{M}}$, $\dl=0$ and the state is connected. Also, from \eqref{eqn:cond-d2g}, we infer that the centre of mass $\dl + \frac{\dr}{2}$ of the free swarm is located at $\frac{1}{2}-\frac{g}{S}$.

For $g>\gc$ there are no equilibria in the form \eqref{eqn:ss-oned}. As shown below, the equilibrium in this case is a delta accumulation at the origin, which is also a global minimizer of the energy. Physically this can be explained by having a threshold value $\gc$ beyond which the gravity is so strong that it pins all mass on the boundary. 

We consider now the two cases: $g<\gc$ and $g>\gc$.

\medskip
{\em \bf Case $\mathbf{g<\gc}$.} As noted above, c.f. \eqref{eqn:cond-d2g}, there exists a family of solutions to \eqref{eqn:equilsup-gen} of the form \eqref{eqn:ss-oned}, parameterized by $\dr \in \bigl( 0, 1 - \sqrt{\frac{2g}{M}} \bigr]$. 
By an elementary calculation, one can compute the values of $\Lambda(x)$ in each component of the support, $\Om_1$ and $\Om_2$, respectively:
\begin{subequations}
\label{eqn:lambda12-g}
\begin{align}
\lambda_1 &= -\frac{M}{24}(1-\dr)^3 + \frac{M}{8}(1-\dr)^2 - \frac{M}{12} + \frac{g^2}{2M} \frac{\dr}{(1-\dr)^2}, \\
\lambda_2 &= -\frac{M}{24}(1-\dr)^3  - \frac{M}{12} - \frac{g^2}{2M}\frac{1}{1-\dr} + \frac{g}{2}.
\end{align}
\end{subequations}
As in the zero gravity case, we find that $\lambda_1>\lambda_2$, unless $\dr = 1-\sqrt{\frac{2g}{M}}$ (or equivalently, $\dl=0$), in which case $\lambda_1 = \lambda_2$. We discuss separately the disconnected and connected states.

\begin{figure}[thb]
  \begin{center}
 \includegraphics[width=0.32\textwidth]{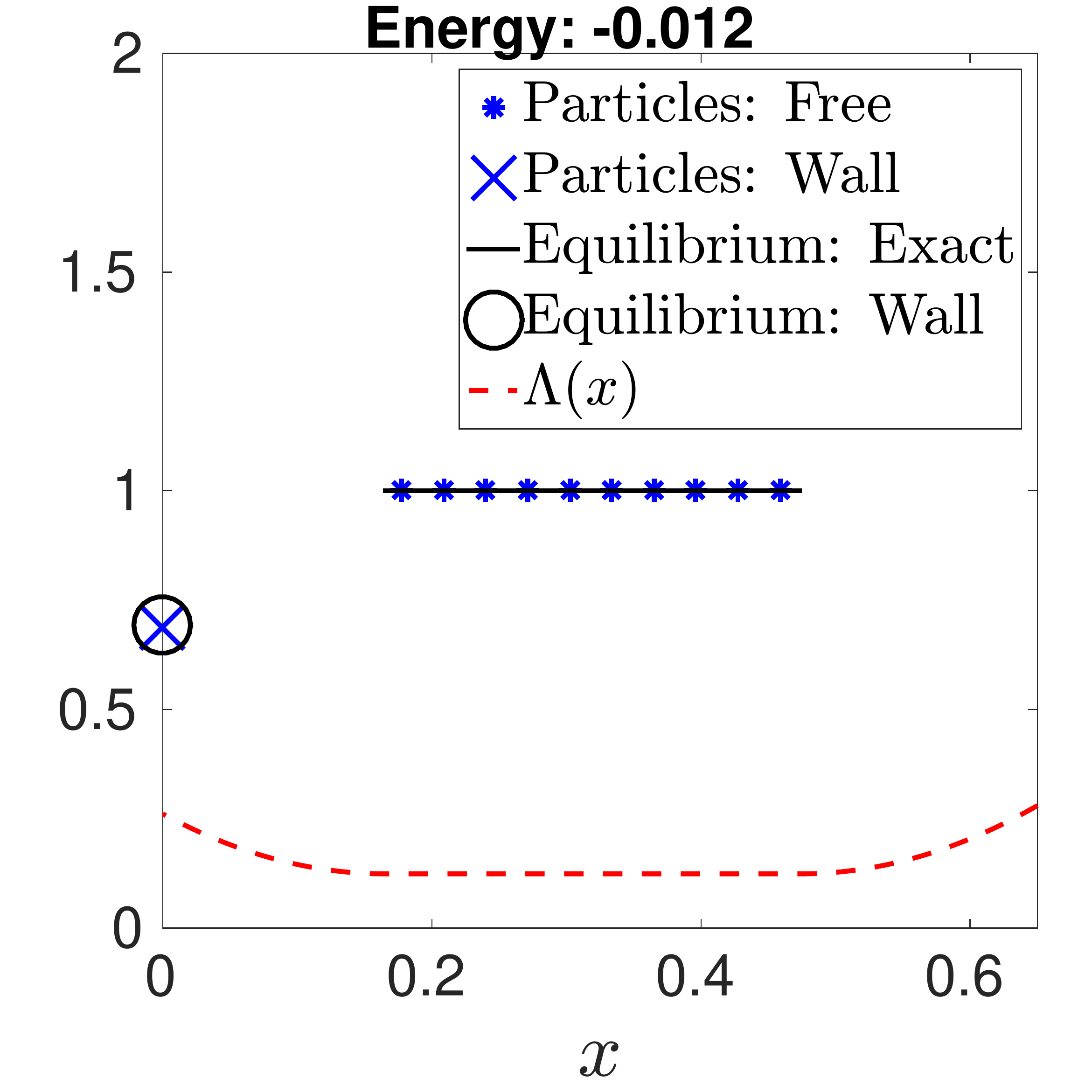} 
 \includegraphics[width=0.32\textwidth]{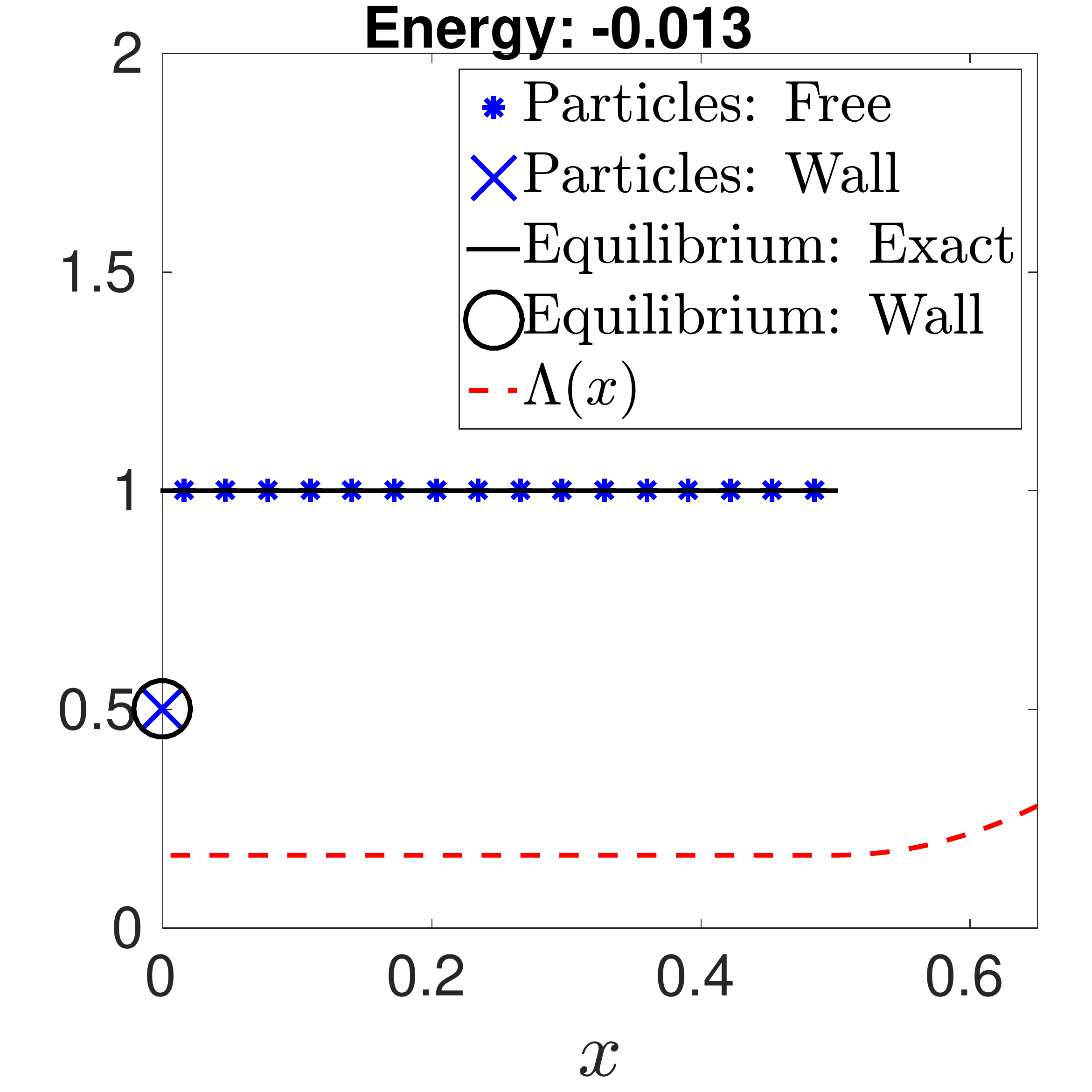} 
 \includegraphics[width=0.32\textwidth]{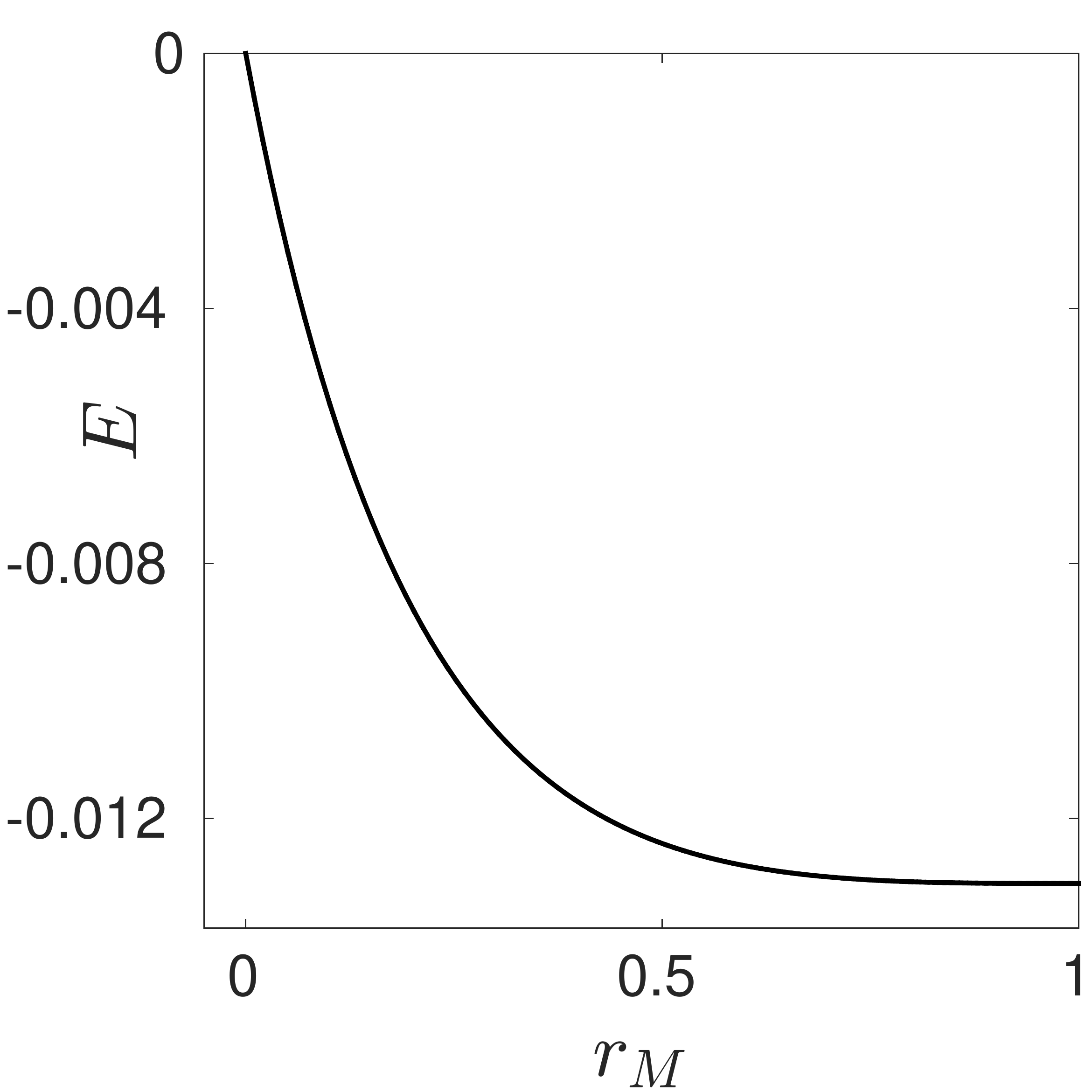} 
\end{center}
\hspace{0.165\textwidth} (a) \hspace{0.28\textwidth} (b) \hspace{0.29\textwidth} (c)
\caption{Equilibria \eqref{eqn:ss-oned} on half-line for $V(x)=gx$ (linear exogenous potential) with $g=0.125$. (a) Disconnected state consisting in a free swarm of constant density and a delta aggregation at the origin. (b) Connected state with a constant density in a segment adjacent to the origin and a delta aggregation at origin. (c) Energy of equilibria \eqref{eqn:ss-oned} as a function of the mass ratio; the lowest energy state corresponds to the connected equilibrium $\Bigl( \mr=\sqrt{\frac{M}{2g}}-1\Bigr)$.}
\label{fig:oned-wg}
\end{figure}

{\em i) Disconnected equilibria  ($\dl>0, \dr<1-\sqrt{\frac{2g}{M}}$).} A generic disconnected solution to \eqref{eqn:Lambda-const} (here $g=0.125$) is shown in Figure \ref{fig:oned-wg}(a); the solid line indicates the constant density in the free swarm and the circle on the vertical axis indicates the strength of the delta-aggregation.  To show that these states are equilibria, one only needs to check the velocity in $\Om_1$,  the boundary of the domain. By a similar argument as in the zero gravity case (attractive and repulsive effects at the origin are only felt through interactions with the free swarm),  the velocity $v(0)$ calculated from \eqref{eqn:vp} reads:
\begin{equation}
\label{eqn:v0-g}
v(0) = P_0 \left( - \int_{\Om_2}  K'(-y)  \barrho(y) dy -g \right).
\end{equation}
By \eqref{eqn:v0-Om2} and \eqref{eqn:cond-d2g},
\[
- \int_{\Om_2}  K'(-y)  \barrho(y) dy = -{g} \frac{\dr}{1-\dr},
\]
and hence, from \eqref{eqn:v0-g} and \eqref{eqn:proj} we find that 
\[
v(0) = P_0 \bigl(\underbrace{-\frac{g}{1-\dr}}_{<0}\bigr) = 0.
\]
The disconnected state is indeed an equilibrium.

By a direct calculation one can show that $\Lambda(x)$ is strictly decreasing in $(0,\dl)$ and strictly increasing in $(\dl+\dr,\infty)$ --- see the dashed line in Figure \ref{fig:oned-wg}(a). We infer  that disconnected equilibria $\barrho$ in the form \eqref{eqn:ss-oned} are {\em not} local minima; again,  \eqref{eqn:equilcomp} is not satisfied near the origin and an infinitesimal perturbation of mass from $\Om_1$ (boundary) would bring it into $\Om_2$ (free swarm). Nevertheless, these equilibria are asymptotically stable to certain perturbations of the free swarm and our numerical explorations indicate, as in the zero gravity case, that such disconnected steady states are very relevant for model \eqref{eqn:modelb}, as they are reached dynamically starting from a wide range of initial densities -- see Sections \ref{subsubsect:1d-CoMDyn} and \ref{subsubsect:1d-NumIniMat}. Figure \ref{fig:oned-wg}(a) shows this particular disconnected equilibrium obtained via particle simulations (stars and cross).

\medskip
{\em ii) Connected equilibria.}  There are two different connected equilibria: one that has all mass at the origin and another that corresponds to the limit case $\dl=0$, $\dr=1-\sqrt{\frac{2g}{M}}$ of the disconnected equilibria in part i) above. 

The first type is a delta-concentration at the origin of strength $M$, as in \eqref{eqn:linpot2}. This can be thought of as a degenerate case of \eqref{eqn:ss-oned} with $\dl=\dr=0$. The calculation of $\Lambda$ from \eqref{eqn:Lambda} yields:
\[
\Lambda(x) = -\frac{1}{2}M|x| + \frac{1}{2} Mx^2 + gx.
\]
Since $\Omrho = \{0\}$, \eqref{eqn:equilsup} trivially holds with $\lambda =0$, while \eqref{eqn:equilcomp} is equivalent to 
\begin{equation}
\label{eqn:1dg-allwall-min}
 \left( -\frac{1}{2}M  + \frac{1}{2} Mx + g \right) x >0, \qquad \text{ for all } x>0.
\end{equation}
The inequality above does not hold when $g < \frac{M}{2}$, hence the equilibrium \eqref{eqn:linpot2} is not an energy minimizer when $g<\gc$.

The other type of connected equilibrium is obtained from the disconnected equilibria in part i) in the limit $d_1 \to 0$; it consists of a delta aggregation at the origin of strength $S=\sqrt{2gM}$ and a constant density $M$ in the interval $\bigl(0,1-\sqrt{\frac{2g}{M}}\bigr)$. The connected equilibrium for $g=0.125$ and $M=1$ is illustrated in Figure \ref{fig:oned-wg}(b) --- see solid line and circle on vertical axis indicating the strength of the delta aggregation. The connected state is a swarm minimizer, as can be inferred from a direct calculation of $\Lambda(x)$ --- see dashed line in Figure \ref{fig:oned-wg}(b).

The energy corresponding to the equilibria \eqref{eqn:ss-oned} in the gravity case can be computed through elementary calculations from \eqref{eqn:energy}, \eqref{eqn:Lambda}, and \eqref{eqn:equilsup-gen}, along with the expressions of $\lambda_1$ and $\lambda_2$ from \eqref{eqn:lambda12-g}. We skip details and list only the final outcome:
\begin{equation}
\label{eqn:energy-ss-g}
E[\barrho] = \frac{M^2}{24} \dr (-3 + 3 \dr -\dr^2) + \frac{g}{2} M \dr - \frac{g^2}{2}\frac{\dr}{1-\dr}.
\end{equation}
The zero gravity calculation \eqref{eqn:energy-ss} can be obtained from \eqref{eqn:energy-ss-g} by setting $g$ to zero. Also as expected, by inspecting the energy in \eqref{eqn:energy-ss-g} we find that among all equilibria in the form \eqref{eqn:ss-oned}, the one that has the lowest energy is the connected state, corresponding to $\dr = 1-\sqrt{\frac{2g}{M}}$. 

\begin{remark}
\label{rmk:mr-oned}
As noted in Remark \ref{rmk:mr-oned-zerog}, the family of equilibria above can be alternatively parametrized by $\mr$, the mass ratio between the mass in the free swarm and the mass on the wall. By \eqref{eqn:cond-d2g}, $\mr$ is given by
\begin{equation}
\label{eqn:mr-1d}
\mr = \frac{M \dr}{S} = \frac{\dr}{1-\dr}.
\end{equation}
The parameter $\dr$ ranges in $\Bigl( 0, 1-\sqrt{\frac{2g}{M}}\Bigr)$ for the disconnected equilibria, while $\dr=0$ and $\dr = 1-\sqrt{\frac{2g}{M}}$ correspond to the two connected equilibria discussed above. Hence, $\mr \in \Big[ 0, \sqrt{\frac{M}{2g}}-1 \Bigr]$, or equivalently $\mr \in \Big[ 0, \sqrt{\frac{\gc}{g}}-1 \Bigr]$. 
\end{remark}

Figure \ref{fig:oned-wg}(c) shows the energy \eqref{eqn:energy-ss-g} of the equilibria in the form \eqref{eqn:ss-oned} for the gravitational potential with $g=0.125$, plotted as a function of the mass ratio $\mr$.  Note the monotonically decreasing profile, with the equilibrium of lowest energy being the connected state shown in Figure  \ref{fig:oned-wg}(b); this equilibrium corresponds to the largest possible value of mass ratio, which in this case is $\mr= 1$. By an argument similar to that from Remark \ref{rmk:1d-gloMin} one can infer in fact that the connected equilibrium is a global minimizer.

\begin{figure}[ht]
  \begin{center}
 \includegraphics[width=0.46\textwidth]{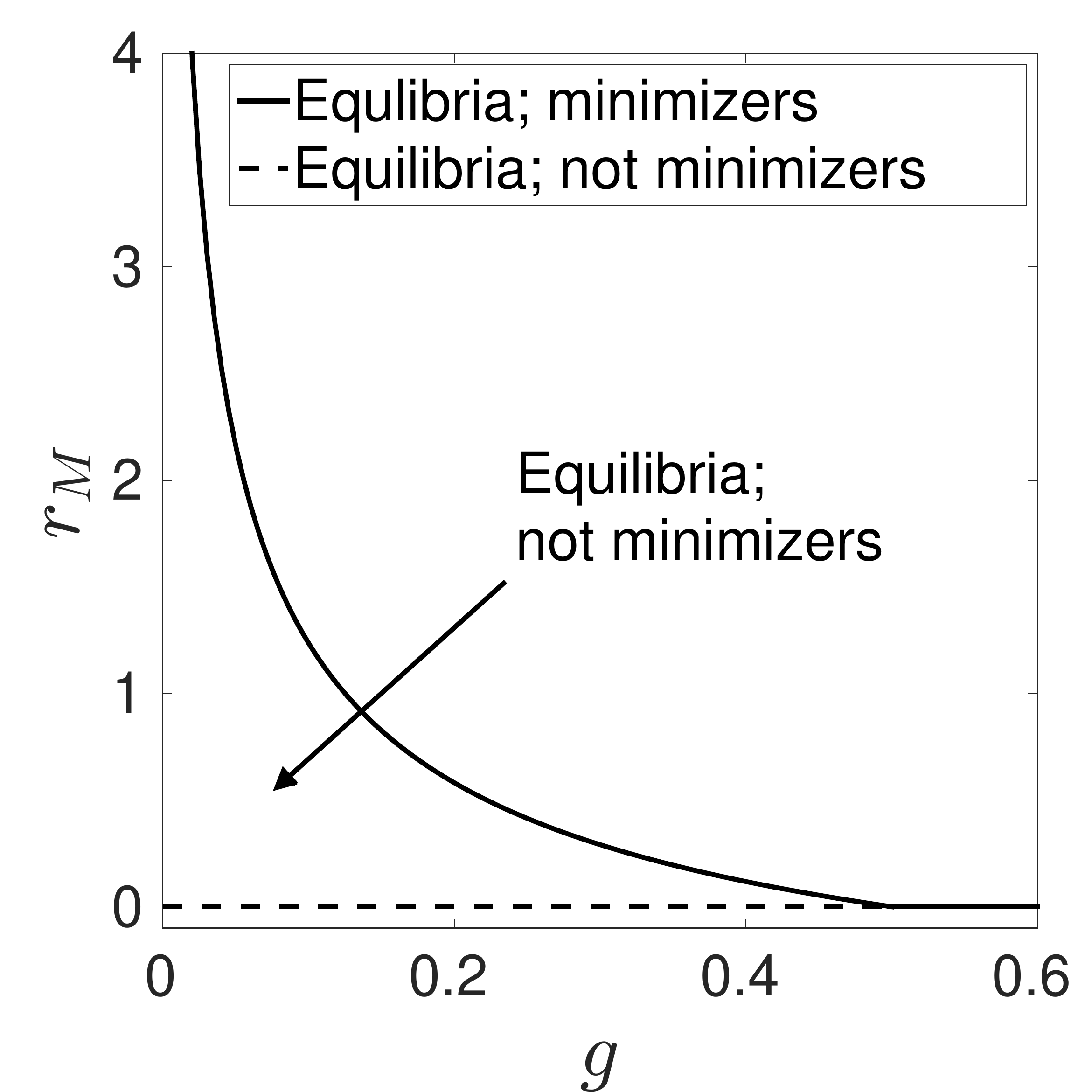} 
 \includegraphics[width=0.46\textwidth]{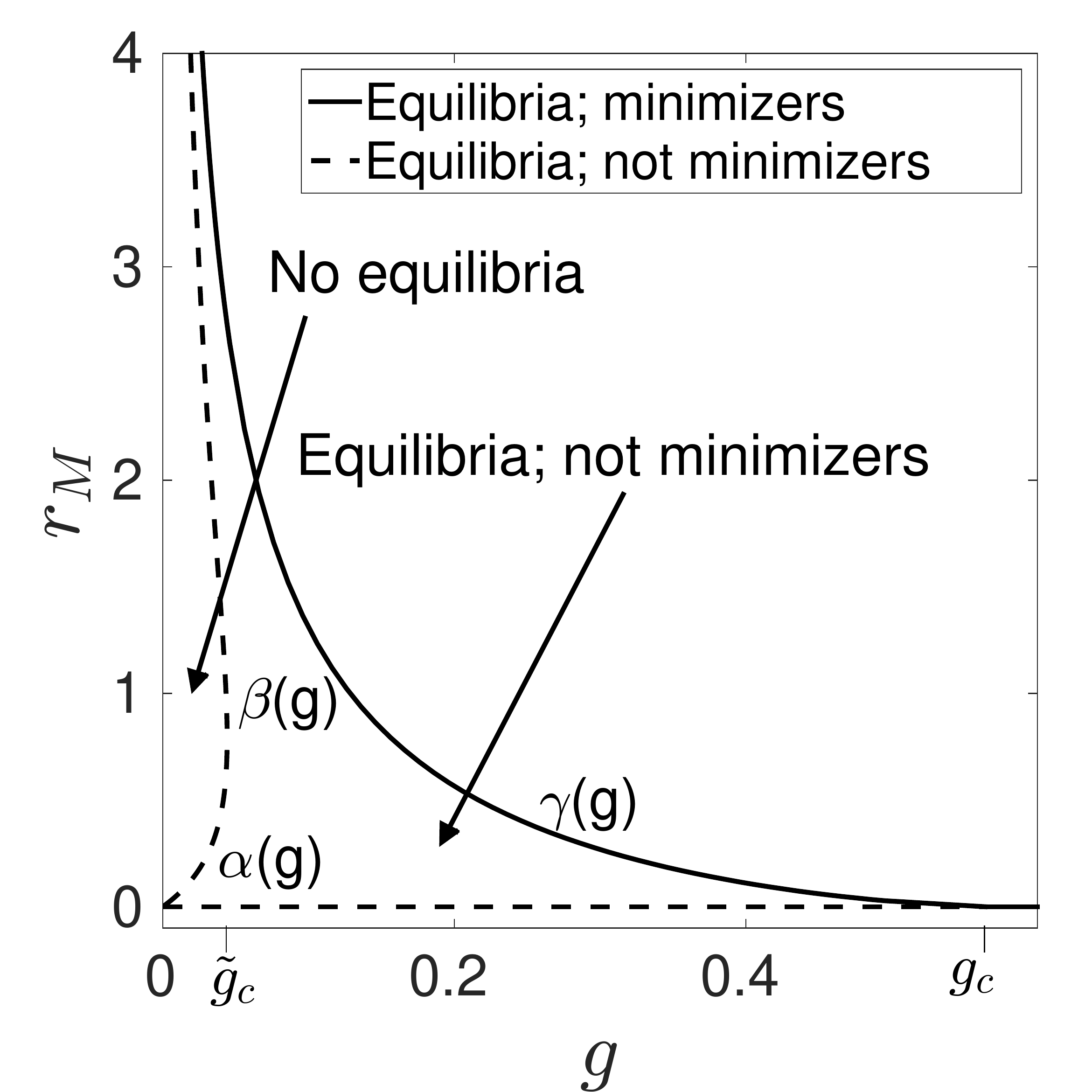} 
\end{center}
\hspace{0.265\textwidth} (a) \hspace{0.425\textwidth} (b)
\caption{Existence and stability of connected and disconnected equilibria. (a) One dimension, $V(x) = gx$, $\gc = 0.5$. For $0<g<\gc$, disconnected equilibria in the form \eqref{eqn:ss-oned} exist for all mass ratios $\mr \in  \Bigl( 0,\sqrt{\frac{\gc}{g}}-1 \Bigr)$; these equilibria are not energy minimizers. The only stable equilibrium is the connected state with $\mr = \sqrt{\frac{\gc}{g}}-1$.  For $g>\gc$, there exists no equilibrium in the form \eqref{eqn:ss-oned}.  The trivial equilibrium where all mass lies at the origin ($\mr=0$) is unstable for $g<\gc$, but it is a global minimizer when $g>\gc$. (b) Two dimensions, $V(x_1,x_2) = g x_1$, $\tgc \approx 0.044$, $\gc \approx 0.564$. For $0<g<\tgc$ disconnected equilibria in the form \eqref{eqn:2obs-equil} exist only for mass ratios $\mr \in (0, \ra(g)) \cup (\rb(g),\rc(g))$, while for $\tgc<g<\gc$ disconnected equilibria exist for {\em all} mass ratios $\mr \in (0,\rc(g))$; none of these disconnected equilibria are energy minimizers. The only stable equilibrium for $0<g<\gc$ is the connected state with $\mr = \rc(g)$. For $g>\gc$, there exists no equilibrium in the form \eqref{eqn:2obs-equil}.  The equilibrium \eqref{eqn:wall-sol} that has all mass on the wall  ($\mr=0$) is unstable for $g<\gc$, but it is a global minimizer when $g>\gc$.}
\label{fig:regions}
\end{figure}

A schematic of the existence and stability of equilibria in one dimension is shown in Figure \ref{fig:regions}(a). Note that the only stable equilibrium for $g<\gc$ is the connected state with $\mr = \sqrt{\frac{\gc}{g}}-1$.
Also, the closer the gravity to the critical value $\gc$, the smaller the range of possible mass ratios; at critical value $g=\gc$ the interval collapses to $\mr=0$ (no free swarm). On the other hand, in the limit of vanishing gravity $g \to 0$, an equilibrium exists for any mass ratio $\mr \in [0,\infty)$ (including infinite mass ratio), as consistent with the zero gravity case studied in Section \ref{subsect:1dnoV} --- see also Remark \ref{rmk:mr-oned-zerog}.

\bigskip
{\em \bf Case $\mathbf{g>\gc}$.} The equilibrium solution in this case consists in a delta-concentration at origin (see \eqref{eqn:linpot2}). As noted above, for such equilibrium,  \eqref{eqn:equilcomp} is equivalent to \eqref{eqn:1dg-allwall-min}, which holds trivially when $g \geq \frac{M}{2}$. We conclude from here that  \eqref{eqn:linpot2} is an energy minimizer. This fact is also illustrated in the schematic from Figure \ref{fig:regions}(a): the only (stable) equilibrium when $g>\gc$ is the configuration with all mass at the origin ($\mr=0$), which is in fact a global minimizer.

\subsection{Dynamic evolution of the aggregation model}
\label{subsect:1d-dyn}
In this section we investigate the dynamics of model \eqref{eqn:modelb}, with a focus on how and how often the equilibria  \eqref{eqn:ss-oned} are reached dynamically. In particular, we determine  under which perturbations the equilibria \eqref{eqn:ss-oned} are asymptotically stable.

\subsubsection{Reduced dynamics and basins of attraction}
\label{subsubsect:1d-CoMDyn}

In this study of the dynamics we assume a fixed amount of mass $S$ on the wall and an arbitrary density profile $\rho_2$ in the interior of $\Om$. We wish to quantify the dynamics of the support of $\rho_2$ and its centre of mass. We achieve explicit expressions defining the support of $\rho_2$ and its centre of mass which will hold up until mass would be transferred onto or off the wall. Furthermore we derive conditions for this transfer to happen, and thus identify when the assumption of having a fixed amount of mass on the wall is violated. 

Consider the evolution in \eqref{eqn:modelb} of a time-dependent density that has two distinct components:
\begin{equation}
\label{eqn:simpDyn-rhoDef}
\rho(x,t) = \rho_1(x) + \rho_2(x,t),
\end{equation}
where $\rho_1(x) = S\delta(x)$ is a delta aggregation at origin (with $S$ fixed) and $\rho_2(x,t)$ is the density profile of the free swarm, with support $\Om_2(t) = [a(t),b(t)]$. Here, $b(t) > a(t) > 0$ holds up until the time when the free swarm touches the wall. 

Let
\begin{equation}
\label{eqn:simpDyn-MassAndMoment}
M_2 = \int_{\Om_2(t)} \rho_2(x,t)\dx \quad \text{ and } \quad \CM(t) = \frac{\int_{\Om_2(t)} x\rho_2(x,t)\dx}{M_2},
\end{equation}
be the mass and the centre of mass of the free swarm, respectively. Note that since the mass on the wall is fixed, $M_2$ does not depend on $t$ and we have $M_2 = M - S$. 

Solutions of form \eqref{eqn:simpDyn-rhoDef} satisfy the equation \eqref{eqn:modelb} in the weak sense. Note that \eqref{eqn:modelb} is an equation in conservation law form and its weak formulation is standard \cite{EvansPDE}. Assume that in the free swarm the solution $\rho_2(x,t)$ is smooth enough so that \eqref{eqn:modelb} holds in the classical sense. By a standard argument \cite[Chapter 3.4]{EvansPDE} one can then derive the Rankine-Hugoniot conditions which give the evolution of the two discontinuities $a(t)$ and $b(t)$. For instance, the evolution of the left end is given by
\begin{equation}
\label{eqn:RH-a}
\frac{\rm{d}a}{\dt} = v(a,t),
\end{equation}
and by \eqref{eqn:vp}, \eqref{eqn:intpot}, \eqref{eqn:phi} we calculate
\begin{align*}
v(a,t) &= - aS + \frac{S}{2} - \int_{\Om_2} \left( a-y + \frac{1}{2} \right) \rho_2(y,t) dy - g \\
&= -Ma + M_2 \left(\CM - \frac{1}{2}\right) + \frac{S}{2} - g.
\end{align*}

By a similar calculation, 
\begin{align*}
 \frac{\rm{d}b}{\dt}&= v(b,t)\\
&= -Mb + M_2\left( \CM + \frac{1}{2} \right) + \frac{S}{2} - g.
\end{align*}

Finally, we derive the evolution of the centre of mass of $\rho_2$ and close the system. 
Multiply \eqref{eqn:contp} by $x$, integrate over $\Om_2$ and use integration by parts in the right-hand-side to get:
\begin{equation}
\label{eqn:simpDyn-FirstMani2}
\int_{\Om_2(t)} x(\rho_2)_t \dx = \left(x\rho_2(K\ast\rho_1 + K\ast\rho_2 + V)_x\right)\Big|_a^b - \int_{\Om_2} \rho_2(K\ast\rho_1 + K\ast\rho_2 + V)_x \dx.
\end{equation} 
By an elementary calculation,
\begin{equation}
\label{eqn:simpDyn-IntboundTerms}
\frac{\rm{d}}{\dt} \int_{\Om_2(t)} x\rho_2(x,t)\dx = \int_a^b x(\rho_2)_t \dx + \rho_2(b,t)b\frac{{\rm{d}} b}{\dt} - \rho_2(a,t)a\frac{{\rm{d}} a}{\dt}.
\end{equation}
Combine \eqref{eqn:simpDyn-FirstMani2} and \eqref{eqn:simpDyn-IntboundTerms} and use the evolution of $a(t)$ and $b(t)$ derived above. The boundary terms cancel and we find
\begin{equation}
\label{eqn:simpDyn-SecondMani}
M_2 \frac{\rm{d}\CM}{\dt}  = - \int_{\Om_2} \rho_2(K\ast\rho_1 + K\ast\rho_2 + V)_x \dx.
\end{equation}
By symmetry of $K$,
\begin{equation*}
\int_{\Om_2} \rho_2(K\ast\rho_2)_x \dx = 0,
\end{equation*}
and with \eqref{eqn:intpot} and $V(x) = gx$ we get
\begin{align*}
(K\ast\rho_1)_x = S\left(x - \frac{1}{2}\right), &\quad V_x = g, \nonumber \\
\int_{\Om_2} \rho_2(K\ast\rho_1 + V)_x \dx &= S M_2 \CM + M_2\left(g - \frac{S}{2}\right).
\end{align*}
Hence, from \eqref{eqn:simpDyn-SecondMani} one can derive the evolution of $\CM$, which together with the evolution of $a$ and $b$, yields the following system of evolution equations:
\begin{subequations}
\label{eqn:simpDyn-diffSoln}
\begin{gather}
\frac{\rm{d} \CM}{\dt} = -S \CM + \left(\frac{S}{2} - g\right), \label{eqn:simpDyn-diffSolnM}\\
\frac{{\rm{d}} a}{\dt} = -Ma+ M_2 \left(\CM - \frac{1}{2} \right) + \frac{S}{2} - g, \label{eqn:simpDyn-diffSolna}\\
\frac{{\rm{d}} b}{\dt} = -Mb + M_2\left( \CM + \frac{1}{2} \right) + \frac{S}{2} - g. \label{eqn:simpDyn-diffSolnb}
\end{gather}
\end{subequations}

It is now an elementary exercise to solve \eqref{eqn:simpDyn-diffSoln} for $\CM(t)$, $a(t)$, and $b(t)$ given initial data $\CM(0), a(0), b(0)$. A first observation is that provided our assumptions hold for all $t \geq 0$ (i.e., the mass on the wall is fixed and $a(t)>0$), the equilibrium solution for \eqref{eqn:simpDyn-diffSoln} corresponds to the disconnected state \eqref{eqn:ss-oned}. Indeed, one can check that at the equilibrium for \eqref{eqn:simpDyn-diffSoln}, $a= \dl$, $b = \dl + \dr$, and $\CM= \dl + \frac{\dr}{2}$, with $\dl$ and $\dr$ given by \eqref{eqn:cond-d2g} in terms of $S$.

Next we wish to use the reduced dynamics to determine under which perturbations the disconnected equilibria \eqref{eqn:ss-oned} are asymptotically stable.  By inspecting the profile of $\Lambda(x)$, we have already observed that these equilibria are unstable under infinitesimal perturbations which move mass off the wall  (see Figures \ref{fig:oned-zerog} and \ref{fig:oned-wg}). Therefore disconnected equilibria can only be (asymptotically) stable with respect to perturbations of the free swarm.  We take such a perturbation and consider the evolution of a density of the form
\begin{equation}
\label{eqn:pert-rho2}
\rho(x,t) = \barrho(x) + \trho_2(x,t), 
\end{equation}
where $\barrho$ is the disconnected equilibrium \eqref{eqn:ss-oned} and $\trho_2$ has support away from the origin and zero mass. Note that density \eqref{eqn:pert-rho2} can also be written in the separated form \eqref{eqn:simpDyn-rhoDef}, where
\begin{equation}
\label{eqn:pert-rho3}
\rho_1(x) = S \delta(x) \qquad \text{ and } \qquad \rho_2(x,t) = M \One_{(\dl, \dl+\dr)}(x) + \trho_2(x,t)
\end{equation}

The reduced dynamics  \eqref{eqn:simpDyn-diffSoln} can be used to track the dynamics of the centre of mass $\CM(t)$ and the support $[a(t),b(t)]$ of $\rho_2(x,t)$, provided: 

(i) no mass leaves the origin, {\em and} 

(ii) no mass transfers from $\rho_2$ to the origin. 

We will quantify now when (i) and (ii)  can happen. To address (i), one needs to inspect the velocity at origin, which computed by \eqref{eqn:vp} and \eqref{eqn:intpot} (see also \eqref{eqn:simpDyn-MassAndMoment}) gives:
\begin{align*}
v(0,t) & = P_0 \left( \int_{\Om_2(t)} \left( y - \frac{1}{2} \right)  \rho_2(y,t) \dy - g \right) \\
& = P_0 \left( M_2\left( \CM(t)- \frac{1}{2}\right) - g \right).
\end{align*} 
We find that no mass leaves the origin ($v(0,t)=0$) provided 
\begin{equation}
\label{eqn:no-off-wall}
\CM(t) \leq \frac{1}{2} + \frac{g}{M_2}. 
\end{equation}
In particular, an initial perturbation $\trho_2(\cdot,0)$ in \eqref{eqn:pert-rho2} must satisfy this restriction at $t=0$.

\begin{figure}[ht]
  \begin{center}
 \includegraphics[width=0.46\textwidth]{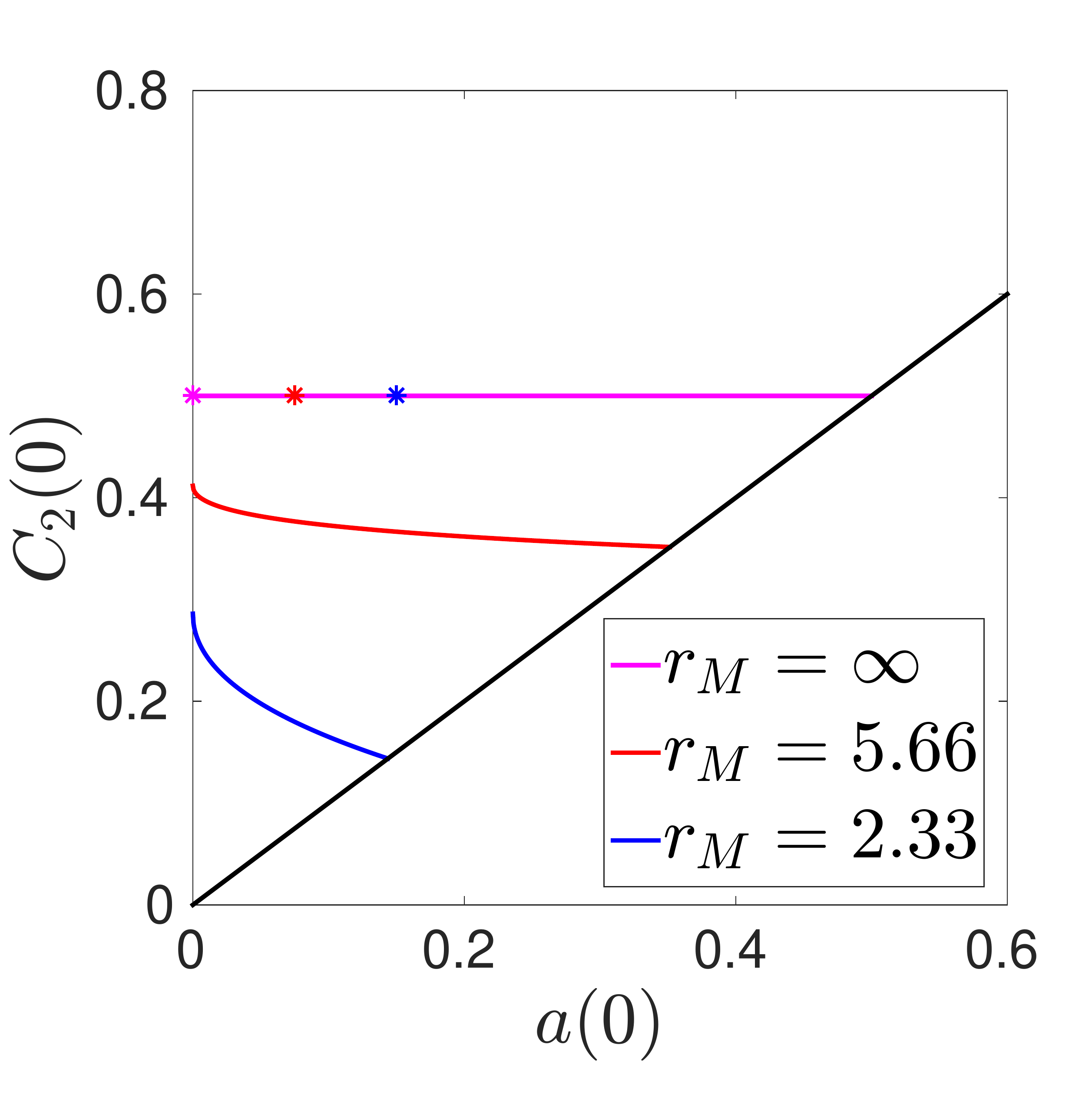} 
 \includegraphics[width=0.46\textwidth]{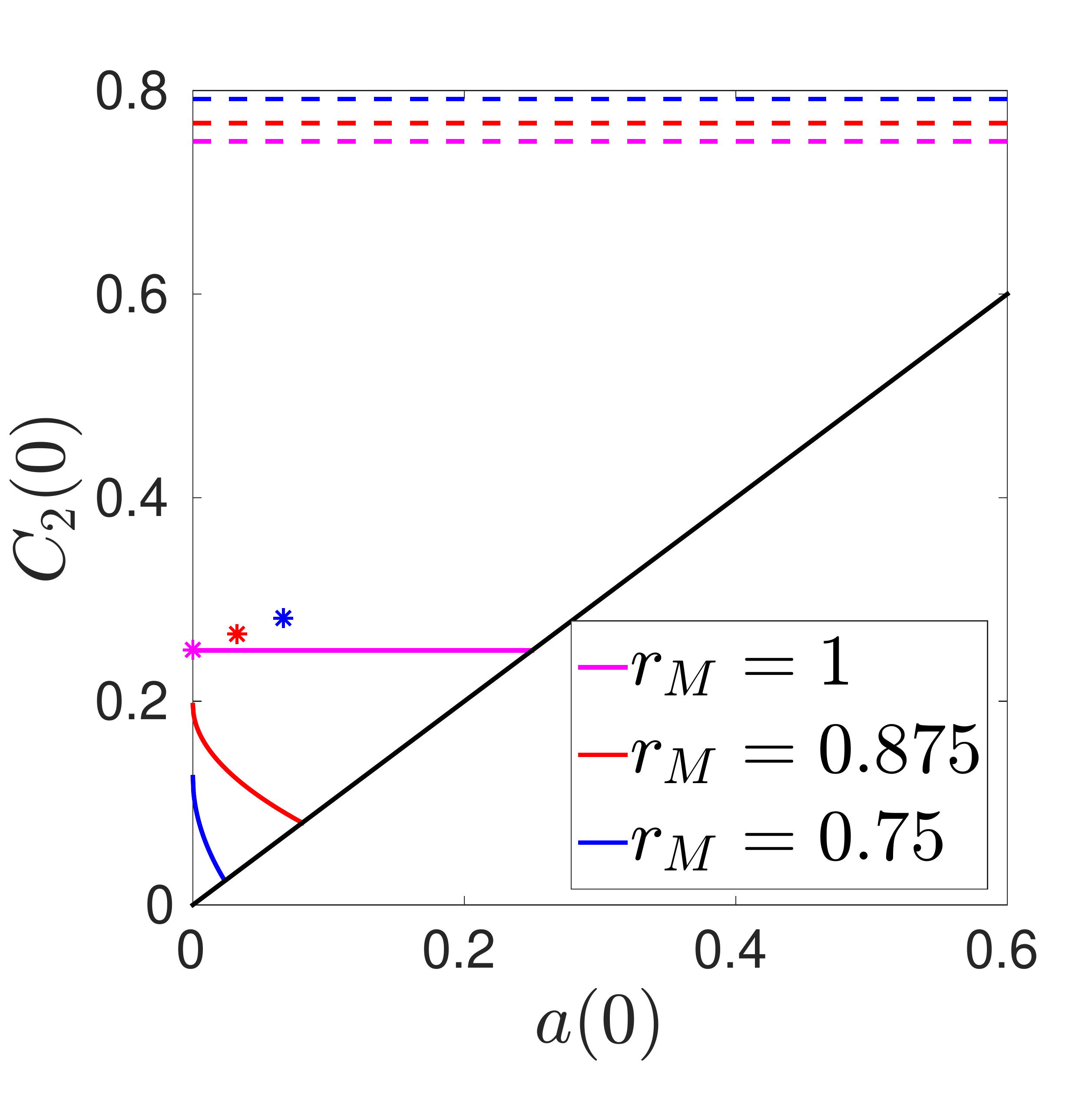} 
\end{center}
\hspace{0.265\textwidth} (a) \hspace{0.425\textwidth} (b)
\caption{Disconnected equilibria \eqref{eqn:ss-oned} are asymptotically attracting certain initial densities of type \eqref{eqn:simpDyn-rhoDef}. Considered are three mass ratios, one of which being the mass ratio of the minimizer (purple), for $g=0$ and $g=0.125$. An initial perturbation of $\barrho$ (see \eqref{eqn:pert-rho2} and \eqref{eqn:pert-rho3}) that has $a(0)$ and $\CM(0)$ in the region above the solid curves and $\CM(0) \leq 1/2 + {g}/{M_2}$,  will evolve dynamically to the disconnected equilibrium of the corresponding mass ratio. An initial condition with $a(0)$, $\CM(0)$ below the curves will evolve dynamically to an equilibrium of a smaller mass ratio. The horizontal dashed lines in figure (b) indicate the thresholds ${1}/{2} + {g}/{M_2}$ above which mass on the wall would lift off. Stars indicate the equilibrium locations for the colour-related line.}
\label{fig:simpDyn-breakDown}
\end{figure}

For (ii), we note that mass transfer occurs when the left end of the support of the free swarm meets the wall and pushes into it. Mathematically, this amounts to have $a=0$ and $\frac{{\rm d} a}{\dt} < 0$ hold simultaneously. Otherwise, no transfer from the free swarm into the wall can take place. Using the exact solution for \eqref{eqn:simpDyn-diffSoln} (note that \eqref{eqn:simpDyn-diffSolnM} and \eqref{eqn:simpDyn-diffSolna} can be solved separately from \eqref{eqn:simpDyn-diffSolnb}), we can locate the initial conditions $a(0)$, $\CM(0)$ for which (ii) holds for all times. Figure \ref{fig:simpDyn-breakDown} illustrates the initial data $a(0)$, $\CM(0)$ with this property, for various mass ratios $\mr$ (or equivalently, for various delta strengths $S$). Specifically, no mass transfer occurs for initial data in the region {\em above} the solid curve(s); also note that necessarily, $\CM(0) > a(0)$. 

Take a disconnected equilibrium $\barrho$ and an initial perturbation $\trho_2$ of the free swarm such that $a(0)$, $\CM(0)$ is in the region which guarantees that (ii) holds for all times; for the mass ratios considered in Figure \ref{fig:simpDyn-breakDown}, this amounts to taking $a(0)$, $\CM(0)$ above the corresponding solid curves. Also choose the initial perturbation such that $\CM(0)$ satisfies \eqref{eqn:no-off-wall}. For zero gravity this simply means  $\CM(0) \leq \frac{1}{2}$, regardless of which equilibrium $\barrho$ we consider perturbations about. For non-zero gravity however, the threshold ${1}/{2} + {g}/{M_2}$ depends on the equilibrium $\barrho$; for the mass ratios in Figure \ref{fig:simpDyn-breakDown}(b) these thresholds are indicated by dashed lines. It is now easy to see from the dynamics \eqref{eqn:simpDyn-diffSolnM} of $\CM$ that once \eqref{eqn:no-off-wall} is satisfied at the initial time, it will be satisfied for all times. Indeed, 
\eqref{eqn:simpDyn-diffSolnM} simply drives $\CM$ monotonically to the equilibrium value at $\frac{1}{2} - \frac{g}{S}$.

These considerations imply that, starting from such an initial perturbation, conditions (i) and (ii) are satisfied for all times, and hence, \eqref{eqn:simpDyn-diffSoln} can be used to track the support and centre of mass of $\rho_2$. As noted above, the equilibrium for \eqref{eqn:simpDyn-diffSoln} recovers the centre of mass and the support of the free component of $\barrho$; in Figure \ref{fig:simpDyn-breakDown} the equilibrium locations are indicated by stars. From \eqref{eqn:evol} we know that at the equilibrium of model \eqref{eqn:modelb}, $\rho_2$ equals $M$ everywhere on its support. Combining these facts, we conclude that the same equilibrium $\barrho$ that we have perturbed about, is reached asymptotically, and hence, it is asymptotically stable with respect to the perturbations $\trho_2$ that have been considered here.

Certain remarks are in order. 
\begin{remark}
\label{rm:basin-attraction} The calculations above do not restrict the size of the perturbations $\trho$ in \eqref{eqn:pert-rho2}; the only restrictions are placed on the centre of mass $\CM(0)$ and the left-end point $a(0)$ of the perturbed free swarm at the initial time. Consequently, the basins of attraction of the disconnected equilibria are considerable in size and highly nontrivial. Section \ref{subsubsect:1d-NumIniMat} will elaborate further on this point. 
\end{remark}

\begin{remark}
\label{rm:connected}
The connected equilibria ($\mr = \infty$ and $\mr=1$) in Figure \ref{fig:simpDyn-breakDown} and their corresponding magenta curves have been included only for illustration. Strictly speaking, we should have shown only mass ratios that correspond to disconnected equilibria, for which the considerations in this subsection hold. Nevertheless, by a continuity argument,  the magenta lines can be thought to correspond to disconnected equilibria that are arbitrarily close to the connected states. In fact, we infer from the figure that  if we perturb the connected equilibrium such that the centre of mass of the free swarm decreases (the centres of mass of the connected equilibria for $g=0$ and $g=0.125$ are at $1/2$ and $1/4$, respectively), then mass will transfer to the wall and result dynamically in a disconnected state.
\end{remark}

\begin{remark}
\label{rmk:not-achieved}
Regarding the solid curves in Figure \ref{fig:simpDyn-breakDown}, we found that there is a minimal mass ratio ($\mr \approx 1$ for $g=0$ and $\mr \approx 0.6$ for $g=0.125$) below which these curves do not cross through the relevant $0 < a(0) < \CM(0)$ region. For such mass ratios, any initial perturbation with $0<a(0)<\CM(0) \leq 1/2$ would dynamically result in (i) and (ii) being satisfied for all times, and hence, equilibrium $\barrho$ being achieved at steady state. On the other hand, this observation also implies that equilibria $\barrho$ with mass ratios below this threshold cannot be achieved dynamically starting from initial densities with different mass ratios (as no mass transfer into the origin occurs below the threshold). This fact is also supported by the numerical simulations in Section \ref{subsubsect:1d-NumIniMat}.
\end{remark}

\subsubsection{Non-trivial initial conditions leading to disconnected equilibria}
\label{subsubsect:1d-NumIniMat}

We show in this section that a wide range of initial conditions can lead to disconnected states. Furthermore we show that the mass ratios of the resultant states follow trends related primarily to the initial centre of mass and secondarily to how close the swarm is to the wall. 

To this end we consider initial states of particles with positions randomly generated from a uniform distribution on $(\dli,\dli+\drj)$, where  $1 \leq i, j \leq 10$, and
\begin{align}
\label{eqn:nonTriv-initCondPairs}
\dli = \frac{1}{20}(i-1), \quad \drj = &\frac{1}{10}j, \quad {\rm for} \quad g=0, \\
\dli = \frac{1}{40}(i-1), \quad \drj = &\frac{1}{20}j, \quad {\rm for} \quad g=0.125.
\end{align}

We ran $50$ particle simulations of $N = 1024$ particles for each interval $(\dli,\dli+ \drj)$, with $1 \leq i \leq 10$ and $1 \leq j \leq 10$. We evolved the particle simulations until the state is steady and calculated the mass ratio of the resultant state. 

For convenience of discussion later, denote the midpoint of the initial interval,
\begin{equation}
\label{eqn:nonTriv-midPt}
\md = \frac{1}{2}\left(\dli + \left(\dli + \drj\right)\right).
\end{equation}
We mention here as well that the centre of mass of the initial swarm will be close to this midpoint as we have drawn particle positions from a uniform distribution. This is particularly important in comparing with the results of Section \ref{subsubsect:1d-CoMDyn}, as the intervals $(\dli,\dli+\drj)$ have been constructed in such a way that for $i = 11 - j$ we have $\md = \frac{1}{2}$ for $g = 0$ and $\md = \frac{1}{4}$ for $g = 0.125$, in consideration of Figure \ref{fig:simpDyn-breakDown} and Remark \ref{rm:connected}. 

\begin{figure}[ht]
  \begin{center}
 \includegraphics[width=0.4\textwidth]{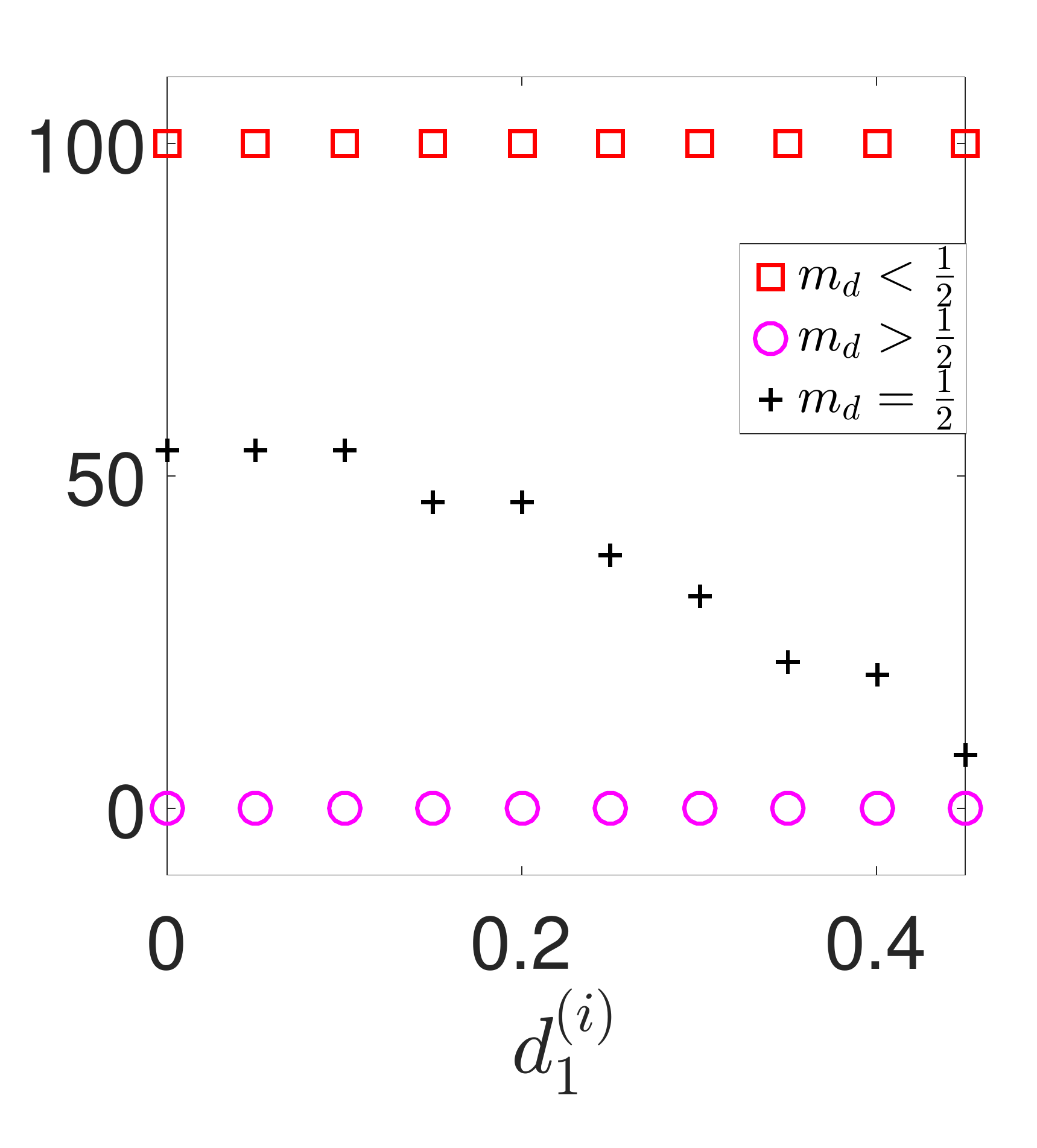} \hspace{1.1cm}
 \includegraphics[width=0.4\textwidth]{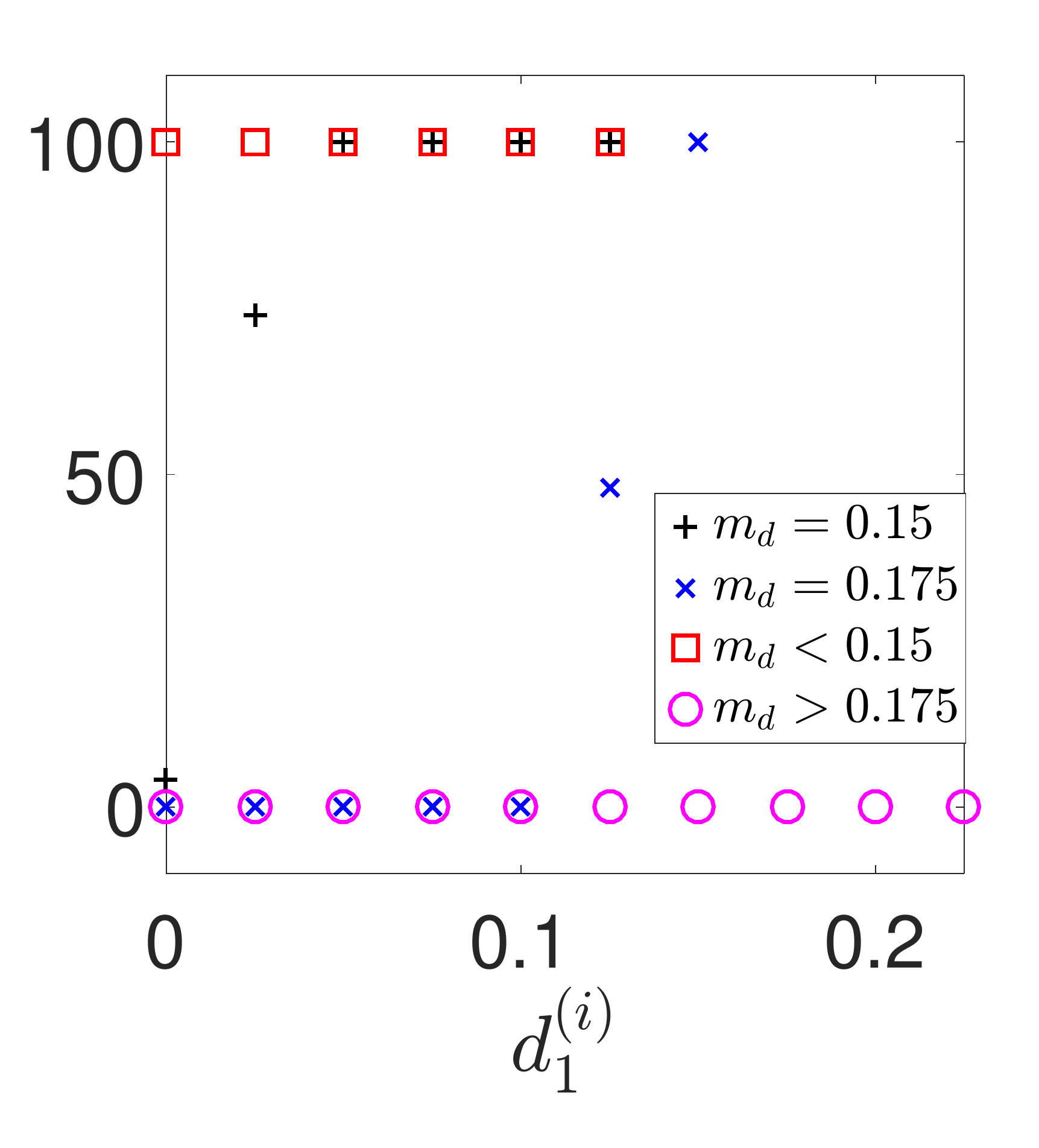} 
\end{center}
\hspace{0.24\textwidth} (a) \hspace{0.44\textwidth} (b)
\caption{Percentage of initial states which resulted in disconnected states for (a) $g=0$ and (b) $g=0.125$. Note that we find disconnected resultant states for a significant set of initial data.}
\label{fig:nonTriv-PercDisc}
\end{figure}

We find that for the $g=0$ case, all initial states with $\md < \frac{1}{2}$ resulted in a disconnected state and all initial states with $\md > \frac{1}{2}$ resulted in a connected state -- see squares and circles in Figure \ref{fig:nonTriv-PercDisc}(a) indicating percentages of disconnected states. This result is consistent with Remark \ref{rm:connected}. Furthermore we see that for $\md = \frac{1}{2}$ some resultant states were disconnected and some were connected, accounting for the fact that due to the random distribution of particles sometimes we have the centre of mass larger than $\frac{1}{2}$ and sometimes smaller -- see Figure \ref{fig:nonTriv-PercDisc}(a). 

For the case of $g=0.125$ we find relatively similar results. Figure \ref{fig:nonTriv-PercDisc}(b) shows that for $\md < 0.15$ we always get disconnected resultant states and for $\md > 0.175$ we always get the connected equilibrium. For $\md = 0.15$ and $\md = 0.175$ the results can be mixed. We suspect in fact that there may be discrete numerical effects and that the true value of $\md$ where we observe variability in disconnected/connected resultant states is closer to $\frac{1}{4}$, in support of Remark \ref{rm:connected}. 
A discrete effect that occurs for $g=0$ for instance is when the correct resultant state has mass on the wall which is greater than zero but less than that of two particles; this case cannot be identified by the particle method as disconnected. Also note that this error, which favours connected states, is more likely to occur for larger $\dli$.

\begin{figure}[ht]
  \begin{center}
 \includegraphics[width=0.4\textwidth]{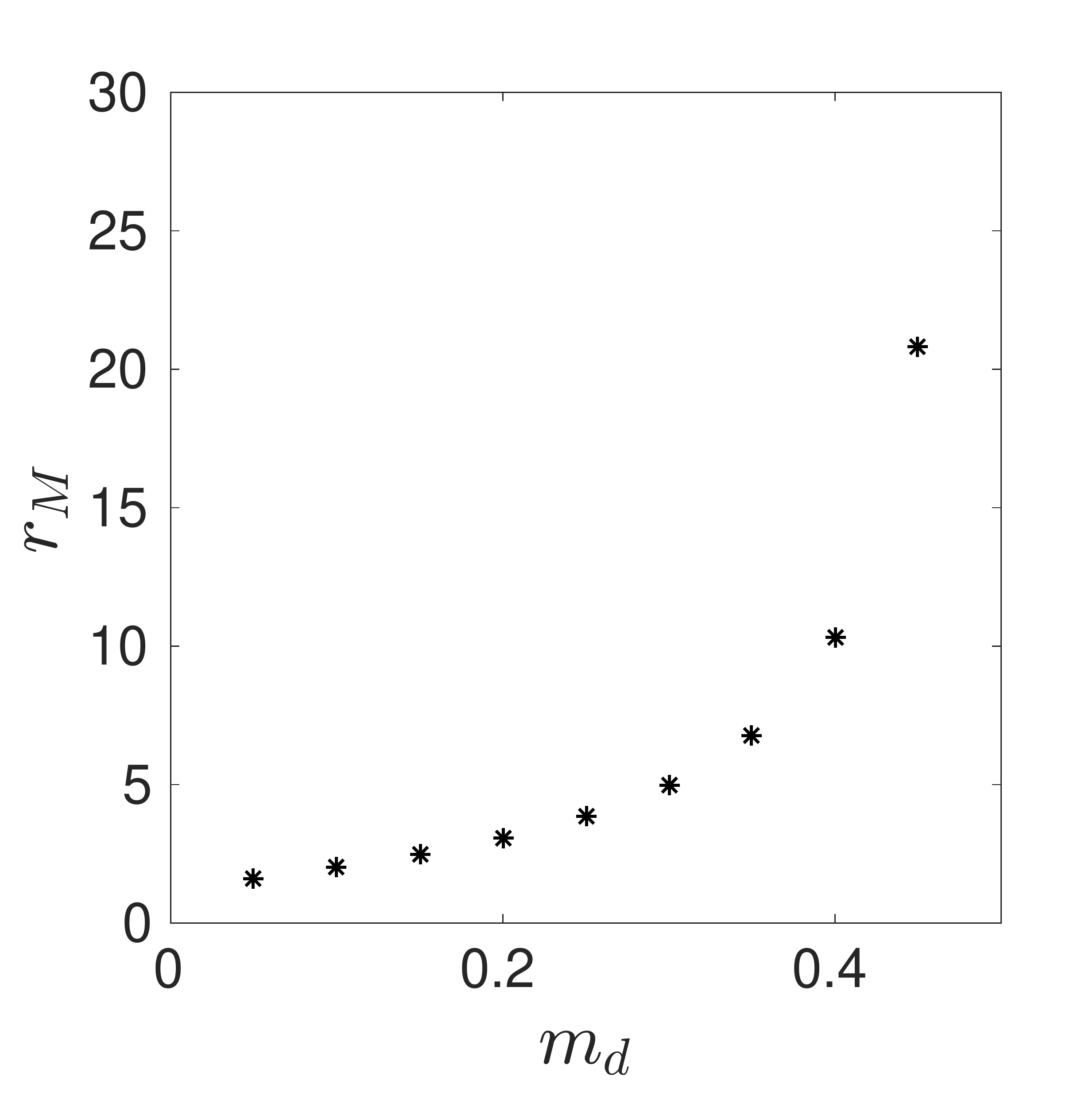} \hspace{1.1cm}
 \includegraphics[width=0.4\textwidth]{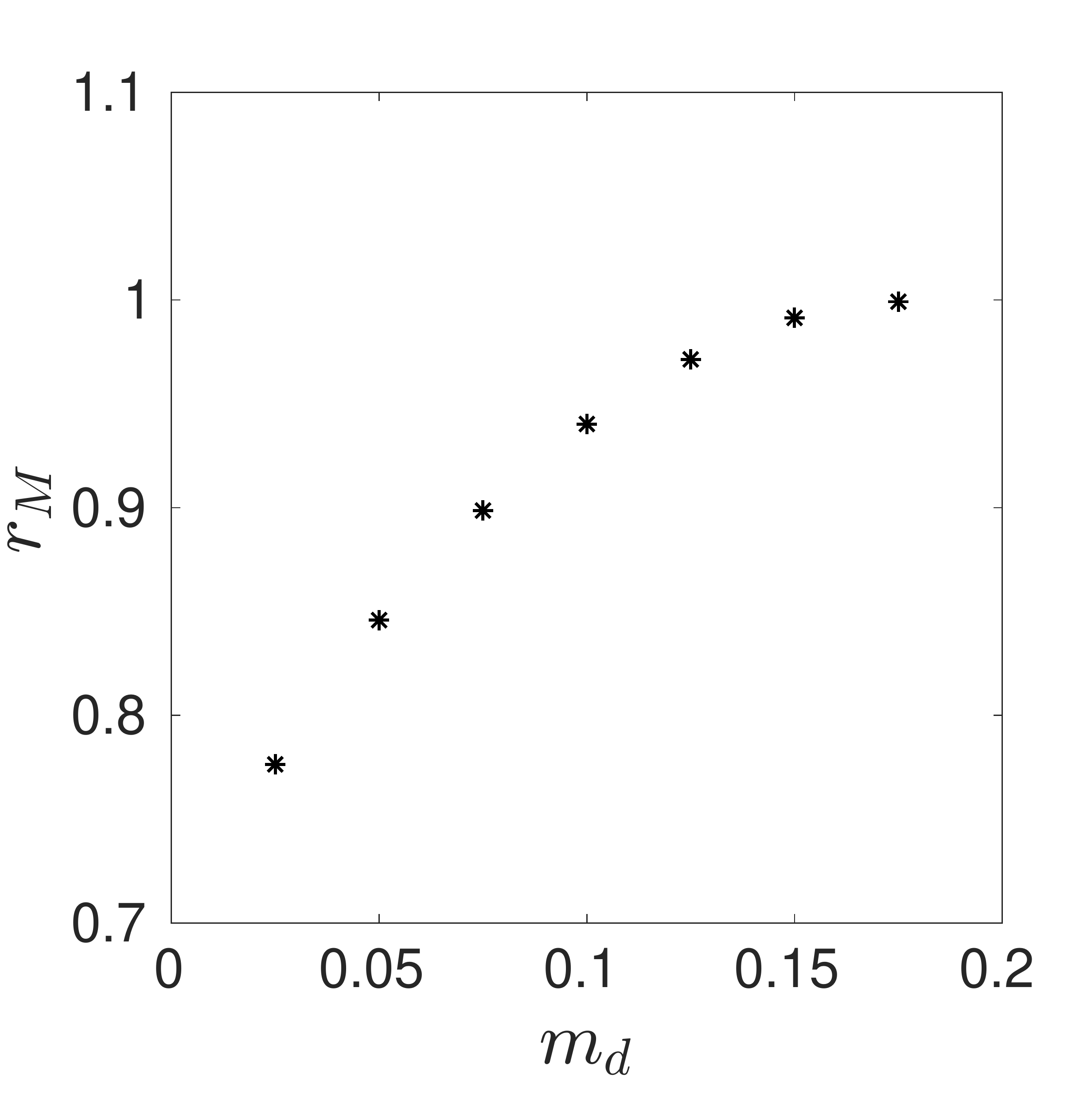} 
\end{center}
\hspace{0.24\textwidth} (a) \hspace{0.44\textwidth} (b)
\caption{Average mass ratios of resultant states for particular $\md$ for (a) $g = 0$ and (b) $g = 0.125$. $\md = \frac{1}{2}$ has been neglected from (a) for clarity as the average is much larger (about $516$).}
\label{fig:nonTriv-MeanrM}
\end{figure}

We also computed the mass ratios of the resultant equilibria in these simulations and averaged over runs that have the same initial midpoint $\md$. In both the gravity and no gravity cases we found that the average resultant mass ratio tends to be smaller for smaller $\md$ -- see Figure \ref{fig:nonTriv-MeanrM}. The results further support that there indeed exists a minimal mass ratio for equilibria that can be achieved dynamically -- see Remark \ref{rmk:not-achieved}.   

\subsubsection{Discrete energy dissipation}
\label{subsubsect:1d-Ediss}
%

We wish to demonstrate that particle simulations which lead to disconnected states obey a discrete-space, discrete-time analog to \eqref{eqn:gflow}. Let $\Delta t$ be the length of time steps taken in a given particle simulation (see Section \ref{subsubsect:ssnumPart} for details on the implementation of the particle method) and further let $s = n\Delta t$ and $t = (n+1)\Delta t$ for $n \geq 0$ be two successive times in \eqref{eqn:gflow}. We then have, after dividing by $\Delta t$,
\begin{equation}
\label{eqn:Ediss-timeStepVer}
\frac{E[(n+1)\Delta t] - E[n\Delta t]}{\Delta t} = - \frac{1}{\Delta t}\int_{n\Delta t}^{(n+1)\Delta t}\int_\Om |P_x(-\nabla K \ast \rho(x,\tau) - \nabla V(x))|^2 \rho(x,\tau) \dx \dtau.
\end{equation}

We now check whether \eqref{eqn:Ediss-timeStepVer} holds in numerical simulations..
To this purpose we transfer to a discrete space analog as per a particle simulation with particles of equal weight, namely $\frac{M}{N}$, where $N$ is the number of particles. Let $x_i$ represent the position of particle $i$. The discrete density is a superposition of delta accumulations at the particle locations:
\[
\rho^N(x,t) = \frac{M}{N} \sum_{i=1}^N \delta (x-x_i(t)),
\]
and the corresponding discrete energy (see \eqref{eqn:energy}) is given by
\begin{equation}
\label{Ediss-EdiscSpace}
E[\rho^N] = \frac{M^2}{2N^2}\sum_{i =1}^N \sum_{j =1}^N K(x_i(t) - x_j(t)) + \frac{M}{N}\sum_{i=1}^N V(x_i(t)).
\end{equation}
Note that $K(0) = 0$ in this context so we do not need to exclude the case of $i=j$ in the double sum representing social interaction. 

\begin{figure}[ht]
  \begin{center}
 \includegraphics[width=0.4\textwidth]{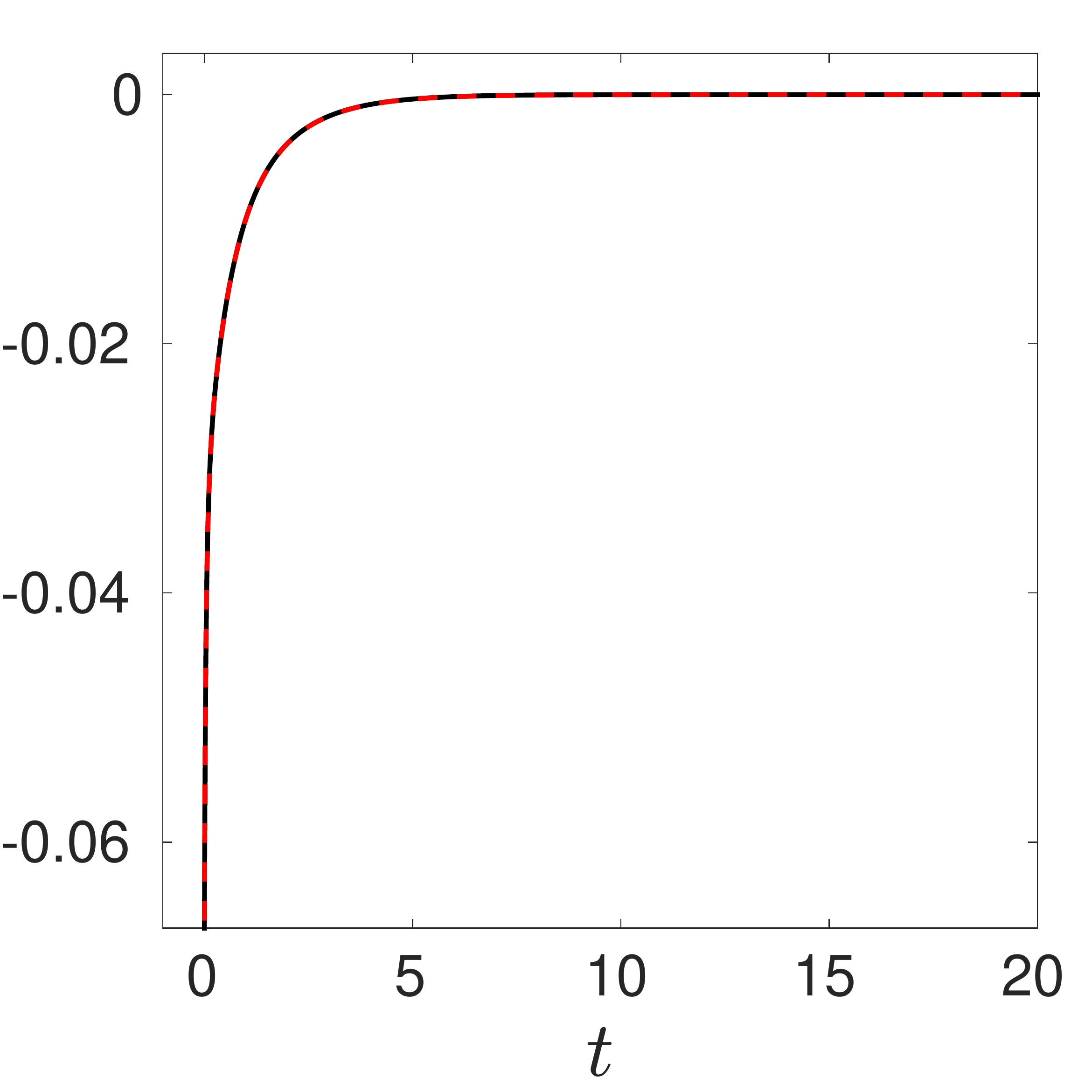} \hspace{1.1cm}
 \includegraphics[width=0.4\textwidth]{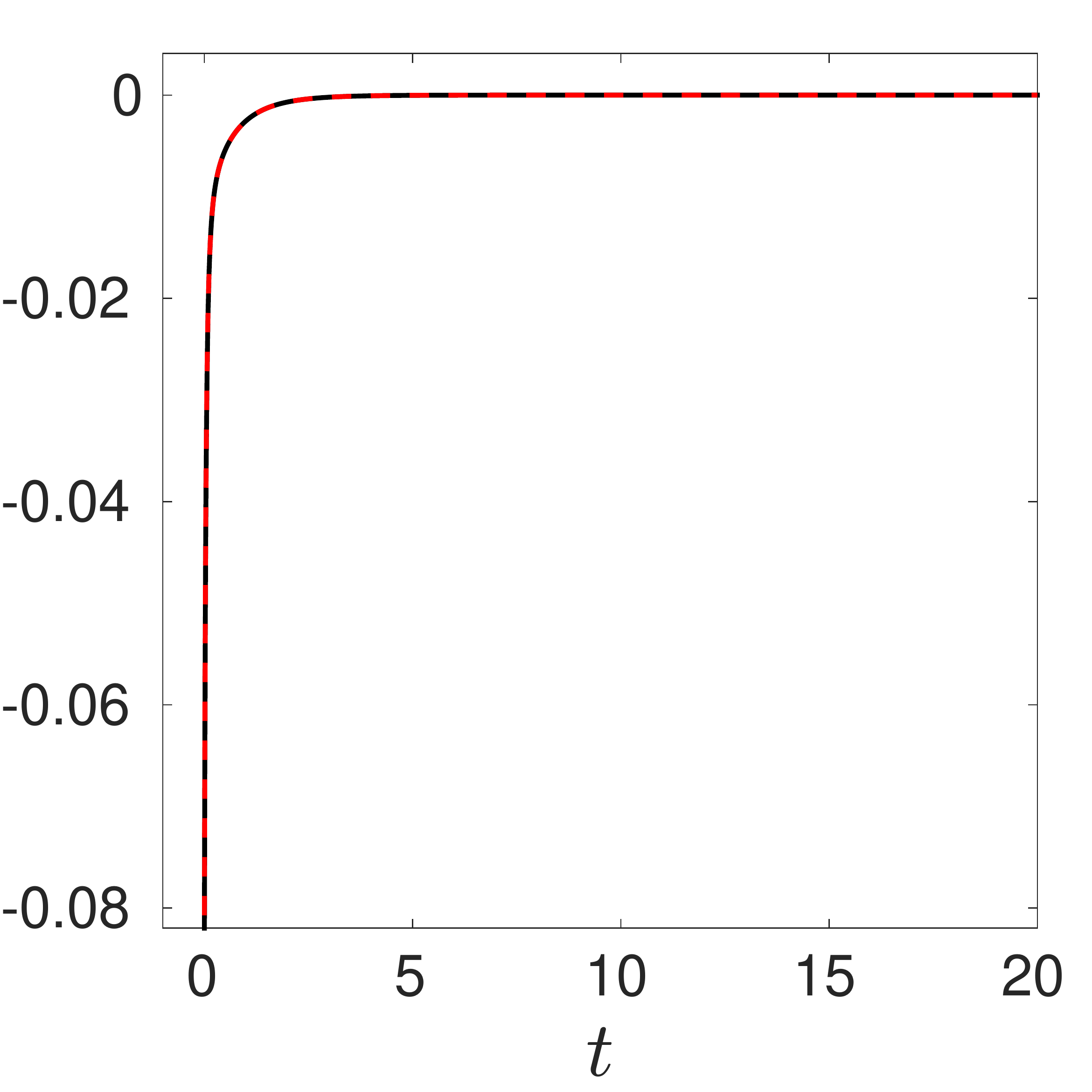} 
\end{center}
\hspace{0.24\textwidth} (a) \hspace{0.44\textwidth} (b)
\caption{Verification of the energy dissipation formula \eqref{eqn:gflow} for particle simulations where more than one particle joins the wall and disconnected states are emerging.  (a)  $g=0$ and (b) $g=0.125$. Approximations of the left-hand-side of \eqref{eqn:Ediss-timeStepVer} (solid lines) and the right-hand-side of \eqref{eqn:Ediss-timeStepVer} (dashed lines) fall within the discretization error, and are indistinguishable at the scale of the figure.}
\label{fig:1d-Ediss}
\end{figure}

Figure \ref{fig:1d-Ediss} shows (solid lines) the left-hand-side of \eqref{eqn:Ediss-timeStepVer} as computed from a particle simulation, with $E$ given by \eqref{Ediss-EdiscSpace}. The simulations correspond to emerging disconnected equilibria. Shown there are $g=0$ and $g=0.125$. Also plotted in the figure (dashed lines), but indistinguishable at the scale of the figure, are discrete-time approximations of the right-hand-side of \eqref{eqn:Ediss-timeStepVer}; we used the trapezoidal rule to approximate the time integration and a discrete analog of the projection operator $P_x$ (c.f., Section \ref{subsubsect:ssnumPart}). The difference between the two computed approximations fall within the discretization error of the particle method and hence, the results show that the energy dissipation formula \eqref{eqn:Ediss-timeStepVer} holds at the discrete level. Note also that the energy decays for all times and levels off as it approaches the equilibrium.

\subsection{Morse potential}
\label{subsect:1d-Morse}

In this subsection we consider the same problem setup as in Section \ref{subsect:1dnoV} except we use the Morse-type potential investigated in \cite{BeTo2011}:
\begin{equation}
\label{eqn:1dm-kernel}
 K(x) = -GLe^{-|x|/L} + e^{-|x|}.
\end{equation}
We consider the case of $G = 0.5$ and $L = 2$ throughout this subsection as these values were one of the cases highlighted in \cite{BeTo2011}. The main point we want to make is that the findings above apply to the Morse potential as well. In particular, there is a one-parameter family of disconnected equilibria to model \eqref{eqn:modelb}, which are not energy minimizers, but are realized dynamically starting from a variety of initial conditions.

One can find explicit forms for the equilibria for the Morse potential in just the same way as we found explicit forms for the potential \eqref{eqn:intpot}-\eqref{eqn:phi}. We assume the solution form
\begin{equation}
\label{eqn:morseSoln}
\barrho(x) = S\delta(x) + \rho_*(x)\One_{(\dl, \dl+\dr)},
\end{equation}
with
\begin{align}
\rho_*(x) = & C\cos(\mu x) + D\sin(\mu x) - \frac{\lambda_2}{\epsilon}, \nonumber \\
\mu = \sqrt{\frac{\epsilon}{\nu}}, & \quad \epsilon = 2(GL^2 - 1), \quad \nu = 2L^2(1 - G).
\label{eqn:morseForm}
\end{align}
The density $\rho_*$ of the free swarm comes from the free space solution found in \cite{BeTo2011}. 

Then we seek to satisfy, c.f. \eqref{eqn:Lambda-const} and \eqref{eqn:massrho},
\begin{equation}
\label{eqn:morseCond}
\Lambda(0) = \lambda_1 \quad \Lambda(x) = \lambda_2 \quad \text{ for } x \in [\dl,\dl+\dr], \quad  S + \int_{\dl}^{\dl+\dr}\rho_*(x)\dx = M.
\end{equation}
The Appendix shows the system of equations that arise from these conditions. We end up with four equations from requiring $\Lambda(x) = \lambda_2$ for $x \in [\dl,\dl+\dr]$, as one can find that the constant term on the left-hand-side is already $\lambda_2$ and the non-constant terms comprise four linearly independent terms in $x$:
\begin{equation}
\label{eqn:morseIndep}
-GL\exp\left(-\frac{x}{L}\right), \quad \exp(-x), \quad -GL\exp\left(\frac{x}{L}\right), \quad \exp(x).
\end{equation}
Together with $\Lambda(0) = \lambda_1$ and the mass constraint, this yields a system of six equations for seven unknowns ($C$, $D$, $S$, $\dl$, $\dr$, $\lambda_1$, and $\lambda_2$), indicating a one-parameter family of equilibria, as in Sections \ref{subsect:1dnoV} and \ref{subsect:1d-linV}.  We solved numerically this system for various $S \in [0,1]$ fixed.

\begin{figure}[thb]
  \begin{center}
 \includegraphics[width=0.32\textwidth]{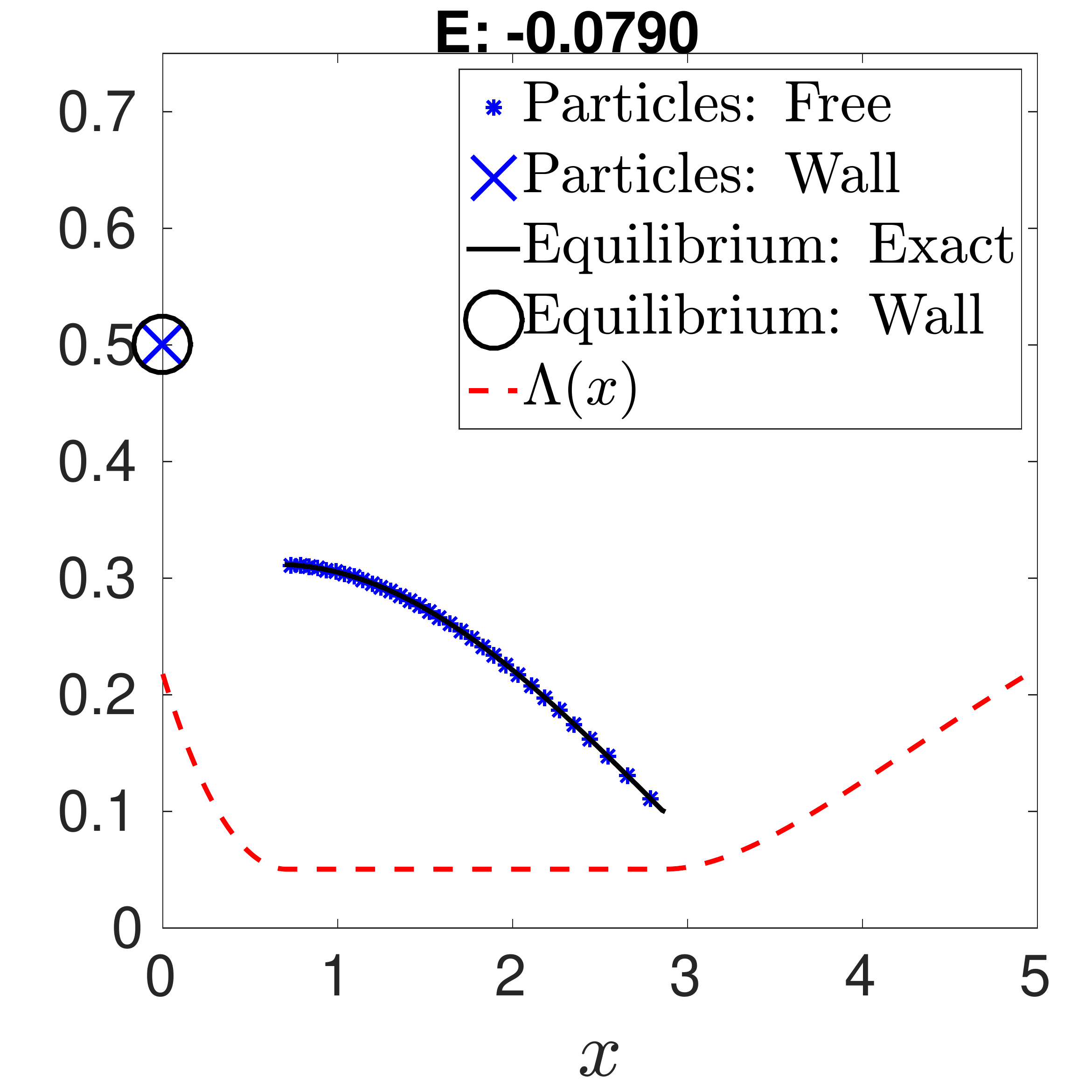} 
 \includegraphics[width=0.32\textwidth]{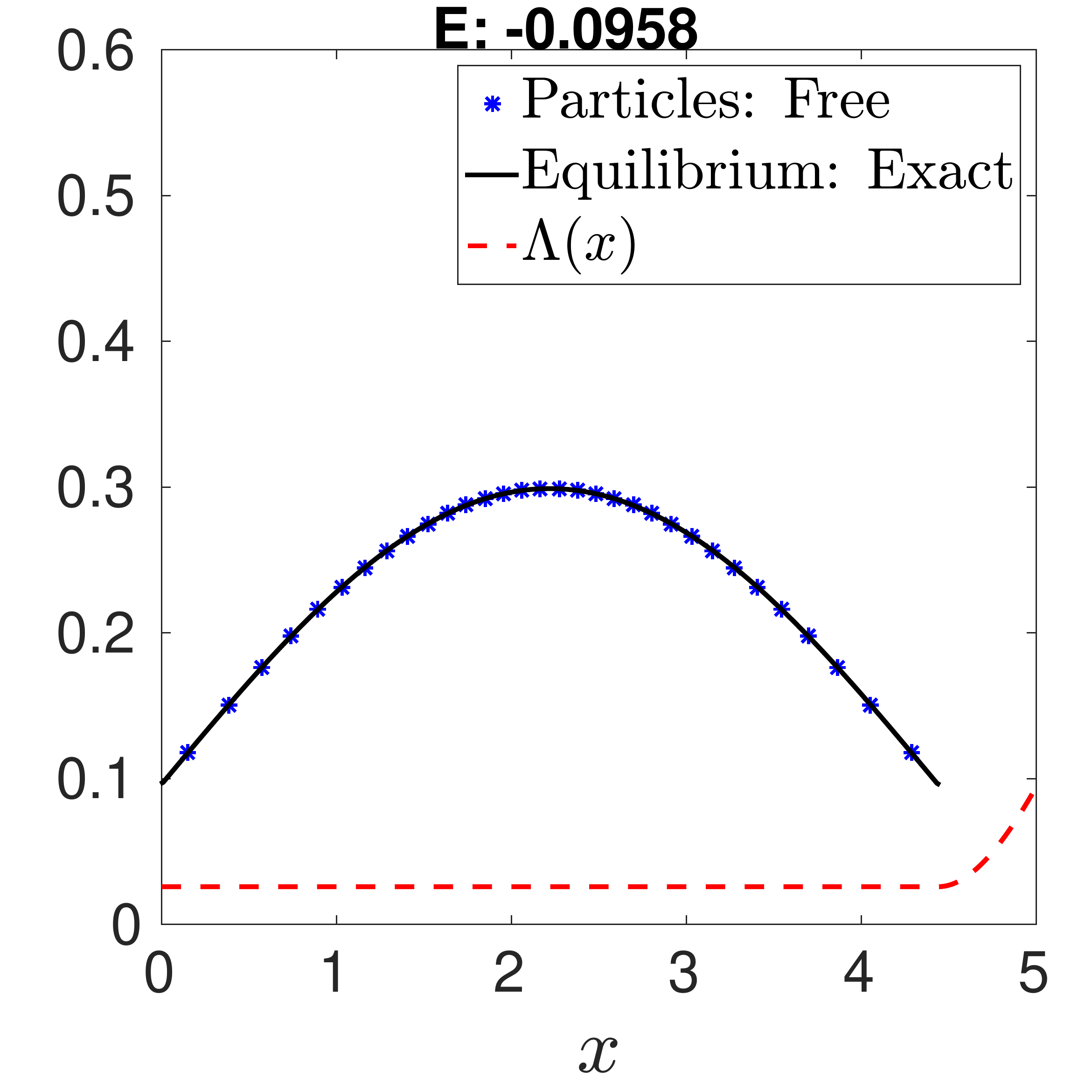} 
 \includegraphics[width=0.32\textwidth]{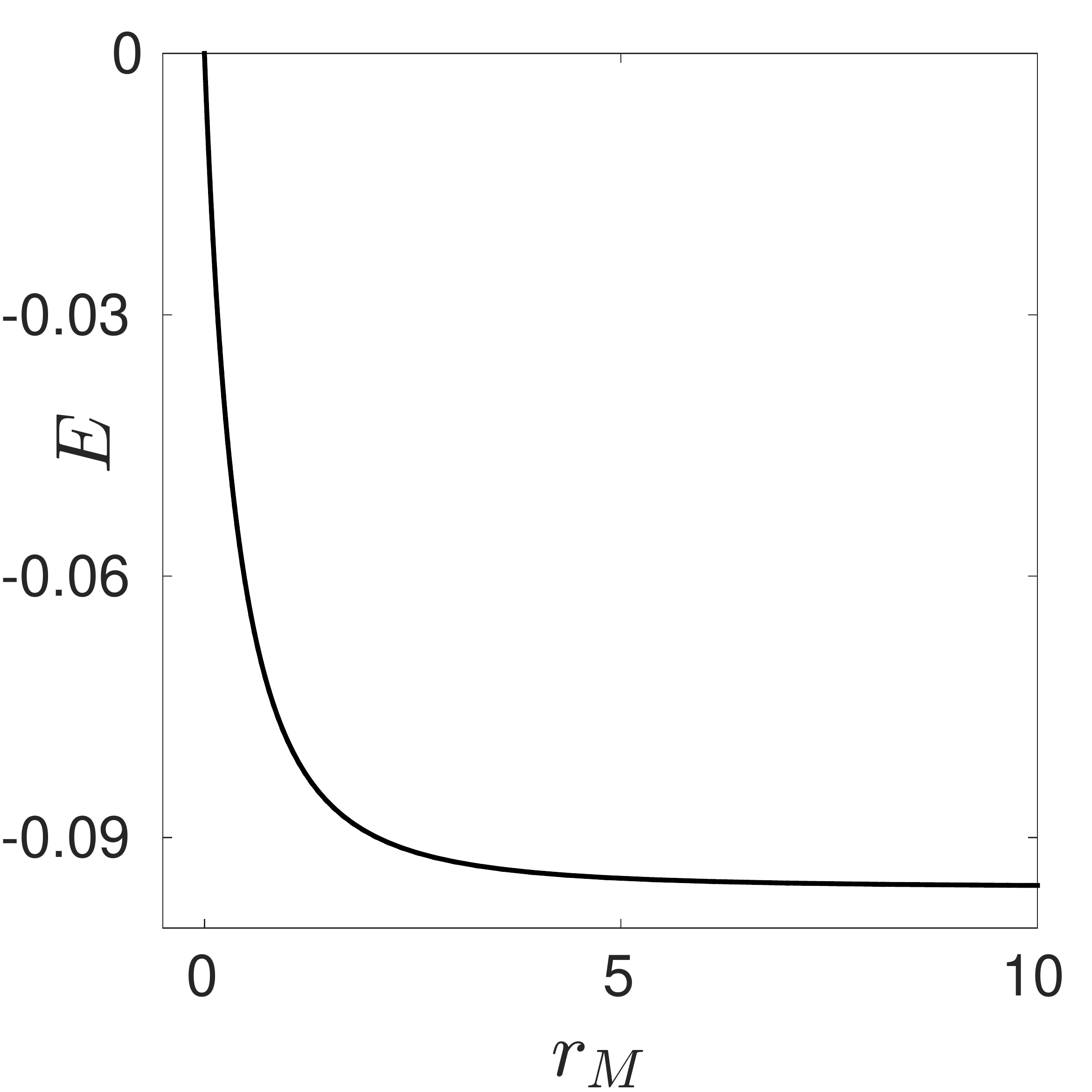}
\end{center}
\hspace{0.165\textwidth} (a) \hspace{0.28\textwidth} (b) \hspace{0.29\textwidth} (c)
\caption{Equilibria \eqref{eqn:morseSoln} on half-line for $V(x)=0$ (no exogenous potential). The interaction potential is given by \eqref{eqn:1dm-kernel}, where $G = 0.5$ and $L = 2$. (a) Disconnected state. (b) Connected state with no aggregation at the origin; this is the same as the free space solution from \cite{BeTo2011}. (c) Energy of equilibria as a function of the mass ratio; the lowest energy state corresponds to the connected equilibrium $r_M = \infty$.}
\label{fig:oned-morse}
\end{figure}

Figure \ref{fig:oned-morse}(a) shows a disconnected equilibrium found by solving \eqref{eqn:morseCond}; the circle indicates the delta strength at origin and the solid line the free swarm. The $\Lambda$ profile (dashed line) shows that such disconnected equilibria are not energy minimizers (as in Section \ref{subsect:1dnoV}).  Figure \ref{fig:oned-morse}(b) shows the connected equilibrium, which is in fact the free space solution from \cite{BeTo2011}. The connected equilibrium is an energy minimizer, as inferred from the $\Lambda$ profile. In both plots note the excellent agreement with the particle simulations (crosses for delta aggregations and stars for free swarms). Finally, Figure \ref{fig:oned-morse}(c) shows the energy of the equilibria \eqref{eqn:morseSoln}, as a function of the mass ratio; as expected, the lowest energy is achieved by the connected state with $\dl=0$ (or equivalently, $\mr =\infty$).

The study in this subsection illustrates that the existence of (attracting) disconnected equilibria which are not minimizers, as well as of a minimizing equilibrium that is the connected state, are not specific properties of the potential \eqref{eqn:intpot}-\eqref{eqn:phi}, but seem to be quite generic. 


\section{Two dimensions: equilibria on half-plane}
\label{sect:twod}

We consider equilibria in two dimensions, in the domain $\Om = [0,\infty) \times \R$. The boundary $\partial \Om =   \{ 0 \} \times \R$ can be interpreted as an impenetrable wall. The interaction kernel is given by \eqref{eqn:intpot}-\eqref{eqn:phi}, i.e., consists of Newtonian repulsion and quadratic attraction.

\subsection{No exogenous potential}
\label{subsect:twod-noV}

As discussed in Section \ref{sect:prelim}, in the absence of boundaries the equilibrium is a constant density in a disk \cite{FeHuKo11}; in two dimensions the value of this constant density is $2M$ -- see \eqref{eqn:Drho-c}. Similar to one dimension, in domains with boundaries we expect to have equilibria that consist in swarms of constant densities {\em away} from the wall together with a possible aggregation build-up {\em on} the boundary.  

Given the considerations above, we search for an equilibrium that consists of a constant density in a bounded domain $\D$ that lies off the wall ($x_1>0$) and a Dirac delta accumulation on $\partial \Om$. For consistency of notations with the study in one dimension, we take the horizontal extent of the free swarm $D$ to be $ \dl<x_1<\dl+\dr$, with $\dl \geq0$, $\dr>0$. Also, we assume symmetry in the vertical direction, and take the vertical extent of  $D$ to be given by the lower and upper free boundaries $x_2=-\g(x_1)$ and $x_2=\g(x_1)$, respectively. 

Specifically,
\[
\D = \{ (x_1,x_2) \mid \dl<x_1<\dl+\dr,  -\g(x_1)<x_2<\g(x_1) \},
\]
and the equilibrium we look for has the the form:
\begin{equation}
\label{eqn:2obs-equil}
\barrho(x_1,x_2) =  \f(x_2) \delta_{\partial \Om}(x_1,x_2) + 2M \, \One_D(x_1,x_2),
\end{equation}
where the density profile $\f(x_2)$ on the wall is assumed to have support $[-\vs,\vs]$.

The support $\Omrho$ of $\barrho$ consists of two components:
\begin{equation}
\label{eqn:2dsupport}
\Om_1 = \{0\}\times [-\vs,\vs] \qquad \text{ and } \qquad \Om_2 = \bar{D},
\end{equation}
where the bar indicates the closure of the set.

As in one dimension, we focus our efforts on solving the necessary equilibrium condition \eqref{eqn:equilsup-gen}. The unknowns in this case are the density profile $\f$ on the wall along with its extent $\vs$, and the free boundary $\g$ along with its horizontal extent given by $\dl$ and $\dr$. It can be immediately noted that this is a highly nonlinear problem and, unlike the one-dimensional case, a solution can only be sought numerically.

Denote the area of $D$ by $\CalA$. By the mass constraint \eqref{eqn:massrho} we find
\begin{equation}
\label{eqn:massc-twod}
\int_{-\vs}^{\vs} \f(x_2) dx_2 + 2M \CalA  = M.
\end{equation}
Calculate $\Lambda(\bfx)$ for $\bfx=(x_1,x_2) \in \Omrho$ using \eqref{eqn:Lambda}, where $K$ is given by \eqref{eqn:intpot}-\eqref{eqn:phi} and $V=0$. A generic point $\bfy=(y_1,y_2) \in \Omrho$ can lie either on the wall or in $\D$, along with its free boundary. Consequently, $\Lambda(\bfx)$ consists of two terms:
\begin{align}
\Lambda(\bfx) &= \int_{-\vs}^{\vs} \left(  -\frac{1}{2\pi} \ln  \sqrt{x_1^2 + (x_2-y_2)^2} + \frac{1}{2} \left( x_1^2 + (x_2-y_2)^2 \right) \right)  \f(y_2) dy_2 \nonumber \\
&\quad + 2M \iint_{\D} \left( -\frac{1}{2\pi} \ln |\bfx-\bfy| + \frac{1}{2} |\bfx-\bfy|^2 \right) \dbfy.
\label{eqn:Lambda-2d}
\end{align}

For an equilibrium,  $\Lambda(\bfx)$ has to be constant in each component of $\Omrho$:
\begin{equation}
\label{eqn:Lambda-const-2d}
\Lambda(\bfx) = \lambda_1 \quad \text{ in } \{0\}\times [-\vs,\vs] ,  \qquad \text{ and} \qquad \Lambda(\bfx) = \lambda_2 \quad \text{ in }  \bar{D}.
\end{equation}
Solving \eqref{eqn:Lambda-const-2d} numerically would be very expensive. First, it requires approximating $\Lambda$ on a two-dimensional numerical grid, where at each grid point we would have to evaluate numerically a convolution integral. Second, it requires a nonlinear solver to solve \eqref{eqn:Lambda-const-2d} at all grid points of this two-dimensional domain. We show below how one can reduce the dimensionality of the problem by making use of specific properties of the interaction potential.

Calculate the Laplacian of $\Lambda$ from \eqref{eqn:Lambda-2d}:
\begin{align}
\Delta \Lambda(\bfx) &=   2  \int_{-\vs}^{\vs} \f(y_2) dy_2 + 2M \left(-1 + 2 \iint_D \dbfy \right) \nonumber \\
&= 2\left(  \int_{-\vs}^{\vs} \f(y_2) dy_2 -M + 2M \CalA \right),
\label{eqn:DeltaLambda}
\end{align}
where for the first equality we used $\Delta \left(-\frac{1}{2\pi} \ln |\bfx| \right) = -\delta$; in particular, the logarithmic term in the single integral $\int_{-\vs}^{\vs}$ is harmonic for $\bfx \in D$.

Using the mass constraint \eqref{eqn:massc-twod}, one can infer from \eqref{eqn:DeltaLambda} that $\Lambda$ is harmonic in $D$:
\begin{equation}
\label{eqn:Lambda-har}
\Delta \Lambda (\bfx) = 0, \qquad \text{ for all } \bfx \in D.
\end{equation}
This observation greatly simplifies the problem of solving \eqref{eqn:Lambda-const-2d}. Indeed, provided $\Lambda(\bfx) = \lambda_2$ is satisfied for $\bfx \in \partial D$ (i.e., only on the boundary of the free swarm), then by \eqref{eqn:Lambda-har}, using standard theory for harmonic functions, $\Lambda(\bfx) = \lambda_2$ for all $\bfx \in  \bar{D}$. Consequently, \eqref{eqn:Lambda-const-2d} reduces to solving
\begin{equation}
\label{eqn:Lambda-const-bdry}
\Lambda(\bfx) = \lambda_1 \quad \text { in }  \{0\}\times [-\vs,\vs],  \qquad \text{ and} \qquad \Lambda(\bfx) = \lambda_2 \quad \text{ on }  \partial D.
\end{equation}

We solve numerically equation \eqref{eqn:Lambda-const-bdry} to find $\vs$, $\dl$, $\dr$, $\lambda_1$, $\lambda_2$, and the profiles $\f$ and $\g$ of the wall aggregation and the free boundary. Approximations for the latter are found on a uniform grid in the vertical, respectively horizontal, directions. Details on the numerical implementation are presented in Section \ref{subsubsect:ssnumLam}; we only reemphasize here that there is a huge computational cost of solving \eqref{eqn:Lambda-const-bdry} versus  \eqref{eqn:Lambda-const-2d}, by having reduced  the dimensionality of the problem. 

Similar to one dimension, we find both disconnected and connected solutions to \eqref{eqn:Lambda-const-bdry}. We present them separately. We note again that in all numerical simulations presented in this paper, the mass $M$ is set to $1$.

\medskip
{\em Disconnected equilibria ($\dl>0$)}. As in one dimension, denote by $\mr$ the ratio of the mass $M_2$ of the free swarm and the mass $M_1$ of the aggregation on wall:
\begin{equation}
\label{eqn:mr}
\mr : = \frac{M_2}{M_1} = \frac{2M \CalA}{\int_{-\vs}^{\vs} \f(x_2) dx_2}.
\end{equation}
Numerical simulations suggest that disconnected solutions to \eqref{eqn:Lambda-const-bdry}  in the form \eqref{eqn:2obs-equil} exist for all $\mr \in (0,\infty)$. In fact, based on numerical explorations, we believe that there is a unique disconnected solution to \eqref{eqn:Lambda-const-bdry} for every $\mr \in (0,\infty)$ fixed. The limiting cases, i.e, the zero and infinite mass ratios, correspond to connected states, where all mass lies either {\em on} the wall or {\em off} the wall, respectively. These solutions will be elaborated below, where connected states are discussed. 

To check whether the disconnected solutions are indeed equilibria, we compute numerically the velocity field at points on $\Om_1$, the part of the support that lies on the wall, and inspect its horizontal component (c.f. Remark \ref{rmk:Lambda-const}). For all the disconnected states we computed, we found that velocity vectors at various points located near and at the edges of the wall aggregation point {\em toward} the interior of the domain $\Om$. Consequently, their projections are not zero (see \eqref{eqn:ss}) and the states we computed are {\em not} equilibria. This result clearly shows that \eqref{eqn:equilsup-gen}  is only a necessary, but not sufficient, condition for equilibrium --- see Remark \ref{rmk:Lambda-const}. 

The conclusion reached above is exactly the opposite of what has been found in one dimension, where all the disconnected solutions to \eqref{eqn:equilsup-gen} were steady states, though not local minima of the energy. The essentially different behaviour is due to the two dimensional geometry. In one dimension the distance from the wall to any point in the free swarm is necessarily parallel to the normal of the wall. In two dimensions this is not the case and some interactions between the wall swarm and the free swarm can be farther than the distance from the wall to the furthest extent of the free swarm in the horizontal direction. This enables more attractive forces to come into play, most noticeably at the edges of the wall swarm $\bfx = (0,\pm L)$. Indeed, the edges of the wall swarm are where one finds the largest velocities normal to and pointing away from the wall, if one finds them at all.

\medskip
{\em Connected equilibria.} The first type of connected equilibria correspond to aggregations that lie entirely on the wall  (no free swarm, $\mr = 0$). This is a degenerate case of densities of form \eqref{eqn:2obs-equil}, where $D$ is the empty set. The equilibrium in this case has the form of a delta-aggregation on the wall:
\begin{equation}
\label{eqn:wall-sol}
\barrho(x_1,x_2) =  \f(x_2) \delta_{\partial \Om}(x_1,x_2).
\end{equation}

\begin{figure}[ht]
  \begin{center}
 \includegraphics[width=0.525\textwidth]{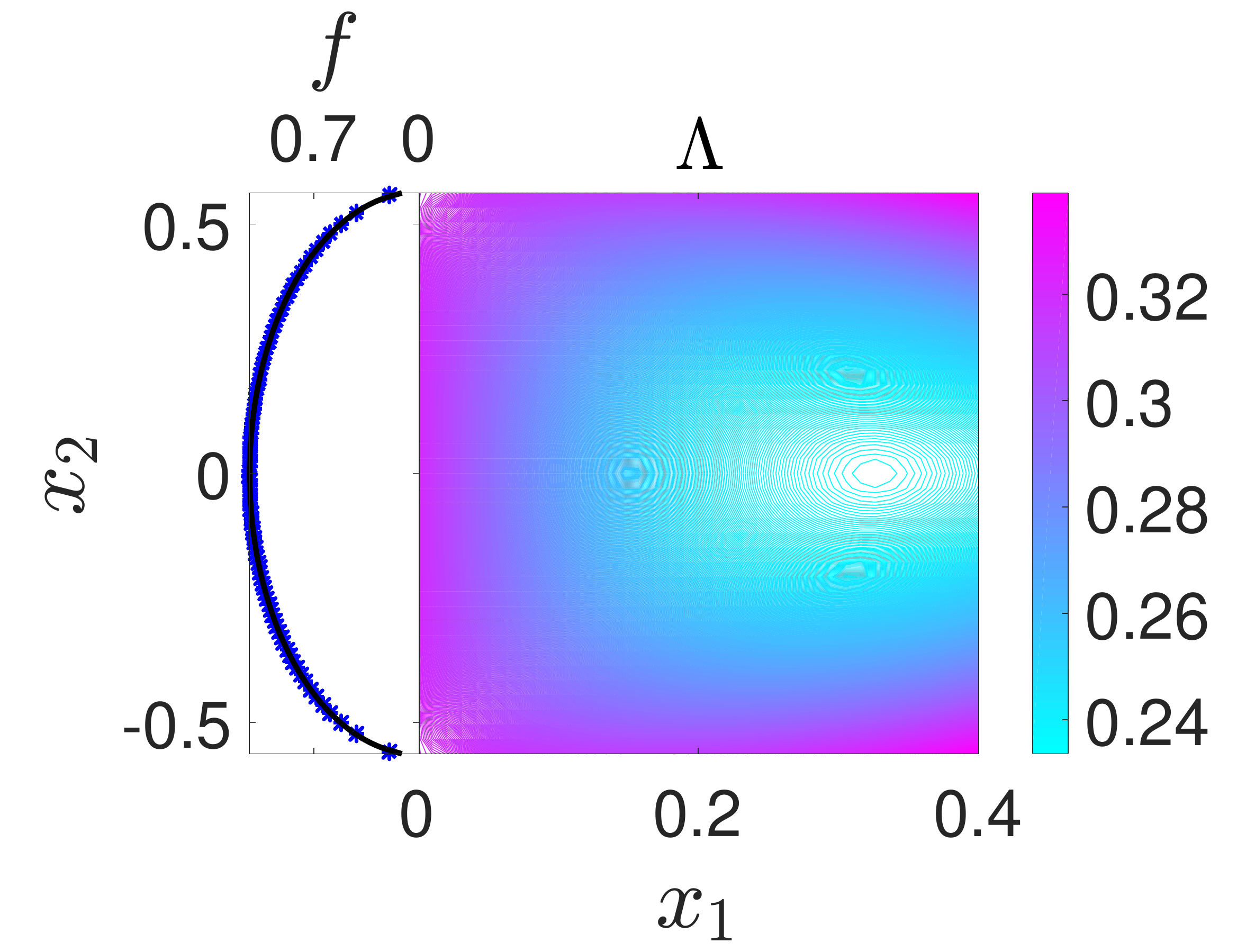} 
 \includegraphics[width=0.445\textwidth]{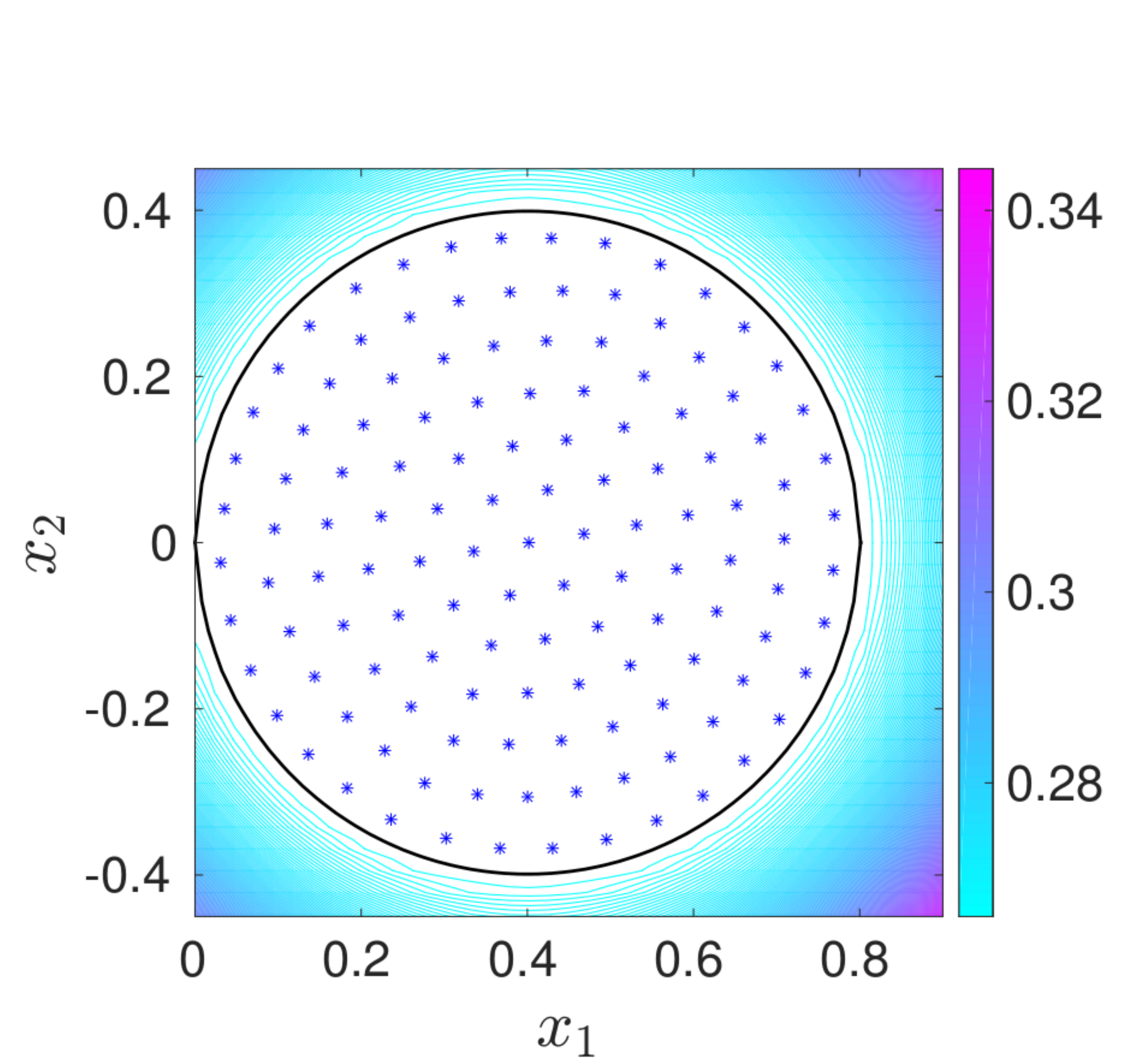}
\end{center}
\hspace{0.285\textwidth} (a) \hspace{0.415\textwidth} (b)
\caption{Equilibria on half-plane in two dimensions for $V=0$ (no exogenous potential).  (a) Equilibrium aggregation that lies entirely on the wall ($\mr = 0$). The solid line represents the density profile $\f$ on the wall as solved from  \eqref{eqn:massc-onew} and \eqref{eqn:equilsup-onew}. Note the excellent agreement with the particle simulations (blue stars). The equilibrium is not an energy minimizer, as indicated by the contour plot of $\Lambda$ (shown on right). (b) Free swarm equilibrium ($\mr = \infty$) of constant density $2M$ in a disk of radius $\frac{1}{\sqrt{2\pi}}$. The contour plot of $\Lambda$, shown in the figure, demonstrates that this equilibrium is an energy minimizer. Note that there are no disconnected equilibria of form \eqref{eqn:2obs-equil} in this case.}
\label{fig:2dconnected}
\end{figure}

We find the density profile $\f(x_2)$ on the wall and its support $[-\vs,\vs]$ by solving numerically \eqref{eqn:equilsup} in $\Omrho = \{0\} \times [-\vs,\vs]$. Note that the mass constraint \eqref{eqn:massrho} implies:
\begin{equation}
\label{eqn:massc-onew}
\int_{-\vs}^{\vs} \f(x_2) dx_2 = M,
\end{equation}
while in our case \eqref{eqn:equilsup} reduces to:
\begin{equation}
\label{eqn:equilsup-onew}
\int_{-\vs}^{\vs} \left(  -\frac{1}{2\pi} \ln  |x_2-y_2|+ \frac{1}{2} (x_2-y_2)^2 \right) \f(y_2) dy_2 = \lambda, \qquad \text{ for all } x_2 \in [-\vs, \vs].
\end{equation}
We solve numerically \eqref{eqn:massc-onew} and \eqref{eqn:equilsup-onew} to find  $\f$, $\vs$ and $\lambda$. The wall profile $f$ is shown in Figure \ref{fig:2dconnected}(a), along with the density profile obtained from particle simulations; note the excellent agreement between the two. We also note here that the only initial configurations that can dynamically evolve into this equilibrium are those with initial support on the wall. 

With this numerically computed solution we then checked \eqref{eqn:equilcomp}, which here reads:
\begin{equation*}
 \int_{-\vs}^{\vs} \left(  -\frac{1}{2\pi} \ln  \sqrt{x_1^2 + (x_2-y_2)^2} + \frac{1}{2} \left( x_1^2 + (x_2-y_2)^2 \right) \right)  \f(y_2) dy_2  \geq \lambda, \qquad \text{ for all } (x_1,x_2) \in \Omrho^c.
\end{equation*}
Note that $\Omrho^c$ is the disjoint union of the two semi-infinite vertical lines $\{0\} \times (-\infty, -\vs)$ and $\{0\} \times (\vs,\infty)$,  with the open half-plane $(0,\infty)\times \R$.  A coloured contour plot of $\Lambda(x)$ is shown (on the right) in Figure \ref{fig:2dconnected}(a). As expected, the inequality above does not hold near the wall and hence,  the equilibrium solution \eqref{eqn:wall-sol} is not an energy minimizer. 

We note in passing that integral equations with logarithmic kernels such as \eqref{eqn:equilsup-onew} arise in the representation of a harmonic function in terms of single-layer potentials. Various  analytical and numerical results have been derived for such equations \cite{SloanSpence1988,YanSloan1988}. These results do not apply to our setting though, as in our problem the extent $\vs$ of the boundary is also an unknown.

The second type of connected equilibria correspond to swarm equilibria in free space, and consist of a constant aggregation of density $2M$ in a disk of radius $\frac{1}{\sqrt{2\pi}}$. A representative equilibrium in this class can be considered the disk tangent to the wall at the origin --- see Figure \ref{fig:2dconnected}(b). This is an equilibrium in the form \eqref{eqn:2obs-equil} where $\f=0$ (no delta aggregation on the wall) and $\dl=0$ (no separation). By taking arbitrary translations that keep the disk within $\Om$, one can then find a family of such constant aggregations.  The constant density equilibria are swarm minimizers, as \eqref{eqn:equilcomp} can be shown to hold; a contour plot of $\Lambda$ is also illustrated in Figure \ref{fig:2dconnected}(b).

To argue that the connected equilibrium in Figure \ref{fig:2dconnected}(b) is a global minimizer one should rule out the existence of minimizers that have other form than \eqref{eqn:2obs-equil}. In one dimension this was done via a simple explicit calculation -- see Remark \ref{rmk:1d-gloMin}. Though such a simple argument does not seem to be available in two dimensions, we believe that the constant density in a disk is a global minimizer; for the problem in free space it was shown that such equilibrium configuration is in fact a global attractor \cite{BertozziLaurentLeger,FeHuKo11}.


\subsection{Linear exogenous potential}
\label{subsect:twod-linearV}

We consider an exogenous gravitational potential $V(x_1,x_2) = g x_1$, with $g>0$. The domain is the same as above, the half-plane $\Om = [0,\infty) \times \R$, so the exogenous forces are acting (horizontally) towards the wall. Note that $\Delta V=0$ and by using this observation in the evolution equation \eqref{eqn:evol} we infer that away from the wall the equilibrium densities are constant (equal to $2M$) on their support. 

As in Section \ref{subsect:twod-noV}, we search for equilibria in the form \eqref{eqn:2obs-equil}, which consist in a delta aggregation on the wall and a constant density free swarm. The same variables and setup from Section \ref{subsect:twod-noV} are being used here as well. In particular, the support $\Omrho$ of the equilibrium is given by \eqref{eqn:2dsupport} and mass conservation leads to \eqref{eqn:massc-twod}. We solve numerically the necessary condition for equilibrium \eqref{eqn:Lambda-const-2d}, with $\Lambda(\bfx)$ given by 
\begin{equation}
\begin{aligned}
\Lambda(\bfx) &= \int_{-\vs}^{\vs} \left(  -\frac{1}{2\pi} \ln  \sqrt{x_1^2 + (x_2-y_2)^2} + \frac{1}{2} \left( x_1^2 + (x_2-y_2)^2 \right) \right)  \f(y_2) dy_2  \\
&\quad + 2M \iint_{\D} \left( -\frac{1}{2\pi} \ln |\bfx-\bfy| + \frac{1}{2} |\bfx-\bfy|^2 \right) \dbfy + gx_1.
\end{aligned}
\label{eqn:Lambda-2dg}
\end{equation}

Since the gravitational potential has zero Laplacian, by the same argument as in the zero gravity case one concludes that $\Lambda$ is harmonic in $D$ (see \eqref{eqn:Lambda-har}). Consequently, the problem reduces to solving \eqref{eqn:Lambda-const-bdry}, with $\Lambda$ given by \eqref{eqn:Lambda-2dg}. We solve this equation numerically to find approximations for $\vs$, $\dl$, $\dr$, and the profiles $\f$ and $\g$ of the wall aggregation and the free boundary.

As in one dimension, there is a critical value $\gc$ such that for $g>\gc$ there is no equilibrium in the form \eqref{eqn:2obs-equil}. The gravity in this case is is too strong and pins all mass on the boundary of the domain; the only equilibrium is a delta aggregation on the wall. For $g<\gc$ however we find genuinely two-dimensional equilibria, which come in two flavours: connected and disconnected. In Section \ref{subsect:gc-twod} we show how $\gc$ can be calculated in the two dimensional problem. We find $\gc = f(0)/2$, where $f$ is the density of the all-on-wall equilibrium (see \eqref{eqn:wall-sol} and Figure \ref{fig:2dconnected}(a)).  In our simulations with $M=1$, $\gc \approx 0.564$. We now proceed in presenting the two cases separately.

\medskip
{\em \bf Case $\mathbf{g<\gc}$.} Solutions to \eqref{eqn:Lambda-const-bdry} of the form \eqref{eqn:2obs-equil} exist only for mass ratios below a maximal value which we denote here by $\rc(g)$. Recall that in one dimension, for subcritical gravities (note that in one dimension $\gc=M/2$), there exists a disconnected equilibrium for any mass ratio in the interval $\big( 0, \sqrt{\frac{\gc}{g}}-1 \bigr)$, where the maximal mass ratio $\sqrt{\frac{\gc}{g}}-1$ corresponds to the connected equilibrium (see Remark \ref{rmk:mr-oned}). 

The subcritical gravity case in two dimensions parallels the findings in one dimension in the fact that for any fixed $g<\gc$, there exists a family of disconnected equilibria which approach, as the separation $d_1$ vanishes, a connected equilibrium supported on both the boundary and the interior of $\Om$. We parametrize this family of disconnected equilibria by $\mr$, the mass ratio defined in \eqref{eqn:mr}. 

The major subtlety in two dimensions is that for low gravities, equilibria in the form \eqref{eqn:2obs-equil} exist only for {\em certain} mass ratios in the interval $(0,\rc(g))$. To illustrate this fact,  we introduce a critical value $\tgc$, with $\tgc <\gc$ (as shown below, $\tgc \approx 0.044$ for our simulations with $M=1$). We find that for gravities  $\tgc <g<\gc$ there exists a disconnected equilibrium for {\em any} mass ratio in $(0,\rc(g))$. On the other hand, for $g<\tgc$, while solutions to  \eqref{eqn:Lambda-const-bdry} in the form \eqref{eqn:2obs-equil} do exist for all $\mr \in (0,\rc(g))$, not all of these solutions are equilibria. As a limiting case, we recover the findings from the zero gravity study, where none of the disconnected solutions to \eqref{eqn:Lambda-const-bdry} were in fact equilibria. We now elaborate on these facts.

\medskip
{\em i) $\tgc<g<\gc$.}

{\em Disconnected equilibria.} For any fixed gravity $g\in(\tgc,\gc)$, we find a family of disconnected solutions to  \eqref{eqn:Lambda-const-bdry} with mass ratios in the interval $(0, \rc(g))$; as mentioned above, $\rc(g)$ denotes the maximal value that the mass ratio of the two components can take for that particular $g$.  In one dimension, $\rc(g) =  \sqrt{\frac{\gc}{g}}-1$ can be calculated explicitly, as discussed in Remark \ref{rmk:mr-oned}. In two dimensions we approximate $\rc(g)$ numerically; see Figure \ref{fig:regions}(b). Note that $\rc(g)$ is strictly decreasing and touches zero at $g=\gc$. This is consistent with the fact that at larger gravities, less mass can end up in the free swarm, and the range of $\mr$ decreases. 

\begin{figure}
\begin{center}
\includegraphics[width=0.46\textwidth]{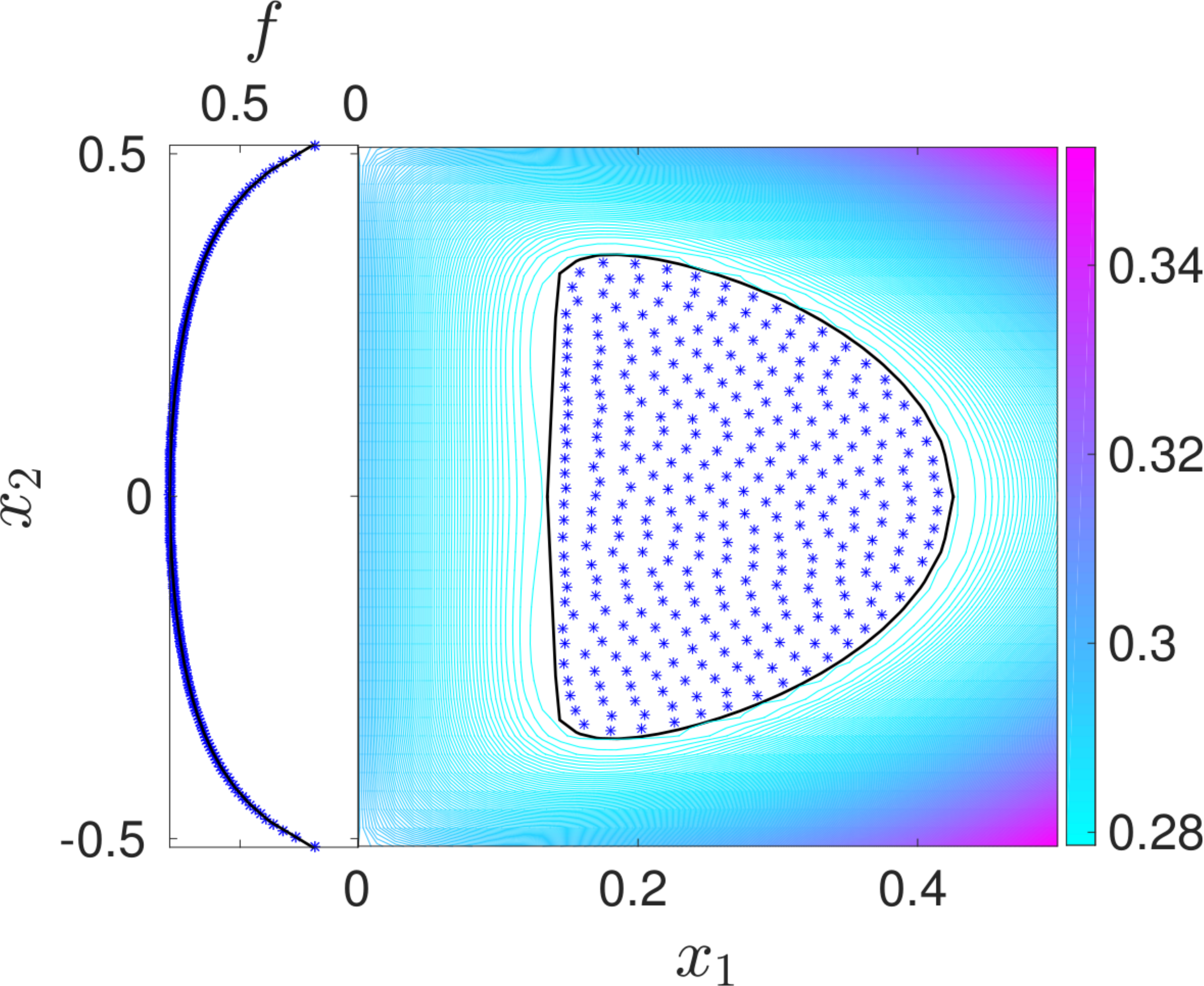}  
\includegraphics[width=0.46\textwidth]{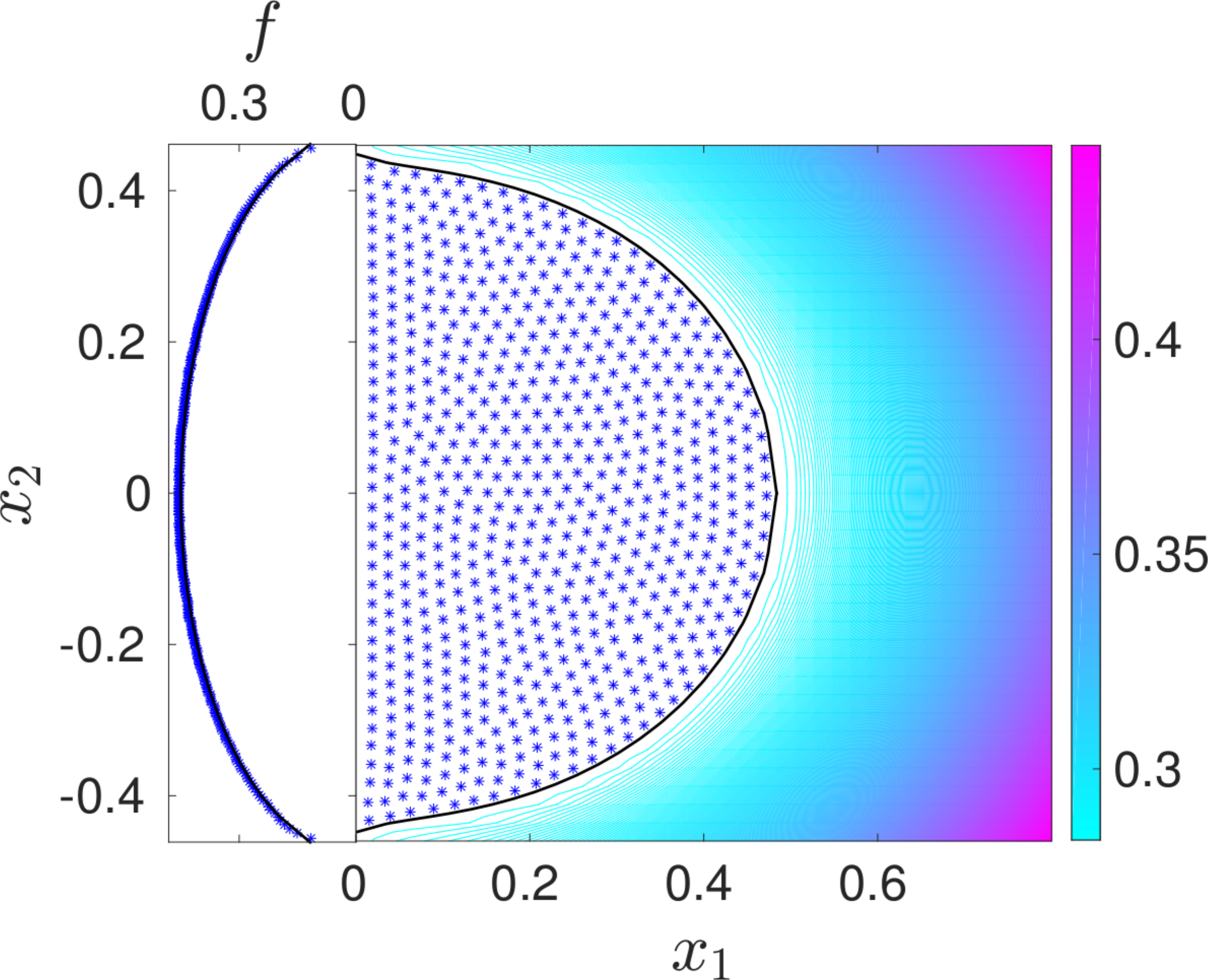}
\end{center}
\hspace{0.285\textwidth} (a) \hspace{0.42\textwidth} (b)
\begin{center}
\includegraphics[width=0.46\textwidth]{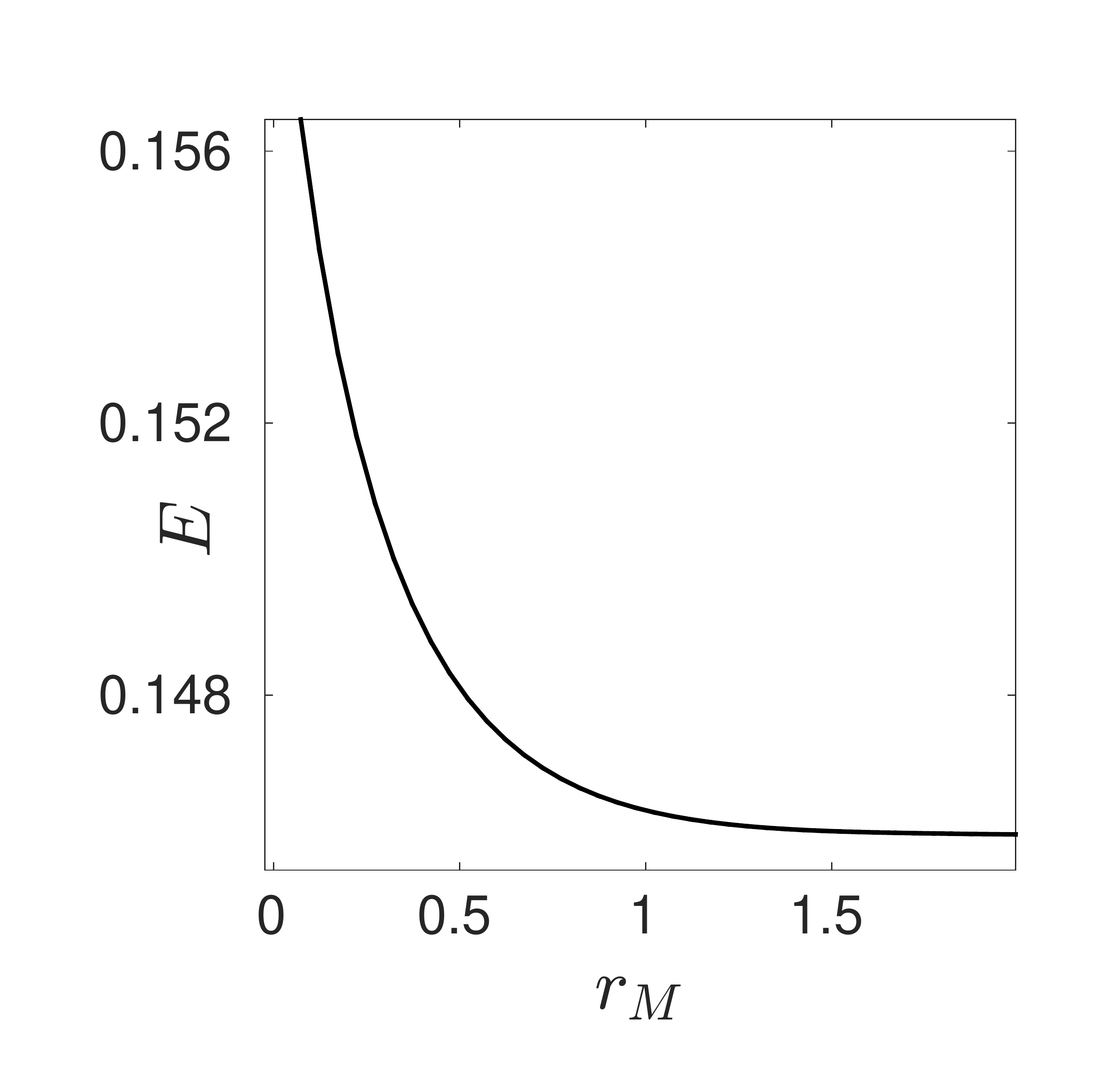}
\end{center}
\hspace{0.515\textwidth} (c)
\caption{Equilibria \eqref{eqn:2obs-equil} on half-plane for $V(x_1,x_2)=gx_1$ (linear exogenous potential) with $g=0.064$. (a) Disconnected state consisting in a free swarm of constant density and a delta aggregation on the wall. (b) Connected state with a constant density in a domain adjacent to the wall and a delta aggregation on the wall. (c) Energy of equilibria  \eqref{eqn:2obs-equil} as a function of the mass ratio; the lowest energy state corresponds to the connected equilibrium with $ \mr=\rc(g)$.}
\label{fig:twod-conng}
\end{figure}

An illustration of a typical disconnected solution to \eqref{eqn:Lambda-const-bdry} is shown in Figure \ref{fig:twod-conng}(a); there $g=0.064$. To check that these solutions are indeed equilibria, we compute numerically the velocity field at points on $\Om_1$, the part of the support that lies on the wall, and inspect its horizontal component (see Remark \ref{rmk:Lambda-const}). Based on these investigations we conclude that all the two-dimensional disconnected states we found here are indeed equilibria. Also, as in the one-dimensional case with subcritical gravity, these equilibria are {\em not} local minima for the energy, as \eqref{eqn:equilcomp-gen} is not satisfied near the wall -- see Figure \ref{fig:twod-conng}(a) for a contour plot of $\Lambda$. Since $\Lambda(\bfx)$ decreases from the wall to the free swarm, an infinitesimal perturbation of mass from the wall into $x_1>0$  would bring the mass into the free swarm. 

To further illustrate the point above, we compute the energy corresponding to the disconnected steady states. Figure \ref{fig:twod-conng}(c) shows the plot of the energy $E[\barrho]$ as function of mass ratio $\mr$, for $g = 0.064$; note that for this value of gravity, the maximal mass ratio is $\rc(0.064) \approx 2.045$. We observe a monotonically decreasing profile that supports what has been noted above: taking mass from the wall and placing into $x_1>0$ would result dynamically into an equilibrium of larger mass ratio, which is more energetically favourable. 

As the mass ratios of the disconnected equilibria increase toward the maximal value $\rc(g)$, the free swarm gets closer and closer to the wall. This behaviour is consistent with the results in one dimension, where explicit calculations show that the separation $\dl$ approaches zero in such limit. Numerical evidence suggests that at $\mr = \rc(g)$, the two-dimensional free swarm touches the wall and it forms a connected state. This aspect will be further discussed in the next paragraphs.

\medskip
{\em Connected equilibria.} There exist two types of connected equilibria: one that corresponds to all mass on the wall ($\mr=0$) and another that corresponds to the maximal mass ratio $\mr = \rc(g)$. The first type can be obtained dynamically by initializing model \eqref{eqn:model} with a density that is supported entirely on the wall. In fact, since the gravitational forces vanish at the wall, the equilibrium density is identical to that computed in the zero gravity case --- see equations \eqref{eqn:wall-sol}-\eqref{eqn:equilsup-onew} and Figure \ref{fig:2dconnected}(a). The only distinction comes in checking the minimization condition \eqref{eqn:equilcomp}, which in the gravitational case reads:
\begin{equation}
\label{eqn:equilsup-min}
 \int_{-\vs}^{\vs} \left(  -\frac{1}{2\pi} \ln  \sqrt{x_1^2 + (x_2-y_2)^2} + \frac{1}{2} \left( x_1^2 + (x_2-y_2)^2 \right) \right)  \f(y_2) dy_2  + g x_1\geq \lambda, \qquad \text{ for all } (x_1,x_2) \in \Omrho^c.
\end{equation}
The gravitational term, $gx_1$, merely translates the minimum of $\Lambda(\bfx)$ shown in Figure \ref{fig:2dconnected}(a) towards the wall. This equilibrium is therefore not an energy minimizer until $g = g_c$ when the minimum of $\Lambda(\bfx)$ becomes degenerate and $\Lambda(\bfx) = \lambda$ for $\bfx \in {0}\times[-L,L]$.

The connected equilibria of second type are supported on both the wall and the interior of $\Om$. As discussed above, they are realized when the free component of the disconnected state touches the wall in the limit $\mr \to \rc(\g)$. Alternatively, one can search for a connected state independently, by assuming an equilibrium of the form \eqref{eqn:2obs-equil}, with a domain $D$ that touches the wall ($\dl=0$). To this purpose we solved \eqref{eqn:equilsup} numerically, with $\Lambda(\bfx)$ given by \eqref{eqn:Lambda-2dg}. Figure \ref{fig:twod-conng}(b) shows the connected equilibrium found by this direct approach, for $g=0.064$, along with a particle simulation that has reached this steady state.  Our numerical investigations indicate that there is a unique such connected equilibrium. Also, the connected state is a local minimizer of the energy, as can be inferred from the contour plot of $\Lambda$ shown in the same figure. Based on our study (see for instance Figure \ref{fig:twod-conng}(c)), we believe in fact that this equilibrium is a global minimizer of the energy, though for a definite conclusion one has to rule out the possibility of having minimizers of a more general form than \eqref{eqn:2obs-equil} -- see Remark \ref{rmk:1d-gloMin} for the one dimensional case.

\begin{figure}[thb]
  \begin{center}
 \includegraphics[width=0.46\textwidth]{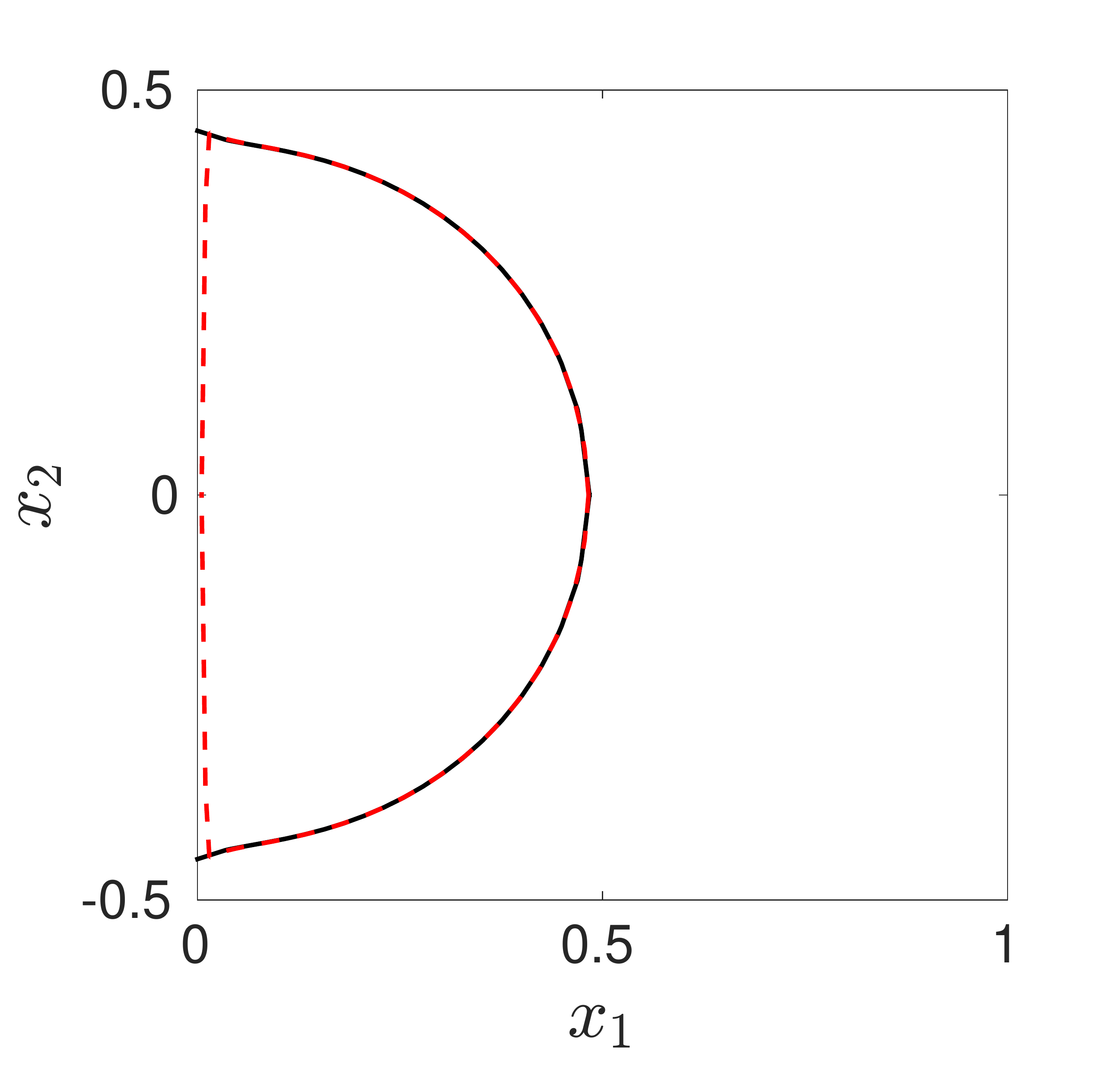} 
 \includegraphics[width=0.46\textwidth]{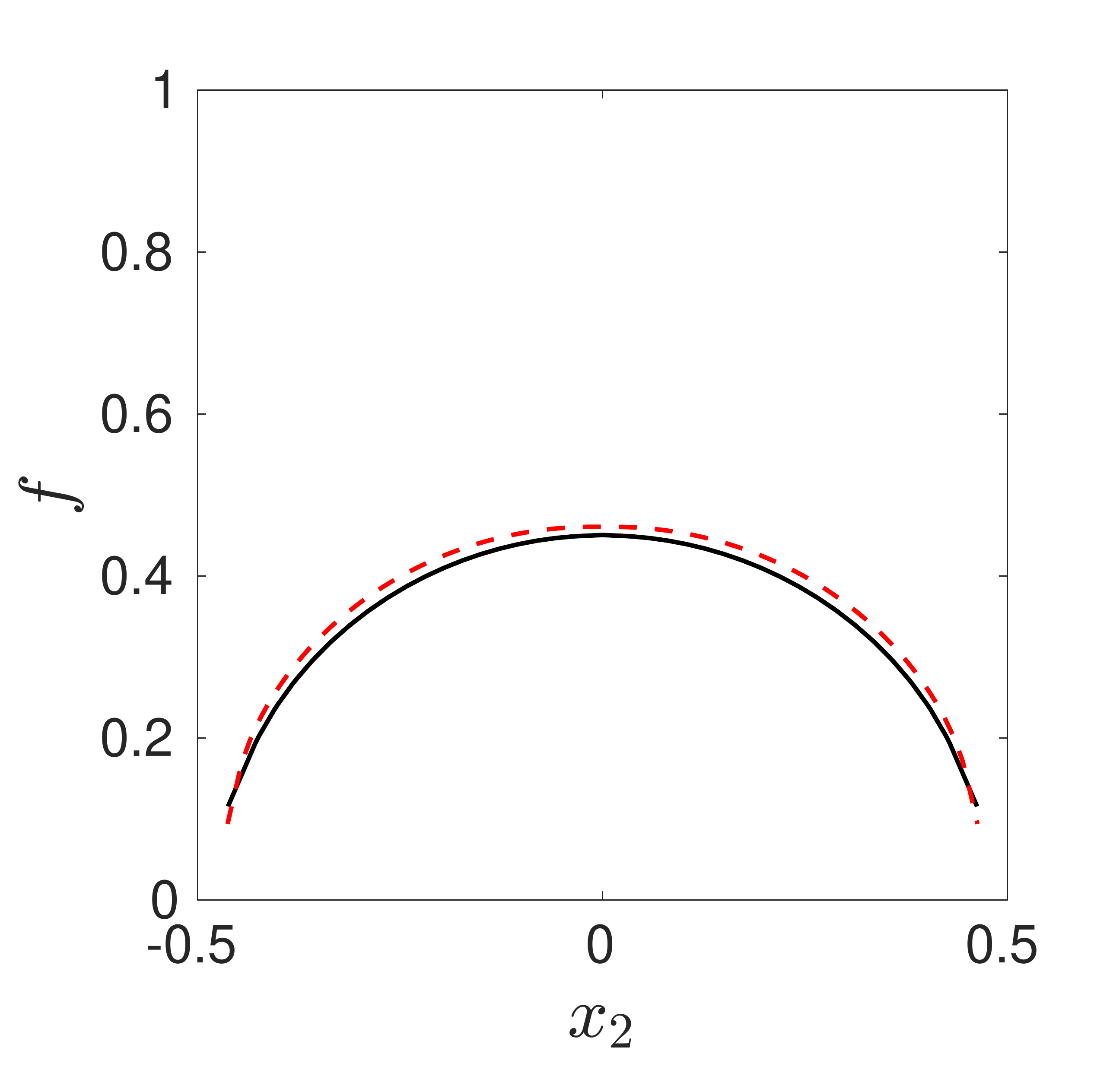}
\end{center}
\hspace{0.265\textwidth} (a) \hspace{0.425\textwidth} (b)
\caption{Equilibria on half-plane in two dimensions for $V(x_1,x_2) = g x_1$, with $g=0.064$: the disconnected equilibria \eqref{eqn:2obs-equil} approach a connected equilibrium state as the separation $\dl$ from the wall approaches $0$ (or equivalently, $\mr$ approaches the maximal mass ratio $\rc(g)$). (a) The solid line represents the connected solution of \eqref{eqn:2obs-equil}. The dashed line shows a disconnected equilibrium with a mass ratio $\mr = 1.873 < \rc(g)$; this is the disconnected state with the largest mass ratio that we were able to obtain in our numerical investigations. (b) Profile $f$ of the density on the wall corresponding to the connected (solid line) and disconnected (dashed line) equilibria shown in plot (a).}
\label{fig:close}
\end{figure}

Figure \ref{fig:close} illustrates the idea that the connected equilibrium is obtained from the disconnected states upon touching the wall. The solid lines in Figure \ref{fig:close}(a) and \ref{fig:close}(b) show the boundary and the wall density profile $f$ of the connected equilibrium, respectively, for the same value of $g$ used above ($g=0.064$). The dashed lines correspond to the disconnected equilibrium with the largest mass ratio $\mr < \rc(g)$ (or equivalently, with the smallest separation $\dl>0$) that we were able to obtain in numerical simulations.  The findings indicate that when the free swarm touches the wall at $\mr = \rc(g)$, it establishes contact with the wall over an {\em entire} vertical segment, and not just at a single point. This suggests that there is a {\em continuous} deformation of the two component equilibrium into the connected equilibrium of mass ratio $\rc(g)$, 

Finally, we remark that in our implementation for computing the connected equilibrium, we do not require $\g(0)=\vs$, that is, we do not ask for the extent of the delta-accumulation on the wall to match the boundary of the constant swarm in $D$. We assume instead the inequality $\g(0)\leq \vs$, which allows the wall accumulation to extend beyond the points where the free boundaries $x_2 = \pm \g(x_1)$ meet the wall. And indeed, the equilibrium we find by solving \eqref{eqn:equilsup} satisfies the {\em strict} inequality $\g(0) < \vs$.

\medskip
{\em ii) $g<\tgc$.} As in case i), solutions to \eqref{eqn:Lambda-const-bdry} of the form \eqref{eqn:2obs-equil} exist for all mass ratios below a maximal value, which is denoted again by $\rc(g)$. The profile of $\rc(g)$ for $g\in (0,\tgc)$ is shown in Figure \ref{fig:regions}(b); note that it connects smoothly at $g = \tgc$ with $\rc(g)$ computed above for $ g \in (\tgc,\gc)$. Also, similar to the one dimensional case, $\rc(g)$ approaches $\infty$ as $g \to 0$. 

To check whether the computed solutions to \eqref{eqn:Lambda-const-bdry} are indeed equilibria (see Remark \ref{rmk:Lambda-const}), we inspect the velocity field on the wall. For an equilibrium, the velocities on the wall (before applying the projection) should point {\em into} the wall, or equivalently, their horizontal components should be negative for all points in the support $[-\vs,\vs]$ of the wall density $\f$. We find that there is an entire range of disconnected solutions to \eqref{eqn:Lambda-const-bdry}, with mass ratios $\mr \in (\ra(g),\rb(g))$ (here $0<\ra(g)<\rb(g)<\rc(g)$), which are {\em not} steady states. Alternatively,  disconnected equilibria exist only for mass ratios $\mr \in (0, \ra(g)) \cup (\rb(g),\rc(g))$. 

\begin{figure}[thb]
  \begin{center}
 \includegraphics[width=0.65\textwidth]{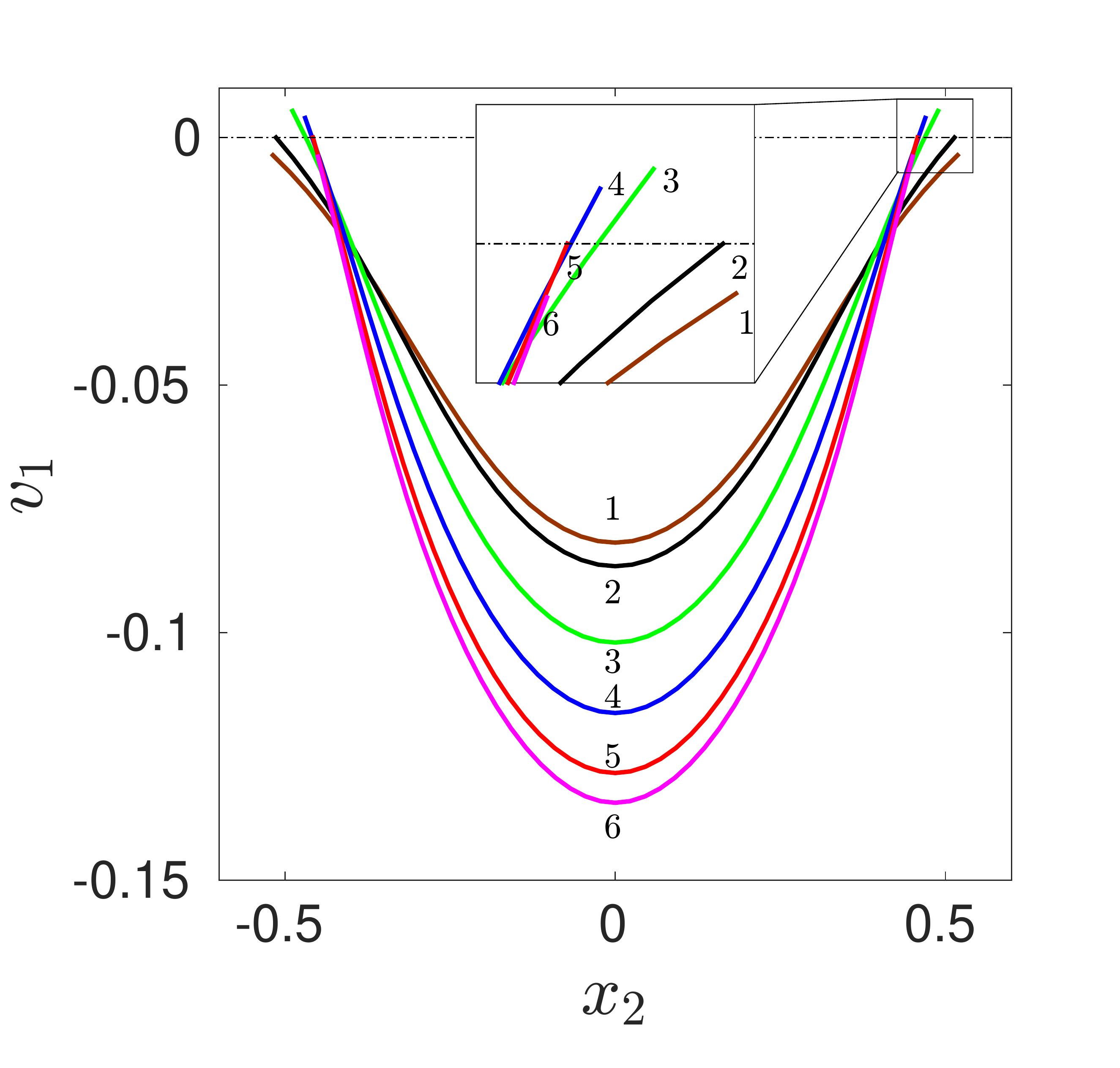} 
\end{center}
\caption{Horizontal velocity (before taking the projection \eqref{eqn:proj}) along wall profile of solutions to \eqref{eqn:Lambda-const-bdry} for $g=0.04<\tgc$ and (1) $r_M = 1.531$, (2) $r_M = \beta(g) \approx 1.379$, (3) $r_M = 1.078$, (4) $r_M = 0.744$, (5) $r_M = \alpha(g) \approx 0.439$, (6) $r_M = 0.362$. Note the positive velocities for $r_M \in (\alpha(g),\beta(g))$, that is, in (3) and (4), indicating that mass would leave the wall and thus these solutions are not steady states. See also Figure \ref{fig:regions}(b).}
\label{fig:2d-ABvelplot}
\end{figure}

Figure \ref{fig:2d-ABvelplot} illustrates the idea above for $g=0.04 < \tgc$ and various mass ratios. At mass ratios $\mr \in (\ra(g),\rb(g))$, the horizontal velocity is positive near the end of the wall profile (see curves labeled (3) and (4)). Hence mass would leave the wall and these solutions to \eqref{eqn:Lambda-const-bdry} are not equilibria. For mass ratios outside this interval, the velocities are everywhere negative, and such solutions are indeed equilibria.

We approximate numerically $\ra(g)$ and $\rb(g)$ and plot their profiles in Figure \ref{fig:regions}(b). Several observations can be inferred. First, the two profiles meet at $\tgc $ indicating a bifurcation of saddle-node type in the qualitative behaviour of solutions to \eqref{eqn:Lambda-const-bdry}. Second, the range $(\ra(g),\rb(g))$ extends to $(0,\infty)$ in the limit $g \to 0$, which is consistent with the findings from the zero gravity case, i.e., no disconnected solution to \eqref{eqn:Lambda-const-bdry} is a steady state. It is expected in fact that as $g$ weakens, the free swarm components of the disconnected solutions tend to be further from the wall, indicating more attractive forces on the wall profile near its ends. Additionally, weaker gravity means less force pushing towards the wall as well, accounting for the widening of the $(\alpha(g),\beta(g))$ region.

On the other hand, the equilibria with mass ratios $\mr \in (0, \ra(g)) \cup (\rb(g),\rc(g))$ have similar properties as those found in case i). Specifically, they are not local minimizers for the energy and they approach a connected equilibrium, with support on both the wall and the interior of $\Om$, in the limit $\mr \to \rc(g)$. The other connected equilibrium corresponds to the limit $\mr \to 0$ and represents an aggregation supported entirely on the wall. The profile of this wall equilibrium does not depend on the particular value of the gravity (see Figure \ref{fig:2dconnected}(a)).

We confirmed all the findings above with particle simulations. In particular, we initialized the particle code at a disconnected state of form \eqref{eqn:2obs-equil}, with mass ratio in the range $(\ra(g),\rb(g))$ and density support provided by the solutions to \eqref{eqn:Lambda-const-bdry}. We observed that indeed, such states are not equilibria, as particles near the end of the wall extent leave the wall and join the free swarm component.

\bigskip
{\em \bf Case $\mathbf{g>\gc}$.} The equilibrium in this case consists in a delta aggregation on the wall, of form \eqref{eqn:wall-sol}, with mass constraint given by \eqref{eqn:massc-onew}. As noted above, the exogenous potential $V$ vanishes on the support $\Omrho=\{0\} \times [-\vs,\vs]$ of $\barrho$, and hence solving \eqref{eqn:equilsup} is identical to the zero gravity case (see \eqref{eqn:equilsup-onew}). We find the same density profile $\f$ illustrated in Figure \ref{fig:2dconnected}(a). By checking \eqref{eqn:equilsup-min} one infers that this equilibrium is a local minimizer of the energy. Given that it is the only equilibrium possible in this case, it is in fact a global minimizer.


\subsection{Calculation of critical gravity $\gc$}
\label{subsect:gc-twod}

To calculate the critical gravity $\gc$ we consider the equilibrium consisting of all mass on the wall -- see \eqref{eqn:wall-sol} and Figure \ref{fig:2dconnected}(a). We then pose the question: How strong does gravity need to be such that a particle placed in the interior of $\Om$ always feels a velocity towards the wall?  We consider such a particle at position $(\epsilon,0)$ with $\epsilon > 0$. We get from \eqref{eqn:vp} that the velocity in the horizontal direction felt by this particle is
\begin{equation*}
\label{eqn:velStart}
v_1 = -\int_{-L}^{L} \left(1 - \left(2\pi(\epsilon^2 + x_2^2)\right)^{-1}\right)\epsilon f(x_2)dx_2 - g.
\end{equation*}

To study the competition between social and gravitational forces one can focus just on the social velocity, defined by
\begin{equation}
\label{eqn:velSocial}
v_1^{s} = -\int_{-L}^{L} \left(1 - \left(2\pi(\epsilon^2 + x_2^2)\right)^{-1}\right)\epsilon f(x_2)dx_2,
\end{equation}
representing the velocity acting on the particle by interaction with the aggregation on the wall. With this notation, the horizontal velocity $v_1$ can be written as
\begin{equation}
\label{eqn:v1-decomp}
v_1 = v_1^s - g.
\end{equation}

We first investigate numerically which $\epsilon$ maximizes the social velocity. To that purpose, we take the approximation of the wall density profile $f(x_2)$ (see Figure \ref{fig:2dconnected}(a)) and evaluate \eqref{eqn:velSocial} for $\epsilon \in (0,1]$. We do not evaluate directly at $\epsilon=0$ because there is a discontinuity there as the particle ceases to feel any velocity in the horizontal direction once $\epsilon=0$. We refer to Section \ref{subsect:ssnum} for details on how $v_1^{s}$ is approximated. 

\begin{figure}[thb]
  \begin{center}
 \includegraphics[width=0.4\textwidth]{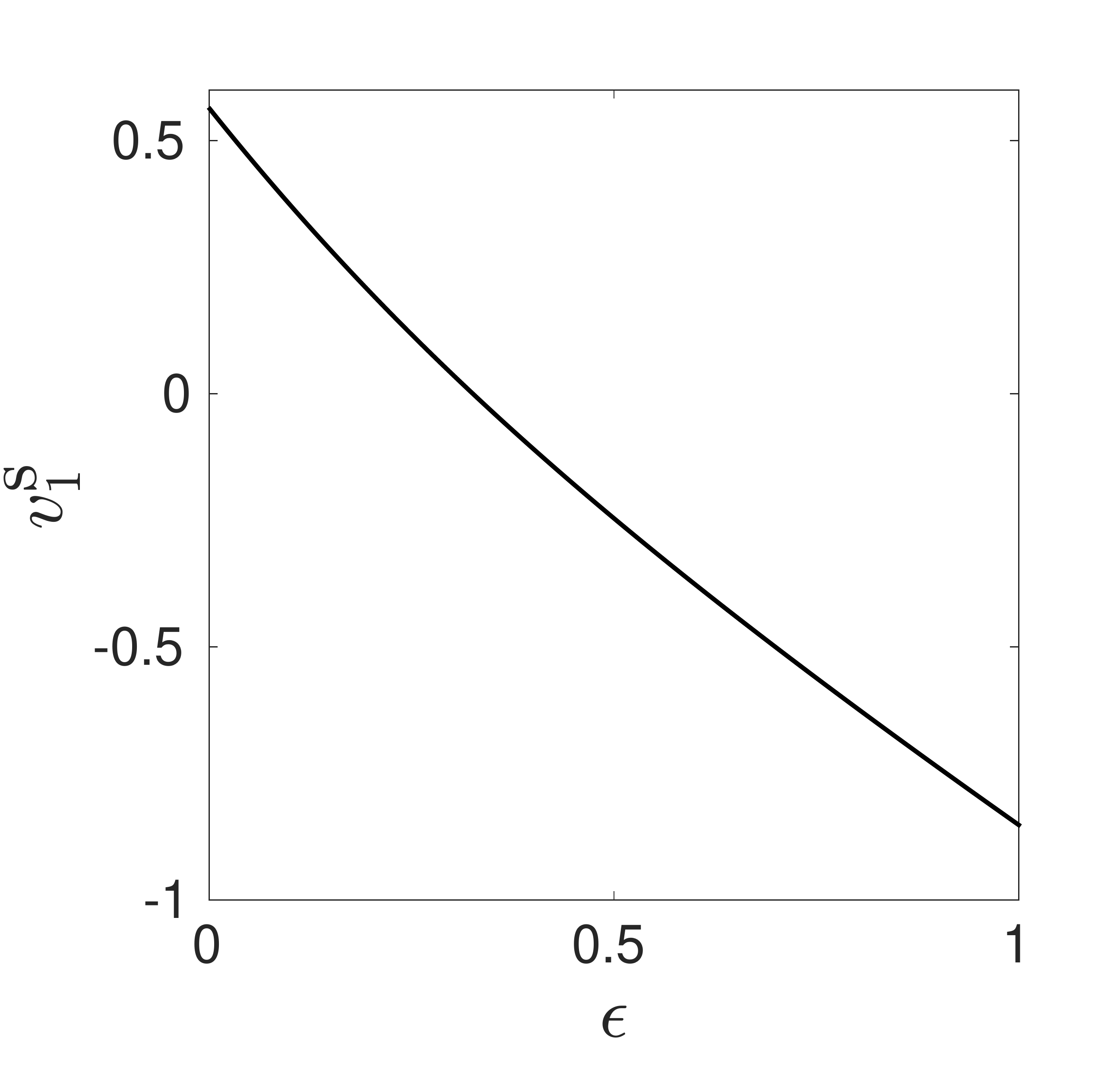} 
\end{center}
\caption{Social velocity \eqref{eqn:velSocial} acting from the aggregation on the wall (see \eqref{eqn:wall-sol} and Figure \ref{fig:2dconnected}(a)) on a particle at position $(\epsilon,0)$.}
\label{fig:velArg}
\end{figure}

The numerical investigation indicates that $v_1^{s}$ is maximized in the limit as $\epsilon \rightarrow 0$ -- see Figure \ref{fig:velArg}. To further cement this evidence, we note that this is expected, as repulsion (corresponding to positive velocity in this case) becomes stronger as distances shrink. Returning to the question we posed at the beginning, by \eqref{eqn:v1-decomp}, gravity is strong enough to yield negative horizontal velocity $v_1$ provided $g$ is larger than the maximal social velocity. Consequently, we set
\begin{equation}
\label{eqn:gc}
\gc = \lim_{\epsilon \to 0} v_1^s.
\end{equation}

In our simulations, an approximation of $\lim_{\epsilon \rightarrow 0}v_1^s$ can already be inferred from Figure \ref{fig:velArg}. A more instructive and explicit formula can be derived however by taking the limit directly in \eqref{eqn:velSocial}:
\begin{align}
\lim_{\epsilon \rightarrow 0}v_1^s 
&= \lim_{\epsilon \rightarrow 0} \int_{-L}^{L} \left(2\pi(\epsilon^2 + x_2^2)\right)^{-1}\epsilon f(x_2)dx_2.
\label{eqn:velLimitCase}
\end{align}
Assume that $f(x_2)$ has a convergent Taylor series centred on $x_2=0$:
\begin{equation}
\label{eqn:velTaylor}
f(x_2) = f(0) + \sum_{n=1}^\infty c_{2n} x_2^{2n}.
\end{equation}
Note that we have used here the symmetry of the wall profile about $x_2=0$. Equations \eqref{eqn:velLimitCase} and \eqref{eqn:velTaylor} then give
\begin{equation}
\lim_{\epsilon \rightarrow 0}v_1^s = f(0) \lim_{\epsilon \rightarrow 0} \int_{-L}^L \frac{\epsilon}{2\pi(\epsilon^2 + x_2^2)}dx_2 + \lim_{\epsilon \rightarrow 0} \int_{-L}^L \sum_{n=1}^\infty \frac{\epsilon c_{2n}x_2^{2n}}{2\pi(\epsilon^2 + x_2^2)}dx_2.
\label{eqn:velLimitAndTaylor}
\end{equation}
Observe that
\begin{align}
\label{eqn:velTermsToZero}
\left|\lim_{\epsilon\rightarrow 0} \epsilon \int_{-L}^L \sum_{n=1}^\infty\frac{c_{2n} x_2^{2n}}{2\pi (\epsilon^2 + x_2^2)} dx_2\right| &\leq \left|\lim_{\epsilon\rightarrow 0} \epsilon \int_{-L}^L \sum_{n=1}^\infty\frac{1}{2\pi}c_{2n} x_2^{2n-2} dx_2\right|, \nonumber \\
&\leq \frac{1}{2\pi} \left|\lim_{\epsilon\rightarrow 0} \epsilon \int_{-L}^L \sum_{n=1}^\infty 2n(2n-1)c_{2n} x_2^{2n-2} dx_2\right|, \nonumber \\
&= 0,
\end{align}
where we have assumed that $f''(x_2)$ has a convergent Taylor series as well. 

Also, by an explicit calculation,
\begin{equation}
\label{eqn:velTerm0}
\lim_{\epsilon \rightarrow 0} \int_{-L}^L \frac{\epsilon}{2\pi(\epsilon^2 + x_2^2)}dx_2 = \frac{1}{2}.
\end{equation}
With \eqref{eqn:velLimitAndTaylor}-\eqref{eqn:velTerm0}, \eqref{eqn:gc} gives
\begin{equation}
\label{eqn:velResult}
\gc = \frac{1}{2}f(0).
\end{equation}
Our numerical simulations yield  $f(0) \approx 1.128$ (cf., Figure \ref{fig:2dconnected}(a)), and hence we find $g_c \approx 0.564$. This value also agrees with Figure \ref{fig:velArg} (as it should).

\subsection{Numerical implementations}
\label{subsect:ssnum}

\subsubsection{Particle method}
\label{subsubsect:ssnumPart}

Consider $N$ particles with positions $x_i$ and velocities $v_i$. In free space, the particle method for model \eqref{eqn:model} is simply implemented by numerically integrating 
\begin{subequations}
\label{eqn:pmfree}
\begin{gather}
\frac{d x_i}{d t} = v_i, \label{eqn:pmfree-x} \\
 \qquad v_i = -\sum_{j\neq i} \nabla K(x_i-x_j) - \nabla V(x_i), \label{eqn:pmfree-v}
\end{gather}
\end{subequations}
with $1 \leq i \leq N$.

In domains with boundaries, one needs to consider the possibility of a particle meeting the boundary within a time step. Let $\Delta t$ denote the time step used in simulations and for simplicity consider an explicit Euler method for time integration. If within a time step, particle $i$ meets the boundary, then, in accordance to \eqref{eqn:vp} and \eqref{eqn:proj}, from the moment of collision it only continues to move in the tangential direction to the boundary.  

For the one dimensional problem on half-line,  this simply means that, had a particle at $x_i$ with velocity $v_i$ reached the origin within a time step $\Delta t$, then it should simply be placed at the origin at the end of the time step. The resulting integrating scheme is then given by
\begin{equation}
\label{eqn:ssnumPart-evolEqn}
x_i(t+\Delta t) = x_i(t) + \Delta t\bar{P}_{x_i} v_i(t),
\end{equation}
where the projection operator $\bar{P}_{x_i}$, which generalizes \eqref{eqn:proj}, is given by:
\begin{equation}
\label{eqn:ssnumPart-1Dproj}
\bar{P}_{x_i} v_i =
\begin{cases}
v_i & \text{ if } x_i + \Delta t v_i \geq 0 \\
-\Delta t^{-1} x_i & \text{ otherwise}.
\end{cases}
\end{equation}

For the two dimensional problem on half-plane we should acknowledge that the vertical velocity of a particle remains unchanged upon colliding with the wall. In this case, a particle $x_i=(x_{i,1},x_{i,2})$ with current velocity $v_i=(v_{i,1},v_{i,2})$ updates its position according to \eqref{eqn:ssnumPart-evolEqn}, except that in two dimensions the discrete projection operator is
\begin{equation}
\label{eqn:ssnumPart-2Dproj}
\bar{P}_{x_i} v_{i} =
\begin{cases}
(v_{i,1}, v_{i,2}) & \text{ if } x_{i,1} + \Delta t v_{i,1} \geq 0 \\
(-\Delta t^{-1} x_{i,1}, v_{i,2}) & \text{ otherwise}.
\end{cases}
\end{equation}
We used several general methods for getting disconnected states:
\begin{itemize}
\item Using initial states that are highly concentrated (very small support) and very close (or adjacent to) the wall. 
\item (1D) Using initial states as in Section \ref{subsubsect:1d-CoMDyn} (in separated form) or as in Section \ref{subsubsect:1d-NumIniMat} (randomly generated from a uniform distribution on a segment in $\Om$).
\item Manually removing particles from the wall and placing them into the free swarm. This allows us to see representations of all disconnected equilibria even if they are not dynamically achievable.
\end{itemize}
Lastly we mention particular issues that can arise with regards to the choice of time step $\Delta t$. There are two phenomena that one can observe:
\begin{itemize}
\item If particles are very concentrated and/or $\Delta t$ is too large then one can observe erratic dynamics where particles are sent far away from the free swarm.
\item Related to the item above, particle methods applied in the manner described here tends to over-approximate the number of particles on the wall in their resultant states, though generally these errors are relatively small and we have investigated the severity through decreasing the time step.
\end{itemize}

\subsubsection{Discretization of the first variation of the energy (equations \eqref{eqn:Lambda-const} and \eqref{eqn:Lambda-const-bdry})}
\label{subsubsect:ssnumLam}

\begin{figure}[ht]
  \begin{center}
 \includegraphics[width=0.46\textwidth]{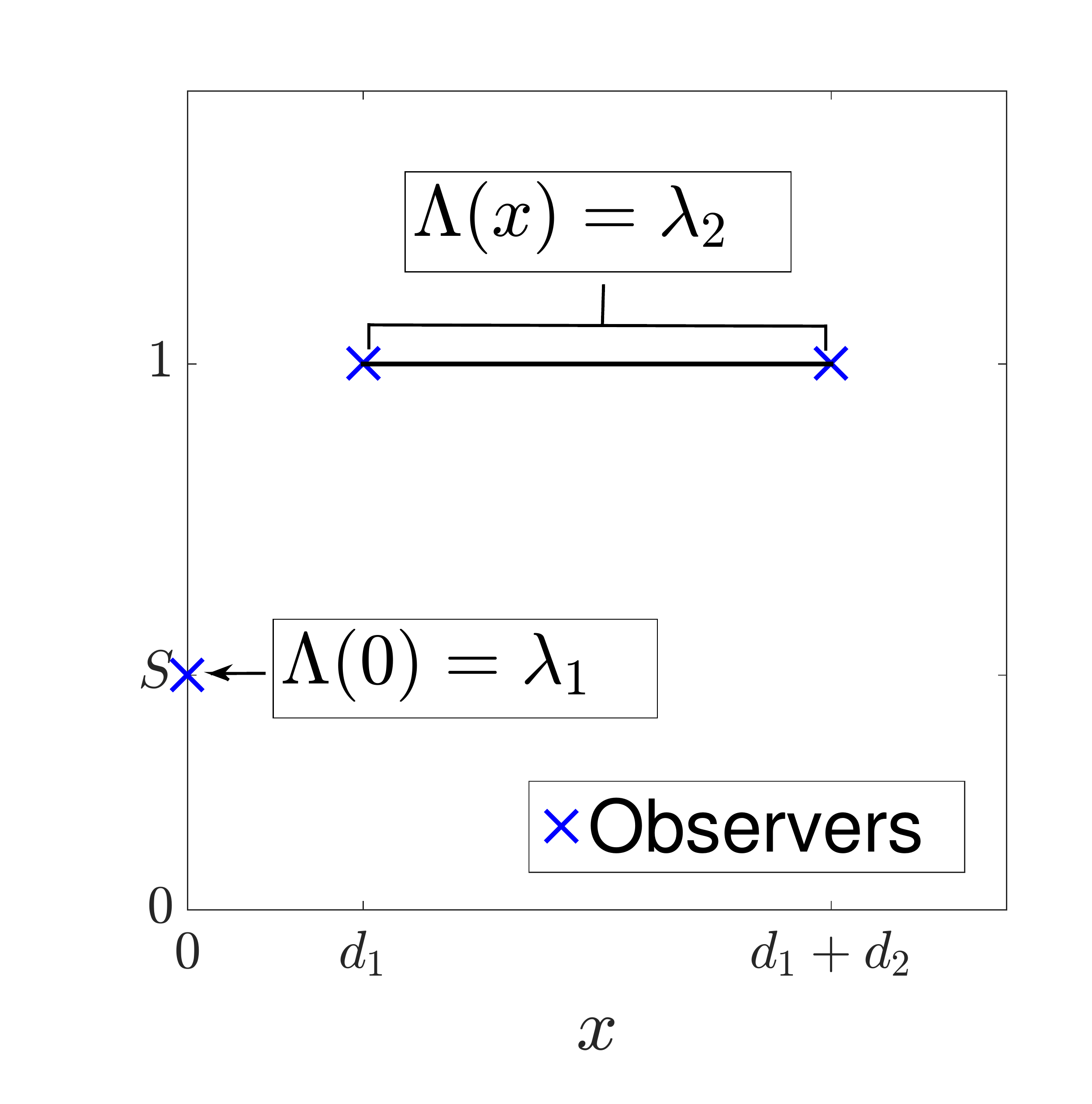} 
\end{center}
\caption{Abstracted solution presumed in the numerical solver in one dimension showing locations of observers where we solve $\Lambda(x)$ to be a constant. Variables for the system are $d_1$, $d_2$, $S$, $\lambda_1$, and $\lambda_2$ as shown in the figure.}
\label{fig:numMethod-example1D}
\end{figure}

{\em $1$D case.} We assume a solution of the form \eqref{eqn:ss-oned} where we now treat $S$, $\dl$, and $\dr$ as variables to be determined by satisfying \eqref{eqn:Lambda-const-bdry}. Furthermore we also focus on the disconnected state, so $\lambda_1$, $\lambda_2$ are also variables. We use the term {\em observers} in this context to describe points along the boundary of $\Omrho$ at which we evaluate $\Lambda(x)$. In one dimension we only require $3$ observers -- see Figure \ref{fig:numMethod-example1D}. Finally, we consider the mass constraint \eqref{eqn:condx2} and the mass ratio \eqref{eqn:mr-1d}; in general we keep $r_M$ and $M$ fixed. 

Then our system of equations encompasses \eqref{eqn:condx2}, \eqref{eqn:mr-1d}, and
\begin{equation}
\label{eqn:numMethod-1DLambda}
\Lambda(0) = \lambda_1, \quad \Lambda(\dl) = \lambda_2, \quad \Lambda(\dl+\dr) = \lambda_2.
\end{equation}
for a total of $5$ equations for $5$ unknowns. We also mention that if one fixes the mass ratio to be that of the minimizer for a given gravity $g$, then the result from solving the system of equations is one with $\left|\lambda_2 - \lambda_1\right|$ and $\dl$ below the tolerance of the solver and so effectively recovering the connected solution.

\medskip
{\em $2$D case - Disconnected.} Recall symmetry about $x_2=0$ so we need only focus on half the space but can extend to the full space using symmetry. We now assume a solution of the form \eqref{eqn:2obs-equil} where $\dl$, $\dr$, $L$, $\lambda_1$, $\lambda_2$, as well as the profiles $f(x_2)$ and $g(x_1)$ need to be determined. We define equidistant vertical and horizontal grids
\begin{equation}
\label{eqn:numMethod-numGrids}
y_i = \frac{L}{N_f}i, \quad 0 \leq i \leq N_f, \quad x_j = \dl + \frac{\dr}{N_g}j, \quad 0 \leq j \leq N_g,
\end{equation}
along with midpoints $y^*_i = \frac{1}{2}(y_{i-1} + y_i)$ for $1 \leq i \leq N_f$ and $x^*_j = \frac{1}{2}(x_{j-1} + x_j)$ for $1 \leq j \leq N_g$. We seek to find the $N_f$ + $N_g$ variables
\begin{equation}
\label{eqn:numMethod-numVars}
f(y^*_i) = f_i, \quad 1 \leq i \leq N_f, \quad g(x^*_j) = g_j, \quad 1 \leq j \leq N_g.
\end{equation}
The profile density $f$ and the free boundary $g$ are then extended with a linear interpolant. 

\begin{figure}[ht]
  \begin{center}
 \includegraphics[width=0.47\textwidth]{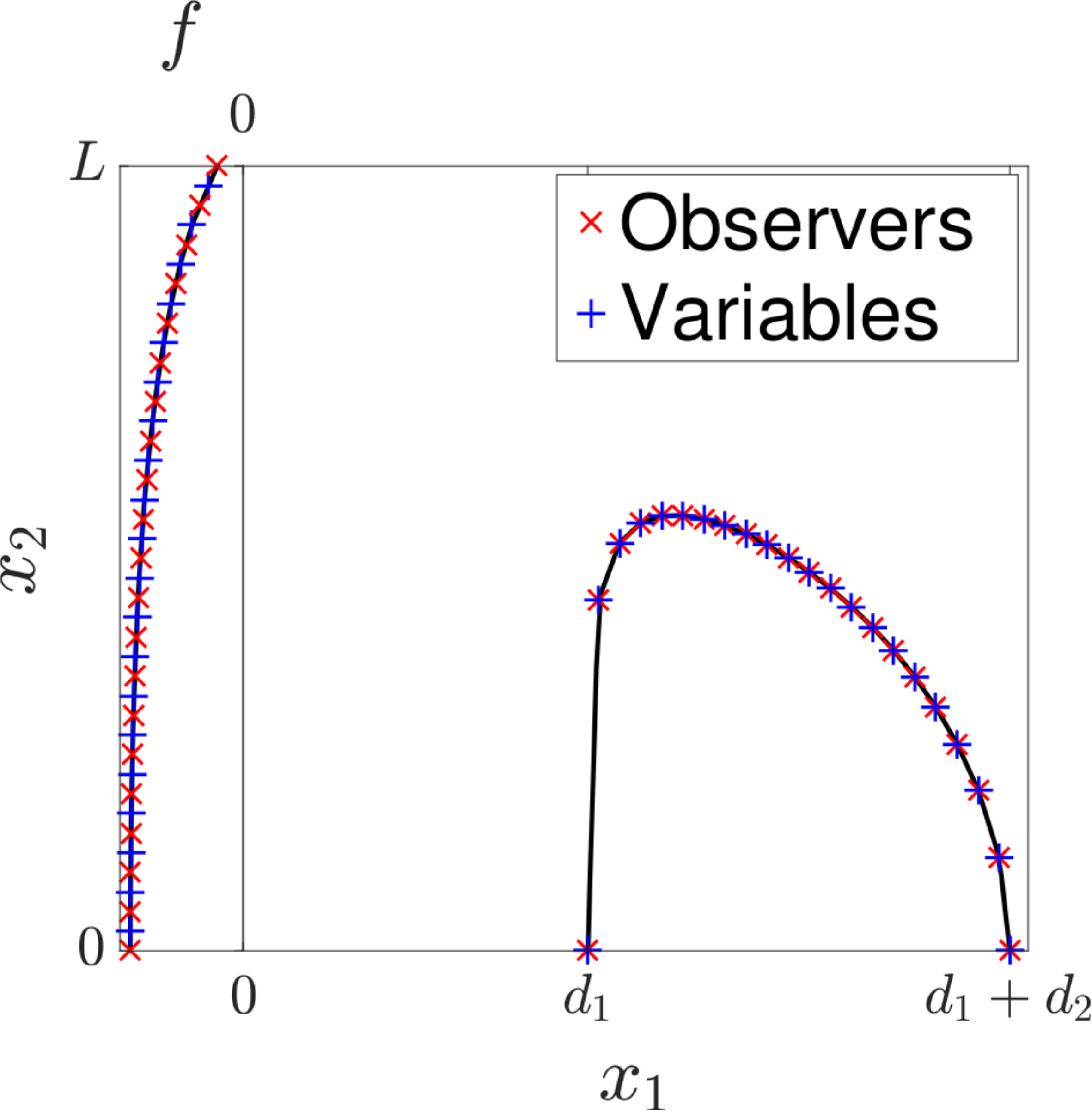} 
 \includegraphics[width=0.45\textwidth]{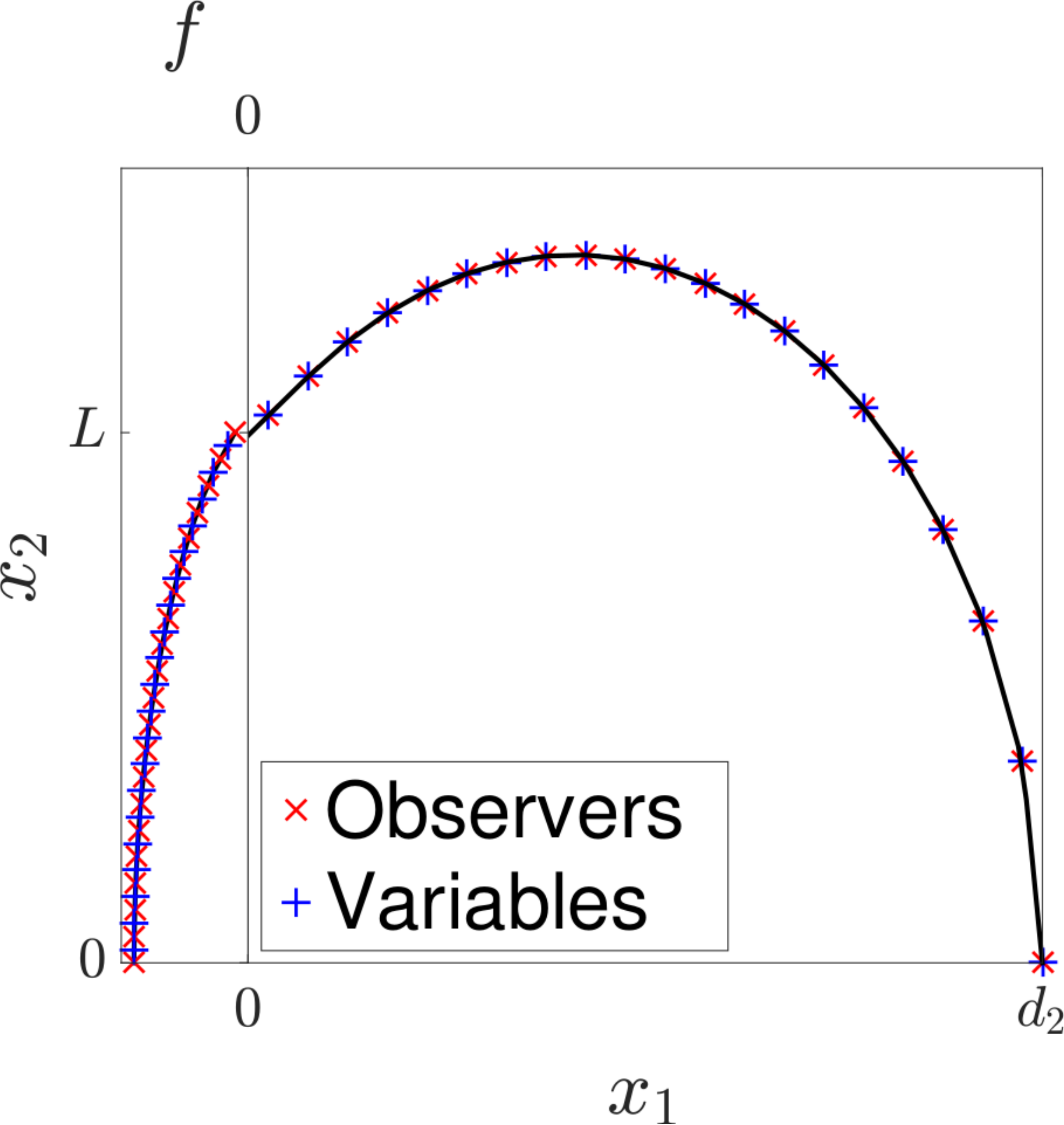} 
\end{center}
\hspace{0.285\textwidth} (a) \hspace{0.435\textwidth} (b)
\caption{Abstracted disconnected (a) and connected (b) solutions presumed in the numerical solver in two dimensions showing locations of observers where we solve $\Lambda(x)$ to be a constant. Variables for the system are $d_1$, $d_2$, $L$, $\lambda_1$, $\lambda_2$, and $f(x_2)$ and $g(x_1)$ evaluated on the numerical grid (see \eqref{eqn:numMethod-numVars}). }
\label{fig:numMethod-example2D}
\end{figure}

To solve for \eqref{eqn:Lambda-const-bdry} we use observers at $(0,y_i)$ for $0 \leq i \leq N_f$, $(x^*_j,g_j)$ for $1 \leq j \leq N_g$, $(\dl,0)$, and $(\dl+\dr,0)$  -- see Figure \ref{fig:numMethod-example2D}(a). We also have the mass constraint \eqref{eqn:massc-twod} and the mass ratio constraint \eqref{eqn:mr}.
Together we have $N_f+N_g+5$ conditions in total and $N_f+N_g+5$ variables.

\smallskip
{\em 2D case - Connected.} The connected state implementation has the same prescription in dealing with $f(x_2)$ but differs for the free boundary $g$. First, we loose the point $(\dl,0)$ as $\dl = 0$ and we do not pin this edge now. Secondly, we drop the mass ratio condition and now we only have $\lambda$ as $\lambda_1 = \lambda_2 =: \lambda$. These are the only differences though and we wind up with $N_f+N_g+4$ conditions and $N_f+N_g+4$ variables -- see Figure \ref{fig:numMethod-example2D}(b).

The system of equations is solved with MATLAB's {\tt fsolve} using default settings and integrals are evaluated with MATLAB's {\tt integral} or {\tt integral2} for $1$D and $2$D integration respectively. When using {\tt integral2} we use the iterated method setting.


\bibliographystyle{plain}
\bibliography{lit}

\newpage

\section{Appendix}
\label{sect:appendix}

We provide below the six equations derived from \eqref{eqn:morseCond}. The four equations that ensure $\Lambda(x) = \lambda_2$ for $x \in [\dl,\dl+\dr]$ are:
\begin{align*}
&\frac{C}{L^{-2} + \mu^2}\exp\left(\frac{\dl}{L}\right)\Big(-\frac{1}{L}\cos(\mu\dl) - \mu\sin(\mu\dl)\Big) + \ldots \nonumber \\
&\qquad \frac{D}{L^{-2} + \mu^2}\exp\left(\frac{\dl}{L}\right)\Big(-\frac{1}{L}\sin(\mu\dl) + \mu\cos(\mu\dl)\Big) + \frac{L\lambda_2}{\epsilon}\exp\left(\frac{\dl}{L}\right) + S = 0, \\
&\frac{C}{1 + \mu^2}\exp\left(\dl\right)\Big(-\cos(\mu\dl) - \mu\sin(\mu\dl)\Big) + \ldots \nonumber \\
&\qquad \frac{D}{1 + \mu^2}\exp\left(\dl\right)\Big(-\sin(\mu\dl) + \mu\cos(\mu\dl)\Big) + \frac{\lambda_2}{\epsilon}\exp\left(\dl\right) + S = 0, \\
&\frac{C}{L^{-2} + \mu^2}\exp\left(-\frac{\dl+\dr}{L}\right)\Big(-\frac{1}{L}\cos(\mu(\dl+\dr)) + \mu\sin(\mu(\dl+\dr))\Big) + \ldots \nonumber \\
&\qquad \frac{D}{L^{-2} + \mu^2}\exp\left(-\frac{\dl+\dr}{L}\right)\Big(-\frac{1}{L}\sin(\mu(\dl+\dr)) - \mu\cos(\mu(\dl+\dr))\Big) + \frac{L\lambda_2}{\epsilon}\exp\left(\frac{-(\dl+\dr)}{L}\right) = 0,\\
&\frac{C}{1 + \mu^2}\exp\left(-(\dl+\dr)\right)\Big(-\cos(\mu(\dl+\dr)) + \mu\sin(\mu(\dl+\dr))\Big) + \ldots \nonumber \\
&\qquad \frac{D}{1 + \mu^2}\exp\left(-(\dl+\dr)\right)\Big(-\sin(\mu(\dl+\dr)) - \mu\cos(\mu(\dl+\dr))\Big) + \frac{\lambda_2}{\epsilon}\exp\left(-(\dl+\dr)\right) = 0. 
\end{align*}


The equation that ensures $\Lambda(0) = \lambda_1$ is
\begin{align*}
&-GL\Big(\frac{C}{L^{-2}+\mu^2}\exp\left(-\frac{y}{L}\right)\Big(-\frac{1}{L}\cos(\mu y) + \mu\sin(\mu y)\Big) + \frac{D}{L^{-2}+\mu^2}\exp\left(-\frac{y}{L}\right)\Big(-\frac{1}{L}\sin(\mu y) - \mu\cos(\mu y)\Big) + ... \\
&\qquad \frac{L\lambda_2}{\epsilon}\exp\left(-\frac{y}{L}\right)\Big)\Big|_{y=\dl}^{y=\dl+\dr} + \Big(\frac{C}{1+\mu^2}\exp(-y)\Big(-\cos(\mu y) + \mu\sin(\mu y)\Big) + \ldots \nonumber \\
& \qquad \frac{D}{1+\mu^2}\exp(-y)\Big(-\sin(\mu y) - \mu\cos(\mu y)\Big) + \frac{\lambda_2}{\epsilon}\exp(-y)\Big)\Big|_{y=\dl}^{y=\dl+\dr} + S(1-GL) = \lambda_1.
\end{align*}

Finally, the mass constraint equation gives
\begin{align*}
S + \frac{C}{\mu}(\sin(\mu(\dl+\dr)) - \sin(\mu\dl)) - \frac{D}{\mu}(\cos(\mu(\dl+\dr))-\cos(\mu\dl)) - \frac{\lambda_2\dr}{\epsilon} = M.
\end{align*}


\end{document}